\input harvmac
\input psfig
\newcount\figno
\figno=0
\def\fig#1#2#3{
\par\begingroup\parindent=0pt\leftskip=1cm\rightskip=1cm\parindent=0pt
\global\advance\figno by 1
\midinsert
\epsfxsize=#3
\centerline{\epsfbox{#2}}
\vskip 12pt
{\bf Fig. \the\figno:} #1\par
\endinsert\endgroup\par
}
\def\figlabel#1{\xdef#1{\the\figno}}
\def\encadremath#1{\vbox{\hrule\hbox{\vrule\kern8pt\vbox{\kern8pt
\hbox{$\displaystyle #1$}\kern8pt}
\kern8pt\vrule}\hrule}}
\def\underarrow#1{\vbox{\ialign{##\crcr$\hfil\displaystyle
 {#1}\hfil$\crcr\noalign{\kern1pt\nointerlineskip}$\longrightarrow$\crcr}}}
% use of underarrow
%A~~~\underarrow{a}~~~B
%
\overfullrule=0pt

%macros
%
\def\tilde{\widetilde}
\def\bar{\overline}

\def\T{{\bf T}}
\def\S{{\bf S}}
\def\R{{\bf R}}

\font\zfont = cmss10 %scaled \magstep1

\def\bigone{\hbox{1\kern -.23em {\rm l}}}
\def\ZZ{\hbox{\zfont Z\kern-.4emZ}}
\def\half{{\litfont {1 \over 2}}}

\def\a{{\alpha}}
\def\b{{\beta}}
\def\g{{\gamma}}
\def\d{{\delta}}

\Title{ \vbox{\baselineskip12pt
\hbox{hep-th/9902098}
\hbox{IFT-P.012/99}
\hbox{HUTP-99/A004}
\hbox{IASSNS-HEP-99-5}}}
{\vbox{\centerline{Conformal Field Theory Of AdS Background }
\bigskip
\centerline{  With Ramond-Ramond Flux }}}
\smallskip
\centerline{Nathan Berkovits}
\smallskip
\centerline{\it 
Instituto de F\'{\i}sica Te\'orica, Universidade Estadual Paulista}
\centerline{\it
Rua Pamplona 145, 01405-900, S\~ao Paulo, SP, Brasil}
\smallskip
\centerline{Cumrun Vafa}
\smallskip
\centerline{\it Jefferson Laboratory of Physics}
\centerline{\it Harvard University, Cambridge, MA 02138, USA}
\centerline{Edward Witten}
\smallskip
\centerline{\it School of Natural Sciences, Institute for Advanced Study}
\centerline{\it Olden Lane, Princeton, NJ 08540, USA}\bigskip

\medskip

\noindent
We review a formalism of superstring quantization with manifest six-dimensional
spacetime supersymmetry, 
and apply it to ${\rm AdS}_3\times {\bf S}^3$ backgrounds with Ramond-Ramond
flux.  The resulting description is a conformal field theory
based on a sigma model whose target space is a certain supergroup
$SU'(2|2)$.

\Date{February, 1999}
%text of paper

\newsec{Introduction}

In the RNS formulation of superstring theory, powerful
conformal field theory methods of computation
are available.  A price one pays is that spacetime supersymmetry
is not visible as a classical symmetry of the worldsheet action.
It can be seen by incorporating spin fields and picture changing
\ref\fms{D. Friedan, E. Martinec, and S. Shenker, ``Conformal Invariance,
Supersymmetry, and String Theory,'' Nucl. Phys. {\bf B271} (1986) 93.}.
On the other hand, the Green-Schwarz description \ref\gs{M. B. Green
and J. H. Schwarz, ``Covariant Description Of Superstrings,''
Phys. Lett. {\bf 136B} (1984) 367.} makes spacetime supersymmetry manifest,
but quantization becomes difficult, except in light cone gauge.

\def\R{{\bf R}}
Though a fully covariant quantization of the Green-Schwarz string
is not known, one does know a partial substitute in that it is possible
to reformulate the RNS description in terms of Green-Schwarz-like variables,
in a way that makes manifest a portion of the spacetime symmetry group.
For example \ref\fourreview{N. Berkovits,  ``Covariant Quantization Of
The Green-Schwarz Superstring In A Calabi-Yau Background,''
Nucl. Phys. {\bf B431} (1994) 258, ``A New Description Of The Superstring,''
Jorge Swieca Summer School 1995, p. 490, hep-th/9604123.}, 
one can exhibit four-dimensional Poincar\'e invariance
with manifest $N=1$ supersymmetry (or, for Type II superstrings, $N=1$
from left-movers and $N=1$ from right-movers), giving a framework in which
one can study superstring compactification on a Calabi-Yau manifold,
with all symmetries more or less manifest.
By a further extension of these ideas \ref\Topo 
{N. Berkovits and C. Vafa,
{\it $N=4$ Topological Strings}, Nucl. Phys. B433 (1995) 123, hep-th/9407190.},
one can exhibit six-dimensional $N=1$ super-Poincar\'e invariance
(or $N=1$ for left-movers and $N=1$ for right-movers), giving a framework
in which all symmetries of superstring compactification on a K3 surface
are visible.
These constructions are hybrids in which the spacetime symmetries
of four or six of the ten dimensions are described in Green-Schwarz-like
variables, while the remaining dimensions are described using RNS variables.
These constructions may thus loosely, but only loosely, be described
as covariant quantization of the Green-Schwarz string in four or six 
dimensions. There also exists a quantization of the ten-dimensional 
superstring which,
after Wick-rotating, preserves manifest $U(5)$ super-Poincar\'e invariance
\ref\ufive{N. Berkovits, ``Quantization of the Superstring with Manifest
U(5) Super-Poincar\'e Invariance'', 
hep-th/99020??.}.

One important property of the RNS 
approach to Type II superstrings is that in the RNS case, some of the
bosonic fields -- those from the Ramond-Ramond sector -- are represented
by spin fields.  As a result, it is exceedingly difficult to describe
and understand backgrounds in which Ramond-Ramond or RR 
fields are present in the
vacuum. At best, one would have to describe such backgrounds by worldsheet
Lagrangians with spin fields added to the action, leading at least
generically to rather untransparent formulas.  In addition, there are several
questions of principle.  A narrow one is simply that one must ask in which
picture or pictures one should take the Ramond-Ramond vertex operators
that are added to the Lagrangian.  A broader question is this:  What is
supposed to replace worldsheet superconformal symmetry?  Superconformal
invariance is usually understood as the guiding principle of the RNS 
description
of perturbative superstrings.   However, in the presence of RR
fields, the superconformal symmetry is violated (as the worldsheet 
supercurrents
are not local with respect to a Ramond-Ramond field), and it is not at 
all clear even in principle what would constitute a satisfactory worldsheet
action with RR fields.

The partially covariant methods mentioned before give an answer
in principle, at least in certain special cases.  If the
RR fields live only in the four or six dimensions whose spacetime
symmetries are manifest in the constructions of \fourreview\ or \Topo,
then their vertex operators are ordinary untwisted worldsheet operators
and can be added to the Lagrangian while sticking within a known framework
in which the rules are all clear.  
For example, to leading order in an $\alpha'$ expansion, a sigma model
containing four-dimensional RR fields has been constructed in
\ref\warrenberk{N. Berkovits and W. Siegel, ``Superspace Effective
Actions For $4-D$ Compactifications Of Heterotic And Type II
Superstrings,'' Nucl. Phys. {\bf B462} (1996) 213.}.  
Although explicit $\beta$-function computations have not yet been done
for this sigma model representing the Type II superstring, 
there have been $\beta$-function computations \ref\sken
{J. de Boer and K. Skenderis, ``Covariant Computation of the
Low Energy Effective Action of the Heterotic Superstring'',
Nucl. Phys. B481 (1996) 129.} in $D=4$ superspace 
for a closely related sigma model representing the heterotic superstring.
The analogous sigma model construction
has not been made in six dimensions either for the heterotic or Type II
superstring.

The problem of describing conformal field theories with RR
fields has acquired greater currency because of
the much-studied dualities \ref\malda{J. Maldacena, ``The Large $N$ Limit
Of Superconformal Field Theories And Supergravity,'' Adv. Theor. Math.
Phys. {\bf 2} (1998) 231.} between string
theories on Anti de Sitter or AdS backgrounds that contain RR fields and
certain conformal field theories.  In the context of these dualities,
it would be highly desireable to be able to construct conformal field
theories representing  backgrounds with RR fields.  In the absence of
an ability to do so, one is limited to studying certain limits of AdS
theories in which supergravity is an adequate approximation.

\def\S{{\bf S}}
The purpose of the present paper is to use the  covariant Green-Schwarz-like
methods mentioned above to construct a conformal field theory description
of a particular AdS background with RR fields.  This will be
the Type IIB compactification on ${\rm AdS}_3\times \S^3\times {\rm K3}$
(or ${\rm AdS}_3\times \S^3\times {\bf T}^4$).  
In fact, we will study the ${\rm AdS}_3\times {\bf S}^3$ examples
with a general mixture of RR and Neveu-Schwarz fluxes.
Though the details are
complicated, the general outline of the answer that we will get
is easily stated.  ${\rm AdS}_3\times \S^3$ will be represented by a sigma
model whose target space is a certain supergroup that we call $SU'(2|2)$;
this sigma model is conformally invariant because of special properties of
that supergroup.  
The sigma model with target space $SU'(2|2)$ 
depends on two parameters, which correspond physically to the
RR and NS fluxes.  For any values of the fluxes, this sigma model
is highly
nonunitary, because for example the worldsheet fermions have a second
order kinetic energy; physical states will be those obeying a certain
fairly elaborate set of constraints.  Except in the special case
that the RR flux vanishes, the $SU'(2|2)$ sigma model must be supplemented
with couplings of matter fields to ghost-like fields. 

\nref\mava{J. Maldacena and A. Strominger, 
``${\rm AdS}_3$ Black Holes And  A Stringy Exclusion Principle,''
hep-th/9804085.}
\nref\gks{A. Giveon, D. Kutasov, and N. Seiberg, ``Comments On String
Theory On ${\rm AdS}_3$,'' hep-th/9806194.}  
\nref\ooguri{J. de Boer, H. Ooguri, H. Robins, and J. Tannenhauser, ``String
Theory On ${\rm AdS}_3$,'' hep-th/9812046.}
We should mention other approaches that give somewhat complementary
information.  The ${\rm AdS}_3\times \S^3$ models have cousins with NS fields
\mava\ that can be described as conformal field theories using RNS variables
\refs{\gks, \ooguri}.  
One important ingredient in this approach is the $SL(2,{\bf R})$ WZW model,
which has been studied from many points of view (for example,
see \ref\umany{
J. Balog, L. O'Raifeartaigh, P. Forgacs and A. Wipf, ``Consistency of
String Propagation on Curved Space-Times: An SU(1,1) Based Counterexample'',
Nucl. Phys. B325 (1989) 225\semi
P.M.S. Petropoulos, ``Comments on SU(1,1) String Theory'', Phys. Lett. B236
(1990) 151\semi 
I. Bars and D. Nemeschansky, ``String Propagation In
Backgrounds With Curved Space-Time,'' Nucl. Phys. {\bf B348} (1991) 89 \semi
M. Henningson, S. Hwang, P. Roberts, and B. Sundborg,
 ``Modular Invariance Of
$SU(1,1)$ Strings,'' Phys. Lett. {\bf B267} (1991) 350 \semi
S. Hwang, ``No-Ghost Theorem for SU(1,1) String Theories'', Nucl. Phys. B354
(1991) 100\semi
I. Bars, ``Ghost-Free Spectrum Of A Quantum String In $SL(2,\R)$ Curved
Space-Time, Phys. Rev. {\bf D53} (1996) 3308, hep-th/9503205,
``Solution Of The $SL(2,\R)$ String In Curved Spacetime,''
in {\it Future Perspectives In String Theory} (Los Angeles, 1995),
hep-th/9511187 \semi
J. Teschner, ``The Minisuperspace Limit Of The $SL(2,{\bf C})/SU(2)$
WZNW Model,'' hep-th/9712258 \semi
J. M. Evans, M. R. Geberdiel, and M. J. Perry, ``The No-Ghost Theorem
And Strings On ${\rm AdS}_3$,'' hep-th/9812252.}).
Also, there are extremely elegant descriptions of
{\rm AdS} backgrounds with RR fields, including 
${\rm AdS}_3\times \S^3$, at the classical
level (for example, see \ref\who{R. R. Metsaev and A. A. Tseytlin,
``Type IIB Superstring Action In ${\rm AdS}_5\times \S^5$ Background,''
Nucl. Phys. {\bf B533} (1998) 109, hep-th/9805028\semi
R. Kallosh, J. Rahmfeld, and A. Rajaraman,
``Near Horizon Superspace,'' hep-th/9805217 \semi
R. Kallosh and J. Rahmfeld, ``The GS String Action On ${\rm AdS}_5\times 
\S^5$,'' Phys. Lett. {\bf B443} (1998) 143, hep-th/9808038 \semi
I. Pesando, ``A k Gauge Fixed Type IIB Superstring
Action On ${\rm AdS}_5\times \S^5$, JHEP 11 (1998) 002, hep-th/9808020\semi
I. Pesando, ``The GS Type IIB Superstring Action On ${\rm AdS}_3\times \S^3
\times {\bf T}^4$, hep-th/9809145\semi
J. Rahmfeld and A. Rajaraman, ``The GS String Action On ${\rm AdS}_3\times 
\S^3$
with Ramond-Ramond Charge,'' hep-th/9809164 \semi
J. Park and S.-J. Rey, ``Green-Schwarz Superstring On ${\rm AdS}_3\times
{\bf S}^3$, hep-th/9812062.}).  Those descriptions
involve strings in superspace, usually a superspace with more odd variables
than will appear in our discussion, and with compensating $\kappa$
symmetry.  There have also been  attempts at ${\rm AdS}_3\times \S^3$
quantization in light cone and temporal gauges \ref\ming{M. Yu and B. Zhang,
``Light-Cone Gauge Quantization Of String Theories On ${\rm AdS}_3$ Space,''
hep-th/9812216\semi A. Rajaraman and M. Rozali, ``On the Quantization
of the GS String on ${\rm AdS}_5\times \S^5$'', hep-th/9902046.}.

This paper is organized as follows.  Because the methods we will use
are comparatively little-known, we begin in sections 2-5 with
an unusually detailed review and summary of previous work.
The reader should {\it not} assume that everything in these sections
(especially sections 2 and 3)
is needed for understanding the rest of the paper.
In section 6, we begin with $\R^6\times {\rm K3}$ (or $\R^6\times {\bf T}^4$)
and describe vertex operators that, locally and to first order,
deform $\R^6$ to ${\rm AdS}_3\times {\bf S}^3$ with RR fields.
Many general lessons become clear from this linearized treatment.
Using those lessons, we construct in section 7
the conformal field theory -- as mentioned before, a sigma model
with target a certain supergroup -- that describes ${\rm AdS}_3\times \S^3$.
Then in section 8, we analyze the necesssary ghost couplings and show
their consistency with the framework reviewed in sections 2-5.
In section 9, we define the conditions which physical vertex operators
have to satisfy in the conformal field theory. 
In section 10, we study the ${\rm AdS}_3\times {\bf S}^3$ models with
NS background in the same framework.  Though this is ``unnecessary,'' in the
sense that conformal field theory descriptions of these backgrounds are
already known, our approach has the virtue of making spacetime supersymmetry
manifest -- at the cost, we     must admit, of a certain complexity.

There is one point of physics that we should perhaps mention here.
One important route to the ${\rm AdS}_3\times \S^3$ backgrounds with
RR flux begins with $\R^6\times {\rm K3}$ (or $\R^6\times {\bf T}^4)$
with parallel  Dirichlet onebranes and fivebranes in $\R^6$ (the fivebranes
being wrapped on K3 or ${\bf T}^4$ to make strings in $\R^6$).  Let 
$Q_1$ and $Q_5$ be the number of such strings or fivebranes. 
(Physically, one can also add threebranes wrapped on two-cycles in K3
or ${\bf T}^4$ to make additional strings, but in this case
the methods we will use are much less effective, as the resulting RR flux does
not ``live'' just in $\R^6$.)  These integer variables determine the
amount of RR flux on the $\S^3$ factor in ${\rm AdS}_3\times \S^3$, and also
the volume of the K3 or ${\bf T}^4$.  In our work, however, we will not
see any quantization of either of those variables.  One way to explain
the reason is as follows.  Let $\lambda$ be the string coupling constant.
The number of RR flux quanta needed to make an ${\rm AdS}_3\times \S^3$ 
background
that remains fixed as $\lambda\to 0$ is of order $1/\lambda$.  Since
(by doing conformal field theory) we are working in the small $\lambda$
limit, the integrality of a quantity that is of order $1/\lambda$
is not readily visible.  For example, the number of wrapped
Dirichlet fivebranes equals the number of RR flux
quanta on $\S^3$.  To see integrality of this number, one should study
$D$-strings on ${\rm AdS}_3\times \S^3$, rather than the elementary strings
that we will in fact be studying.  

By contrast, integrality of the NS flux can be probed by elementary
strings.  If one considers on $\R^6\times {\rm K3}$ NS fivebranes
as well as Dirichlet fivebranes, one gets an ${\rm AdS}_3\times \S^3$ model
with NS as well as RR flux.  The NS flux appears as a Wess-Zumino
coupling in the $SU'(2|2)$ sigma model, which is quantized for topological
reasons.  Integrality of the number of NS fivebranes is thus manifest
in our formalism.

Some of the massless scalar moduli in $\R^6\times {\rm K3}$ (or  similarly
$\R^6\times {\bf T}^4$) get masses when one goes to ${\rm AdS}_3\times {\bf 
S}^3
\times {\rm K3}$.  In fact, one scalar in each tensor multiplet gets
mass.  This is a consequence of the structure of the ${\rm AdS}_3\times \S^3$
supermultiplets, and hence will automatically be true in the framework
we are developing here.  However, we will not describe the vertex
operators for disturbances in ${\rm AdS}_3\times {\bf S}^3$ in the present
paper and so will not see the phenomenon explicitly.

\def\e{{\epsilon}}
\def\s{{\sigma}}
\def\N{{\nabla}}
\def\half{{1\over 2}}
\def\p{{\partial}}
\def\pb{{\bar\partial}}
\def\t{{\theta}}
\def\Gtp{{\tilde G^+}}
\def\Gtm{{\tilde G^-}}
\def\tb{{\bar\theta}}

\newsec{$N=2$ and $N=4$ Topological Strings}

The aim of this section is to recall the definition of $N=2$
\ref\witnet{E. Witten,``On the Structure of the Topological Phase of
Two-Dimensional Gravity,''
 Nucl. Phys. {\bf B340} (1990) 281.}
 and $N=4$ \Topo\
 topological  strings.  Then we will proceed in  section 3
 to review how the superstring can be formulated in the language of
$N=4$ topological string theory, and in section 4 to review
 how this formalism can be used to give a description of the superstring
 with manifest $D=6$ spacetime supersymmetry.
In section 5, we will review the massless vertex operators in this
spacetime-supersymmetric description of the superstring.

\subsec{$N=2$ Topological Strings}
The $N=2$ topological string is modeled after ordinary
$N=0$ bosonic strings.
Thus to motivate it, we first recall what general ingredients
are needed for defining a bosonic string theory.

In bosonic string perturbation theory, the key elements are a two-dimensional
conformal field theory with certain additional structures.\foot{We 
will concentrate on left-movers
and will denote right-movers with barred notation.}  One requires
a spin 1 fermionic BRST current $j_{BRST}$; the charge associated
to it is the BRST operator
$$Q_{BRST}=\oint j_{BRST}$$
and obeys $Q_{BRST}^2=0$.  $j_{BRST}$ has charge one with respect to a 
``ghost number'' symmetry, which is
generated by a ghost current $j_{ghost}$ that
has an anomaly $3g-3$ on a Riemann surface of genus $g$.
In addition there is a spin $2$ antighost field $b$ with ghost number $-1$,
which satisfies 
\eqn\brsm{\{ Q ,b(z)\}=T(z)}
where $T$ is the total energy momentum tensor of the system
 (including both the matter and ghost system).  $T$ has central charge 0.
Moreover  $(b)^2\sim 0$ (in the sense that there is no short distance
singularity in the $b\cdot b$ operator product).
The {\it physical fields} are defined relative to $Q_{BRST}$ cohomology.
A physical field is defined to be a ghost number 1 field $\Phi^+$ that obeys
$$\{ Q_{BRST},\Phi^+ \}=0,$$
with the gauge equivalence relation being
$$\Phi^+ \sim \Phi^+ +\{ Q_{BRST}, \Lambda \}. $$
   For conventional bosonic
strings, one can prove from the structure of the BRST cohomology that
every physical field, modulo the gauge equivalence relation,
has a representative that is a primary field of the Virasoro algebra of
dimension 0.
We will assume in general that $\Phi^+$ has this property.
Let $b_{-1}\Phi ^+$
denote the single pole in the OPE of $b$ with $\Phi^+$. (In what follows,
in general, if $O$ is an
operator  of dimension $d$, 
$O_n \Phi$ will signify the pole of order $d+n$ in the OPE of $O$ with
$\Phi$.)
Then the operator $\int b_{-1}\Phi^+$ 
has dimension one and so is a marginal deformation that can be used
to give a deformation of the action.
By virtue of the properties explained above, 
the operator $\int b_{-1}\Phi^+$ has ghost number zero, is
BRST invariant modulo a surface term,
and commutes with $b$.  So adding it to the action 
 preserves the structure we have assumed for the definition
of the theory, namely the conformal invariance, BRST invariance,  ghost
symmetry, and the existence of $b$.

Now, let us consider the theory formulated on
 a genus $g$ Riemann surface. The moduli space of
such surfaces is a complex $3g-3$ dimensional space
${\cal M}_g$.  The tangent vectors on it are identified
with Beltrami differentials $\mu_{\bar z}^z$
on the corresponding Riemann surface. Beltrami differentials
are equivalent if they differ by ${\bar \partial} v^z$
where $v^z$ is a globally defined vector field on the Riemann surface.
Modulo this relation, the space of 
 Beltrami differentials has dimension $3g-3$ (for $g>1$).
Note that since $b$ is a spin $2$ current, i.e., it is
a quadratic differential on the Riemann surface, the integral
$$b(\mu)=\int b_{zz}\mu^z_{\overline z}$$
makes sense on the Riemann surface.  The $N$-point genus $g$ bosonic string
scattering amplitude is defined by
\eqn\bpar{F_g=\int_{{\cal M}_g}\prod_i\langle \big|\prod_{i=1}^{3g-3}
b(\mu_i)
\big|^2
\prod_{r=1}^{N} \int b_{-1} \bar b_{-1} \Phi_{(r)}^{++} 
 \rangle}
where 
the $\mu_i$ fix the volume element on ${\cal M}_g$ and 
$\Phi_{(r)}^{++}$ satisfies the physicality conditions with respect to both 
left and right-movers.  
It is easy
to check that this definition is consistent with the notion of physical
states defined above.  In particular, if we add a term
$Q\Lambda$ to one of the $\Phi_{(r)}^+$, this does not
affect the genus $g$ partition function.  Also note that in order for
$F_g$ not to be trivially zero, the ghost current needs to have an anomaly.
In particular, there must be $3g-3$ units of ghost charge generated
by the path-integral measure in order to balance the $3g-3$ anti-ghost
$b$ fields in the definition of $F_g$.

Now we are ready to discuss $N=2$ topological strings.
Start with any $N=2$ superconformal field theory.  This will have currents
 $T$, 
$G^{\pm}$, and $J$
 of spins $2,{3\over 2},$ and $1 $,
respectively, where $T$ is the energy momentum tensor, $J$ is the $U(1)$
current, and $G^\pm$ are the two supercurrents with charge
$\pm 1$ under $J$. The energy-momentum tensor can be modified or
``twisted'' by
$$T\rightarrow T'=T-\half \partial J.$$
(or the same formula with $+\half \partial J$ instead).
One can show that $T'$ generates a Virasoro algebra.
Passing from $T$ to $T'$ shifts the dimension of every field by $-1/2$
its $U(1)$ charge.
So with
respect to $T'$,  $G^+$ has spin 1 and $G^-$ has spin 2.
Moreover, the modified energy momentum tensor 
has zero central charge no matter what was the central
charge $\hat c$ of the underlying $N=2$  theory that we started with.
Now we see we have exactly the same structure as needed in the definition
of bosonic string with the following identifications:
$$T\rightarrow T'$$
$$j_{BRST}\rightarrow G^+$$
$$b\rightarrow G^-$$
$$j_{ghost}\rightarrow J.$$
In particular, \bpar\ now takes the form
\eqn\netp{F_g=\int_{{\cal M}_g}\prod_i\langle \big|\prod_{i=1}^{3g-3}
G^-(\mu_i)
\big|^2
\prod_{r=1}^{N} \int G_{-1}^- \bar G_{-1}^- \Phi_{(r)}^{++} 
 \rangle}
Translating the conditions for the physical fields, we see that
they correspond to chiral primary fields with charge $+1$ and dimension
$0$ (they arise from operators that have charge 1 and dimension
 $\half $ before twisting).  
 
 Note also that nonvanishing of the partition
function puts a strong constraint on the anomaly of the $U(1)$ current.
Unless the anomaly of the $N=2$ $U(1)$ current
is $3g-3$ at genus $g$, the partition function trivially
vanishes and we must consider correlation functions
of suitable physical states instead.  If the anomaly is precisely $3g-3$,
the partition function will generally not vanish.
 The anomaly arises from a two point function $J\cdot T'= J\cdot
 (T-\half \partial J)$.  Since the $J\cdot T$ two point function vanishes
(by virtue of the underlying $N=2$ algebra), the anomaly comes
from $J\cdot \partial J$, and hence is proportional to the central charge
$\hat c$ of the $N=2$ algebra.
 The anomaly is in fact given precisely by
by ${\hat c}(g-1)$, so ${\hat c}=3$ is the case for which
the partition function
need not a priori vanish.  An important example of an $N=2$ model
with ${\hat c}=3$
is the supersymmetric sigma model with target space a Calabi-Yau threefold.

\subsec{$N=4$ topological strings}
Let us consider an $N=4$ superconformal theory and see how one can
make a string theory out of that,  as we have just done for $N=2$
superconformal theories.
By an $N=4$ theory, we mean a superconformal
theory with the small $N=4$ algebra, 
which has one spin 2 energy momentum tensor $T$, four spin ${3\over 2}$
fermionic currents $G^\alpha$, and three spin 1 currents $J^a$ forming an
$SU(2)$
current algebra.  We write the supercurrents as $G^{a,i}$, $a,i=1,2$,
obeying in a unitary theory
a reality condition $G^{a,i}=\epsilon^{ab}\epsilon^{ij} G^*_{b,j}$.
The $SU(2)$ symmetry that is part of the $N=4$ algebra acts on the
first index of $G^{a,i}$.  Acting on the second index is an additional
$SU(2)$ symmetry, which we call $SU(2)_{outer}$, that acts by outer
automorphisms on the $N=4$ algebra.  Some $N=4$ theories have $SU(2)_{outer}$
as a symmetry (for instance, the sigma model with target the flat
hyper-Kahler manifold ${\bf R}^4$) but many others do not (for
instance, a sigma model with target a  K3 surface).

In order to define the $N=4$ topological string,
it is convenient to first view the $N=4$ superconformal
theory as an $N=2$ superconformal theory.
To pick an $N=2$ subalgebra in an $N=4 $ theory, we proceed as follows.
First we choose, from among the $SU(2)$ currents $J^a$, a linear
combination that we will regard as  the $U(1)$ current
$J$ of an $N=2$ algebra.
 However, even when $J$ is given, an $N=2$ subalgebra of the $N=4$
algebra is not uniquely determined.
With respect to $J$, there are two supercurrents of charge 1, which
we may denote as $G^{+,i}$, $i=1,2$.
We may take any linear combination of these to be the 
 $G^+$  generator of an $N=2$ algebra
 ($G^-$ will then be fixed as the hermitian
conjugate of 
$G^+$).
The first choice, namely the choice of $J$, is inessential in the sense
that the different choices give equivalent theories;
they differ by $SU(2)$ rotations.  
However, 
the choice of which linear combination of the $G^{+,i}$ we call $G^+$ contains
essential information.
  The choices can indeed be rotated to each other by $SU(2)_{outer}$,
but in general, as we have noted, $SU(2)_{outer}$ is not a symmetry
of the theory.  So the different choices are inequivalent.

If we denote the $G^{+,i}$ as  $G^+$ and ${\tilde G}^+$, then 
we can take the $G^+$ generator in the $N=2$ algebra to be
$$\widehat{{G}^+}(u)=u_1 { G}^+ +u_2 { \tilde G^+}$$
for any complex constants $u_1, $ $u_2$.  If the $u_i$ are multiplied
by a common factor, we get the same $N=2$ algebra, so the choices
are parametrized by a copy of ${\bf CP}^1$.  We can think of this
${\bf CP}^1$ as a two-sphere that is a homogeneous space of the group
$SU(2)_{outer}$; it can be identified as 
$${SU(2)_{outer}\over U(1)}.$$
As we have stressed, since the $SU(2)_{outer}$ is an outer automorphism
of the algebra and is not necessarily a symmetry of the theory,
the different $u$'s will give {\it inequivalent} imbeddings
of the  $N=2$ superconformal algebra in $N=4$.
For any choice of the $u_i$, normalized to $|u_1|^2+|u_2|^2=1$,
we pick the basis
$$\widehat{{G}^+}(u)=u_1 { G}^+ +u_2 { \tilde G^+}$$
$$\widehat {\tilde G^-}(u)=-u_2{G^-}+u_1 { \tilde G}^-$$
$$\widehat {{G}^-}(u)=u_1^*{G}^- + u_2^*{\tilde G^-}$$
\eqn\rot{\widehat {\tilde G^+}(u) =-u_2^* { G^+}+u_1^*{\tilde G}^+ .}
These formulas have been obtained as follows.  
We have regarded $(u_1, u_2)$ as the
top row of an $SU(2)_{outer}$ matrix
$$M=\left(\matrix{u_1 &u_2 \cr -u_2^* & u_1^*}\right).$$
Then we have transformed the 
two $SU(2)_{outer}$ doublets, namely
$(G^+,{\tilde G}^-)$ and $({\tilde G^+},G^-)$, by this
group element to get the above basis.
 
 Either  $(G^+,G^-)$
or $({\tilde G^+} ,{\tilde G^-})$, or their transforms by $M$,
 can be used to form
the two supercurrents of an $N=2$ algebra.
Moreover, the OPEs
$$G^+{\tilde G}^- \sim 0 \qquad G^-{\tilde G^+}\sim 0$$
have no singularities.  

We now construct an $N=2$ twisted
theory by modifying the energy momentum tensor
as explained before: $T\to T-\half \partial J$.  
Since this operation shifts the dimension of every field
by $-\half$ its $U(1)$ charge, the twisted theory has the following
symmetry currents.   There are two bosonic spin 2 fields, namely the
modified
energy momentum tensor $T'$ and the $SU(2)$ current
$J^{--}$, and two fermionic spin 2 currents $G^-,{\tilde G}^-$.
Moreover, there are two spin 1 fermionic currents $G^+,{\tilde G}^+$
and a bosonic spin one current $J$.  In addition, there is 
a spin 0 bosonic current of $SU(2)$, namely $J^{++}$.

Now we have to decide which fields in the twisted theory
we would like to identify as ``physical.''  In the
case of $N=2$ topological theories, we identified the physical
spectrum with  the cohomology
 of $G^+$, i.e. with the space of fields annihilated 
by $G^+$ modulo addition of fields of the form $G^+\Lambda$.
Now we have in principle {\it two} BRST charges, because
both $G^+$ and ${\tilde G}^+$ are spin 1 fermionic currents 
and give rise to nilpotent fermionic charges $G^+_0$ and $\tilde G^+_0$.
Moreover we have
\eqn\comg{\{ G^+_0,\tilde G^+_0 \}=0}
by virtue of the $N=4$ algebra.
It is natural to construct a theory in which a physical field
is required to be annihilated by both $G^+_0$ and $\tilde G^+_0$.
Any field of the form
$[G^+_0,[{\tilde G}^+_0 ,\Lambda ]]$ (where $[~,~]$ denotes
the appropriate commutator or anticommutator) obeys this condition;
we consider such fields to be trivial.
In other words, we identify the physical fields as Virasoro primary
fields $\Phi^+$ satisfying
\eqn\nefp{[G^+_0,\Phi^+ ]=[{\tilde G}^+_0, \Phi^+]=0, \qquad \Phi^+ \sim
\Phi^+ +[G^+_0,[{\tilde G}^+_0 ,\Lambda^- ]].}
As before, to the physical field
$\Phi^+$, we associate the
deformation of the Lagrangian
 $\int G_{-1}^- \Phi^+$.  
 
 One would like to know under what conditions
 this deformation preserves the $N=4$ structure.  A simple sufficient
 criterion, which will be convenient when we discuss massless
 vertex operators, is that $\int G_{-1}^-\Phi^+$ should preserve the $SU(2)$
 symmetry.  (This condition suffices since $N=2$ plus $SU(2)$ gives
 $N=4$.)    In particular, acting with the raising operator $J^{++}$
 (and recalling that, after twisting, its zero mode becomes
$J^{++}_1$),
 we get $J^{++}_1G^-_{-1}\Phi^+=G^-_{-1}J^{++}_1\Phi^++\tilde G^+_0\Phi^+
 =G^-_{-1}J^{++}_{1}\Phi^+.$  A sufficient criterion for
 vanishing of $J^{++}_1 G_{-1}^-\Phi^+$ is thus
 $J^{++}_1\Phi^+=0$.  In terms of the untwisted theory, this means
 that $\Phi^+$, having $U(1)$ charge $1/2$ and being annihilated
 by $J^{++}$, is the top component of an $SU(2)$ doublet.  
As explained in section 2 of reference \Topo,
the condition that the zero mode of $J^{--}$
 annihilates the deformation is that $G^-(J^{--}_{-1}\Phi^+)=0$, in
 other words the bottom component of the doublet containing $\Phi^+$ is
 antichiral from the $N=2$ point of view.

Suppose the cohomology of $\tilde G^+_0$ is trivial, i.e.
$\tilde G^+_0 V=0$ implies $V=\tilde G^+_0 W$ for some $W$. Then 
if we were able to consider a {\it reduced} Hilbert space
$\tilde{\cal H}\subset {\cal H}$ where $\tilde{\cal H}$ is the
subspace of ${\cal H}$ killed by
$\tilde G^+_0$,
the physical state condition \nefp\
acting on this reduced Hilbert space would be
the same as the condition for an $N=2$
twisted physical field\foot{In other words, consider the $ G_0^+$
cohomology in $\tilde {\cal H}$.  It is generated by operators $\Phi\in
\tilde {\cal H}$ obeying $[G_0^+,\Phi]=0$; such states, being in 
$\tilde {\cal H}$, obey also $\tilde G_0^+\Phi=0$.  $\Phi$ is considered
trivial if it is of the form $G_0^+\Psi$ for some $\Psi\in {\tilde {\cal H}}$;
if the $\tilde G^+$ cohomology is trivial, we can write $\Psi=[\tilde G_0^+,
\Lambda]$, and then the equivalence relation on $\Phi$ is $\Phi\sim \Phi
+[G_0^+,[\tilde G_0^+,\Lambda]]$.}
and we could use the same rules of computation
as discussed in the previous subsection.
So the question is how to
do this reduction. Using the fact that $(\tilde G^{+})^2=0$, this can
be done by simply inserting
an operator $\oint \tilde G^+$ around each $a$-cycle on the Riemann surface,
or in a manifestly
modular-invariant way by combining with right-movers
and integrating $\int d^2 z
 \tilde G^+ \overline {\tilde G^+}$
over the surface (when $\tilde G^+$ is holomorphic,
this surface integral reduces to integrals over the cycles in the
usual way).
We are therefore led to considering the $N$-point scattering
amplitude on genus
$g$ defined by the measure over moduli space ${\cal M}_g$
$$\langle |G^-(\mu_1)...G^-(\mu_{3g-3})|^2\big[\int \tilde G^+\tilde {\bar
G}^+ \big]^{g}
\prod_{r=1}^{N} \int G_{-1}^- \bar G_{-1}^- \Phi_{(r)}^{++} 
\rangle $$

However, this guess requires some modification, because
this amplitude is identically zero.  To see this, use
$\oint \tilde G^+ J=-\tilde G^+$ to replace one of the $\tilde G^+$'s
by the contour integral of $\oint \tilde G^+ $ around $J$, and pull
the contour of $\tilde G^+$ off the surface.  Since the $G^-$'s and
$\tilde G^+$'s have no singularity with it, we get zero!
There is another reason why the above formula is
not what we want.  From \nefp\ we wish a deformation to be trivial
only if it can be written as $[G^+_0,[\tilde G^+_0, \Lambda ]]$.
However, in the above definition, it is easy to see that
 $[\tilde G^+_0 ,\Lambda]$ would already be topologically
trivial since we can pull
the $\oint \tilde G^+$ contour
off of $\Lambda$ and get zero by the same reasoning as above.

So instead, we will define the topological $N$-point scattering
amplitude to be
\eqn\defpp{F_g=\int_{{\cal M}_g}\langle |G^-(\mu_1)...G^-(\mu_{3g-3})|^2\big[
\int \tilde G^+\overline {\tilde
G^+} \big]^{g-1}\int J\bar J
\prod_{r=1}^{N} \int G_{-1}^- \bar G_{-1}^- \Phi_{(r)}^{++} 
\rangle }
which is no longer zero. Note that the contour
of $\oint\tilde G^+$ can no longer be pulled off the surface since it hits
the $J$ and gives back $\tilde
G^+$ as the residue. For the same reason,
adding $\oint \tilde G^+ \Lambda$
to the action may change the partition function.  However, 
if we consider adding $\oint G^+\oint \tilde G^+ \Lambda$ to the action, then
the $\tilde G^+$ contour can be pulled off of $\Lambda$
and converts the $J$ to a $\tilde G^+$.  Now pulling the $G^+$ contour
off of $\Lambda$, we encounter no residues from $\tilde G^+$. From the
$G^-$'s, we get residues which are the energy momentum tensor,
thus giving us total derivatives in the moduli which at least formally
(barring anomalies) integrate to zero.  Thus \defpp\
has the correct topological symmetry.

Actually, because of the possibility of making an $SU(2)_{outer}$ rotation
of the $N=2$ embedding,
there is a whole family of topological scattering amplitudes which
can be defined by
\eqn\defp{\hat
F_g(u^*_1,u^*_2,{\bar u}^*_1,{\bar u}^*_2)=\int_{{\cal M}_g}\langle |\widehat
G^-(\mu_1)...\widehat G^-(\mu_{3g-3})|^2\big[
\int \widehat{\tilde G}^+\widehat{\overline {\tilde
G}}^+ \big]^{g-1}\int J\bar J}
$$
\prod_{r=1}^{N} \int \widehat G_{-1}^- \widehat{\bar G}_{-1}^- \phi_{(r)}^{++}
\rangle $$
where $\widehat{\tilde  G}^+$ and $\widehat G^-$ are
defined in \rot.
(Note that
$\bar u_1$ and $\bar u_2$ refer to the right-movers, and are
{\it not} the complex conjugates of $u_1$ and $u_2$, which are denoted
as $u_1^*$ and $u_2^*$.)
Since $\hat F_g$ is a homogeneous
polynomial of degree $4g-4+N$ in $(u^*_1,u^*_2)$
and degree $4g-4+N$ in $({\bar u}^*_1, {\bar u}^*_2)$, we can
decompose it
to each individual
component of the polynomial and obtain $(4g-3+N)^2$ independent
scattering amplitudes $F_g^{(m,n)}$  defined by
\eqn\homp{
F_g(u^*_1,u^*_2,{\bar u}^*_1,{\bar u}^*_2)=}
$$\sum_{m=-2g+2-N}^{2g-2}
\sum_{n=-2g+2-N}^{2g-2}
{(4g-4+N)!\over (2g-2+N+m)!(2g-2-m)!}
{(4g-4+N)!\over (2g-2+N+n)!(2g-2-n)!}$$
$$F_g^{(m,n)} (u^*_1)^{2g-2+N+m} (u^*_2)^{2g-2-m}
({\bar u}^*_1)^{2g-2+N+n} ({\bar u}^*_2)^{2g-2-n}
.$$

Comparing with \defp,
one sees that $F_g^{(m,n)}$ is computed
in the same manner as $F_g^{(2g-2,2g-2)}$ of \defpp\ but with
some of the $G^-$'s and $\tilde G^+$'s
switched to $\tilde G^-$'s and $G^+$'s. Using contour integral manipulations
similar to those described above, 
it is straightforward to show that switching
any $p$ $G^-$'s to $\tilde G^-$'s and any $q$ $\tilde G^+$'s to $G^+$'s
in \defpp\ computes $F^{(2g-2-p-q,2g-2)}$ up to contact terms.
The choice of which
$G^-$'s and $\tilde G^+$'s are switched is irrelevant up to
contact terms.

For example, consider the amplitude below where 
one of the $G^-$'s in the measure has been switched to a $\tilde G^-$:  
\eqn\defrr{
F_g=\int_{{\cal M}_g}\langle 
\tilde G^-(\mu_1)\bar G^-(\bar\mu_1)
| G^-(\mu_2)\bar G^-(\bar\mu_2) ... G^-(\mu_{3g-3})|^2\big[
\int {\tilde G}^+{\bar{\tilde
G}}^+ \big]^{g-1}\int J\bar J}
$$
\prod_{r=1}^{N} \int G_{-1}^- \bar G_{-1}^- \phi_{(r)}^{++}
\rangle .$$
Writing $\tilde G^-(\mu_1)=[\oint G^+,J^{--}(\mu_1)]$,
one can pull the $G^+$ contour off the $J^{--}(\mu_1)$ until it hits
the $J$ to turn it into a $G^+$.
(We are ignoring contact terms coming from when the $G^+$ hits the
$G^-$'s.) One can now write one of the $\tilde G^+$'s from the
cycles as
$\tilde G^+ = [\oint \tilde G^+, J]$, and pull the $\tilde G^+$
contour off until it hits the bare $J^{--}(\mu_1)$,
turning it into a $G^-(\mu_1)$. Finally, one pulls the $G^+$ off one of
the cycles until
it hits the $J$ on the other cycle to give the amplitude
\eqn\defqqp{
F_g=\int_{{\cal M}_g}\langle |
G^-(\mu_1)... G^-(\mu_{3g-3})|^2\big[
\int {\tilde G}^+\bar{\tilde G}^+
\big]^{g-2}\int G^+\bar{\tilde G}^+ \int J\bar J}
$$
\prod_{r=1}^{N} \int G_{-1}^- \bar G_{-1}^- \phi_{(r)}^{++}
\rangle .$$
So we have shown that the amplitude is the same (up to
contact terms) if one switches
a $G^-$ to $\tilde G^-$ or if one switches a $\tilde G^+$ to
$G^+$.

\subsec{Critical $N=2$ String as an $N=4$ Topological String}

One class of  $N=4$ theories that is important for our purposes
is the following.
One can construct a small $N=4$ superconformal theory from
any $\hat c=2$ $N=2$ superconformal theory by defining the three
SU(2) currents to be $J$, $e^{\int J}$ and $e^{-\int J}$, and defining
the four fermionic generators to be $G^+$, $G^-$, $[e^{-\int J},G^+]$
and $[e^{\int J}, G^-]$. Note that $\hat c$ must equal 2 (i.e. the
associated $N=2$
string is critical) in order that
$e^{\pm\int J}$ has spin 1. 
\foot{An important example of this construction is the following.  In general,
the supersymmetric sigma model with target a Calabi-Yau $n$-fold
is an $N=2$ superconformal model with $\hat c=n$.  To get $N=4$
supersymmetry, the target should be hyper-Kahler.  The condition
$\hat c=2$ amounts to $n=2$.  A Calabi-Yau two-fold is always
hyper-Kahler,
and that is why the $\hat c=2$ case automotically gives $N=4$ superconformal
symmetry.}

It is interesting to ask what $F^{(m,n)}_g$
is computing for these $N=4$ strings. 
As was shown in reference \Topo,
$F_g^{(m,n)}$ computes the $N=2$ scattering amplitude
on a surface of left-moving instanton number $m$ and right-moving
instanton number $n$. 
In other words, after introducing the usual set
of $N=2$ super-reparameterization ghosts, with $c=-6$, one can construct
BRST-invariant vertex operators for the critical $N=2$ string
which are in one-to-one correspondence
with the physical vertex operators of the $N=4$ topological string.
Furthermore, one can compute 
$N=2$ correlation functions of these vertex operators
on an $N=2$ super-Riemann surface of genus $g$ and
left and right-moving instanton
number $m$ and $n$, and after integrating over the $N=2$ super-moduli of
this Riemann surface (including the fermionic and $U(1)$ moduli), 
one recovers precisely the 
$N=4$ topological formula $F_g^{(m,n)}$ for the appropriate $N=4$
physical vertex operators. 

\newsec{The Superstring as an $N=4$ Topological String}

We will now review how the usual superstring (which is conventionally
described as a critical $N=1$ string) can also be described as
an $N=2$ string with $\hat c=2$, and therefore
also as an $N=4$ topological string. The advantage of the $N=4$ topological
description of the superstring
is that, after a field-redefinition to Green-Schwarz-like
variables, some of the spacetime supersymmetry
can be made manifest. In this paper, we will be mainly
interested in Type IIB superstring compactifications
on K3, which admit sixteen unbroken spacetime supersymmetries.   (We can also
consider compactification on ${\bf T}^4$, but in that case our construction
exhibits only some of the unbroken supersymmetries.)
As
will be reviewed below, eight of these sixteen supersymmetries will be
manifest in the formalism (i.e. act geometrically on the target superspace), 
while the other eight can be exhibited
in terms of ordinary vertex operators (as opposed to spin operators)
but do not arise just from geometrical symmetries of the target superspace.

We shall first construct the $N=4$ superconformal generators for
the RNS superstring and prove that the physical $N=4$ vertex operators
are in one-to-one correspondence with the usual BRST-invariant
RNS vertex operators. We shall then show that the $N=4$ topological
amplitude $F_g^{(m,n)}$ defined in the previous section computes the 
RNS $g$-loop superstring amplitude where the total left and
right-moving picture of the
vertex operators is $m$ and $n$.

\subsec{N=4 Superconformal Generators and Physical Vertex Operators}

As discussed in \ref\emb{N. Berkovits, 
``The Ten-Dimensional Green-Schwarz Superstring is a
Twisted Neveu-Schwarz-Ramond String'', Nucl. Phys. B420 (1994) 332,
hep-th/9308129\semi
N. Berkovits and C. Vafa,
``On the Uniqueness of String Theory'',
Mod. Phys. Lett. A9 (1994) 653, hep-th/9310170\semi
N. Ohta and J. Petersen, ``$N=1$ From $N=2$ Superstrings'', Phys. Lett.
B325 (1994) 67, hep-th/9312187.}, any critical $N=1$ string can
be embedded into a critical $N=2$ string. This embedding 
allows $N=1$ string scattering amplitudes
to be computed using $N=2$ string methods. 

The critical $N=2$ superconformal generators are constructed from the
$N=1$ matter and ghost fields as follows:
\eqn\embed{ T= T_{N=1} + \half\partial (bc +\xi\eta),}
$$G^+=
\gamma  G_m+
c( T_m -{3\over 2}
\beta\p\gamma-{\half\gamma\p\beta}-b\p c )-\gamma^2 b +\partial^2 c
+\p (c\xi\eta),$$
$$G^- = b,$$
$$J= cb+\eta\xi$$
where $T_{N=1}=T_m + T_g$ is the stress-tensor of the original $N=1$ matter and
ghost fields, and the $(\beta,\gamma)$
super-reparameterization ghosts have been bosonized as
$(\beta=i e^{-\phi}\p\xi, \gamma=-i\eta e^\phi)$.\foot{The factors of $i$
have been put in the bosonization formula so that $\beta$ is real, i.e.
$\beta^* = -i \p\xi e^{-\phi} $= $i e^{-\phi}\p\xi =\beta$.}
For instance, the zero mode of $G^+$ is the $N=1$ BRST operator $Q$ and
after twisting (which in this case amounts to
removing the term $\half \partial(bc+\xi\eta)$ from $T$), $G^+$ 
has conformal weight $1$ while $G^-$ has conformal weight 2.
Furthermore, $J$ is related to the usual ghost current
$j_{ghost}=cb +\gamma\beta$ by 
$$J= j_{ghost} - j_{picture}$$ where $j_{picture}$ 
is the picture-current  
defined as $j_{picture}=-\p\phi + \xi\eta $. Note that $P=
\oint j_{picture}$ commutes with the 
$\beta$ and $\gamma$ ghosts, but not with $\eta$ or $\xi$. 
Moreover, we can write 
\eqn\iko{\oint J=Q_{ghost}-P}
where $Q_{ghost}=\oint j_{ghost}$ is the ghost number.

Since one now has a critical $N=2$ string, one can ask what does
the $N=4$ topological scattering amplitude defined in the previous
section compute? To answer this, first note that the 
additional $N=4$ generators (constructed as in subsection 2.3) are given by 
\eqn\tildegen{\tilde G^+=\eta,}
$$\tilde G^-= b (e^{\phi}G_m+\eta e^{2\phi}\partial b-c
\partial\xi)
 -\xi(T_m-{3\over 2}\beta\partial\gamma-\half\gamma
\partial\beta + 2b\partial c-c\partial b)+\partial^2 \xi,$$
$$J^{++}=c\eta ,\quad J^{--}=b\xi .$$

So physical $N=4$ topological vertex operators $\Phi^+$ are defined by
the conditions \nefp, i.e. 
$$Q \Phi^+ = \eta_0 \Phi^+ = (Q_{ghost} -P-1)\Phi^+ =0, 
\quad \Phi^+ \sim\Phi^+ + Q \eta_0\Lambda^-.$$ 
But these are just the physical conditions for a BRST-invariant
vertex operator in the standard RNS formalism. Note that the condition
$\eta_0\Phi^+=0$ implies that $\Phi$ is independent of the $\xi$
zero mode (i.e. it can be written in terms of the unfermionized
$\beta$ and $\gamma$
ghosts) and, similarly, the triviality condition $\Phi^+=Q\eta_0\Lambda^-$
implies that $\Phi^+ = Q\Omega$ for some $\Omega$ which is independent
of the $\xi$ zero mode (i.e. $\eta_0\Omega=0$). Furthermore, 
$(G-P-1)\Phi^+=0$ is the 
usual ghost-number condition for physical
fields.\foot{For NS states at zero picture, it implies that $G=1$ as
desired. For R states at picture $-1/2$, it implies that $G=1/2$
as desired. All RNS physical states at other pictures are related to
these pictures by multiplication by the picture-raising operator
$Z=\{Q,\xi\}$
or the picture-lowering operator $Y=c\p\xi e^{-2\phi}$. But the operators
$Z$ and $Y$ commute with $\oint J$, so these other physical states also satisfy
the condition $(\oint J-1)\Phi^+=0$. }

\subsec{$N=4$ Topological Amplitudes as RNS Superstring Amplitudes}

In terms of the RNS variables, the scattering amplitude
defined in \defpp\ (or in \defp\ at $u_2 =\bar u_2=0$) is 
\eqn\deqq{F_g^{(2g-2,2g-2)}=
\int_{{\cal M}_g}\langle |b(\mu_1)...b(\mu_{3g-3})|^2\big[
\int \eta\overline {\eta}
 \big]^{g-1}\int (bc+\xi\eta)(\bar b\bar c +\bar\xi\bar\eta)}
$$
\prod_{r=1}^{N} \int b_{-1} \bar b_{-1} \Phi_{(r)}^{++} 
\rangle .$$

Note that if we are in the ``large'' RNS Hilbert space,
i.e. including the $\xi$ zero mode, then the only $\xi$ zero mode in this
expression
comes from the $J$ current $bc+\xi\eta$. 
After integrating out the $\xi$ zero mode,
one is left with the integration 
$$[\int d^2 z (\eta
\bar{\eta})]^g=[\sum_{i=1}^g
(\int_{a_i}dz\eta
\int_{b_i}d\bar z\bar{\eta}
-\int_{b_i}dz\eta
\int_{a_i}d\bar z\bar{\eta})]^g$$
$$=
g!\prod_{i=1}^g
(\int_{a_i}dz\eta
\int_{b_i}d\bar z\bar{\eta}
-\int_{b_i}dz\eta
\int_{a_i}d\bar z\bar{\eta}).$$
Although these $\eta$ integrations over the cycles look strange, 
they can be understood
as coming from the integration over the $(\beta,\gamma)$ ghosts
in the bosonized form of the standard RNS model \ref\Carow{E. Verlinde and
H. Verlinde,``Multiloop Calculations in Covariant
Superstring Theory'', Phys. Lett. B192 (1987) 95\semi 
U. Carow-Watamura, Z. Ezawa, K. Harada, A.
Tezuka and S. Watamura, ``Chiral Bosonization of
Superconformal Ghosts on Riemann Surface and
Path Integral Measure'', Phys. Lett. B227 (1989) 73. }
\foot{As discussed in \Carow, one needs to also insert 
$\delta$-functions in the momentum of the $\phi$ field to 
restrict the picture of states propagating in the internal loops. 
These $\delta$-functions give rise to unphysical poles 
coming 
from the multiloop correlation function for 
the negative-energy field $\phi$ which, at 
the present time, are not well understood. 
So our scattering amplitudes are
only trustworthy when these $\delta$-functions of $\p\phi$ do not
contribute, i.e. when the 
integrands of the scattering amplitudes are independent
of the locations of the picture-changing operators and there
are no unphysical poles. 
Besides tree and one-loop amplitudes,
examples of such multi-loop scattering amplitudes are the
$D=4$ and $D=6$ ``topological'' scattering amplitudes discussed in \ref\taulor
{I. Antoniadis, E. Gava, K. Narain and T. Taylor, 
``Topological Amplitudes in String Theory'', Nucl. Phys. 
B413 (1994) 162 hep-th/9307158\semi
M. Bershadsky, S. Cecotti, H. Ooguri and C. Vafa, 
``Kodaira-Spencer Theory of Gravity and Exact Results for Quantum
String Amplitudes'',
Comm. Math. Phys. 165 (1994) 311
hep-th/9309140\semi
M. Bershadsky, S. Cecotti, H. Ooguri and C. Vafa, 
``Holomorphic Anomalies in Topological Field Theories'', Nucl. Phys. B405
(1993) 279.} and \Topo. }.
The rest of \deqq\ is just the standard RNS scattering amplitude
when the total (left,right)-moving picture of the vertex operators 
$\Phi^{++}_{(r)}$ is
$(2g-2,2g-2)$.

To obtain the scattering
amplitude when the vertex operators are in other pictures, note that
switching $G^-$ with $\tilde G^-$ in the definition of the integrated
vertex operator raises its picture by $+1$, i.e. 
\eqn\picturedis{\int\tilde G^-_{-1}\Phi^+
=\int [G^+, J^{--}_{-1}]\Phi^+ = \int [Q , (\xi b)_{-1}]\Phi^+
= \int [Q , \xi_0 b_{-1}]\Phi^+
=  \int b_{-1} Z\Phi^+}
where  
$Z=\{Q,\xi_0\}$ is the
picture-raising operator and we are assuming that
$(\xi b)_{-1}\Phi^+=\xi_0 b_{-1}\Phi^+$, i.e. 
$\xi$ has no poles with $\Phi^+$ and $b$ has no double poles or higher
with $\Phi^+$. 
It will now be shown that computing $F^{(2g-2,2g-2)}$ in \defp\ with one of
the integrated
vertex operators replaced by $\int\tilde G^-_{-1}\bar G^-_{-1}\Phi^+$ 
is equivalent
to computing $F^{(2g-3,2g-2)}$ in \defp\ where 
all of the vertex operators are of the form 
$\int G^-_{-1}\bar G^-_{-1}\Phi^+$. So if $F^{(2g-2,2g-2)}$ computes
the RNS superstring amplitude when the total (left,right)-moving
picture of the vertex operators
is $(2g-2,2g-2)$, then   
$F^{(2g-3,2g-2)}$ computes
the RNS superstring amplitude when the total (left,right)-moving
picture of the vertex operators
is $(2g-3,2g-2)$.    
Similarly, one can show that
$F^{(m,n)}$ computes the scattering
amplitude when the total picture of the vertex operators is
$(m,n)$.

To prove the above claim (which is a special case of the
claim for $F^{(m,n)}$ made above \defrr) , consider the amplitude  
\eqn\defqq{ 
F_g=\int_{{\cal M}_g}\langle |
G^-(\mu_1)... G^-(\mu_{3g-3})|^2\big[
\int {\tilde G}^+{\bar{\tilde
G}}^+ \big]^{g-1}\int J\bar J}
$$
 \int \tilde G_{-1}^- \bar G_{-1}^- \Phi_{(1)}^{++} 
\prod_{r=2}^{N} \int G_{-1}^- \bar G_{-1}^- \Phi_{(r)}^{++} 
\rangle .$$
Writing $\tilde G^-=[\oint G^+,J^{--}]$ in the vertex operator
$\int\tilde G^-_{-1}\bar G^-_{-1}\Phi^+$,  
one can pull the $G^+$ contour off the $J^{--}$ until it hits
the $J$ to turn it into a $G^+$.
(We are ignoring contact terms coming from when the $G^+$ hits the
$G^-$'s.) One can now write one of the $\tilde G^+$'s from the
cycles as 
$\tilde G^+ = [\oint \tilde G^+, J]$, and pull the $\tilde G^+$
contour off until it hits the bare $J^{--}$ on the vertex operator,
turning it into a $G^-$. Finally, one pulls the $G^+$ off one of
the cycles until
it hits the $J$ on the other cycle to give the amplitude
\eqn\defqqp{ 
F_g=\int_{{\cal M}_g}\langle |
G^-(\mu_1)... G^-(\mu_{3g-3})|^2\big[
\int {\tilde G}^+\bar{\tilde G}^+ 
\big]^{g-2}\int G^+\bar{\tilde G}^+ \int J\bar J}
$$
\prod_{r=1}^{N} \int G_{-1}^- \bar G_{-1}^- \Phi_{(r)}^{++} 
\rangle .$$
But up to contact terms, this is what
$F^{(2g-3,2g-2)}$ computes. So we have shown (up to contact terms)
that the amplitude computed by $F^{2g-3,2g-2}$ is equal to the
usual superstring amplitude when the sum of the
pictures of the vertex operators is $(2g-3,2g-2)$.

\newsec{Manifest Six-Dimensional Spacetime-Supersymmetry}

To compute superstring
 scattering amplitudes  keeping spacetime supersymmetry
manifest, it is convenient to perform a field-redefinition from the
RNS variables to Green-Schwarz-like variables. At the present time,
it is not possible to perform a field-redefinition which preserves
ten-dimensional super-Poincar\'e
covariance; however, it is possible to preserve
either four-dimensional 
\fourreview, six-dimensional \Topo, or $U(5)$ \ufive\ super-Poincar\'e
invariance.
The four-dimensional formalism is useful for
describing compactifications of the superstring on a Calabi-Yau three-fold;
a review can be found in \fourreview.  Since we will be interested in this
paper in compactifications of the superstring on a Calabi-Yau two-fold, 
we will use the six-dimensional formalism \Topo\ which will
be reviewed in the following two sections. 

After reviewing the six-dimensional supersymmetry algebra, we shall
introduce a field-redefinition which allows the supersymmetry to be made
manifest. For compactifications which preserve six-dimensional
supersymmetry, this field redefinition maps the RNS matter and ghost 
variables into a set of six-dimensional superspace variables plus
an $N=2$ $\hat c=2$ superconformal field theory describing the 
compactification. 
Under the field redefinition, the $N=2$ $\hat c=2$ superconformal
generators of \embed\ get mapped into a sum of
$N=2$ $\hat c=0$ superconformal
generators constructed from the six-dimensional superspace variables
and $N=2$ $\hat c=2$ superconformal generators constructed from the 
compactification variables. Also, the six-dimensional supersymmetry
generators get mapped under this
field-redefinition into simple compactification-independent operators.

\subsec{Review of six-dimensional supersymmetry }

Our six-dimensional notation will use the
fact that the rotation group $SO(6)$ is locally isomorphic
to $SU(4)$; likewise, with Lorentz signature, the Lorentz group $SO(1,5)$
is locally a real form of $SU(4)$.
Under the identification of the rotation group with $SU(4)$,
spinors of $SO(6)$
transform as $\bf {4}$'s or $\bf{\bar 4}$'s of SU(4).
Spinors which transform as ${\bf 4}$'s will be denoted with a raised index,
$\xi^a$ for $a$=1 to 4, and spinors
which transform as ${\bf \bar 4}$'s 
will be denoted with a lowered index $\xi_a$.
In this notation, vectors $x^m$ for $m=0$ to 5 can be denoted as
anti-symmetric bispinors
$x^{ab}=(\sigma_m)^{ab} x^m$
where $(\sigma_m)^{ab}$ and $(\s_m)_{ab}$ are the 
SO(5,1) Pauli matrices, which satisfy the algebra
$$(\sigma_m)^{ab}(\sigma_n)_{ac} +
(\sigma_n)^{ab}(\sigma_m)_{ac} =2\eta_{mn}.$$
Also, $(\sigma_m)_{ab}$ is defined to be 
$(\s_m)_{ab}=\half\e_{abcd} (\s_m)^{cd}.$

In $N=1$ supersymmetry in six dimensions, the supersymmetry generators
consist of a pair of ${\bf \bar 4}$'s, say $q_a^\a$ where $\a=\pm$.  
(CPT invariance
together with the pseudoreal nature of the ${\bf \bar 4}$ representation
of $SO(5,1)$ makes it impossible to have a supersymmetric theory with 
supersymmetries
transforming as a single copy of the ${\bf \bar 4}$.) To exhibit 
manifest supersymmetry, one would like a Green-Schwarz style
description in terms of strings moving in a superspace with coordinates
$x^{m}$ and $\theta_a^\a$.   It is not known how to accomplish this much;
what is known is only how to introduce a single multiplet $\theta_a$
of string coordinates, so that half of the supersymmetries act
geometrically.  This is reminiscent of harmonic superspace, but  
the worldsheet analog of the harmonic variables is not yet
understood. The other half of the spacetime supersymmetries
is realized by vertex operators that we
will describe below.   All statements in this paragraph apply to open
strings or to left or right-movers of a closed string.  If supersymmetric
left-movers and right-movers are combined to make a Type II string, then
the amount of supersymmetry is doubled -- both manifest and non-manifest.
Then
there are separate $q_a^\a$'s carried by both left and right-movers,
and separate left and right-moving $\theta$'s.

The $N=1$ $D=6$ supersymmetry algebra is
\eqn\susy{\{q_a^\a,q_b^\b\}= \half\e^{\a\b} \e_{abcd} P^{cd}}
where $P_m$ is the translation generator. 
It is useful to define $SU(2)$ generators, $R^{\a\b}$, which rotate 
these susy generators as 
\eqn\rgen{[R^{\a\b}, q_a^\g] = \half(\e^{\b\g} q_a^\a +\e^{\a\g} q_a^\b).}
This $SU(2)$ group acts by outer automorphisms of the $N=1$ supersymmetry
algebra.

In the RNS formalism, the spacetime-supersymmetry generators can
appear in arbitrary semi-integer picture. To construct these generators,
one first needs to construct 
the six-dimensional spinor fields, $S_a$ and $S^a$, 
out of the six RNS $\psi^m$ fields.  One writes
$$S^a=e^{\half\int^z
(\pm \psi^0\psi^1\pm\psi^2\psi^3\pm\psi^4
\psi^5)}$$ 
with an even number of $+$ signs. $S_a$ is defined 
by the same formula but with an odd number of
$+$ signs. The simplest picture for the susy generators  
is the $-1/2$ picture where 
\eqn\picture{q_a^\pm= \oint e^{-\half\phi} S_a 
e^{\pm{i\over 2} H_C^{RNS}}}
and $J_C^{RNS}=i \p H_C^{RNS}$ 
is the $U(1)$ current of the $N=2$ $\hat c=2$ superconformal
field theory representing the compactification manifold. 

However, as is well known,
the susy generators defined in this picture do not
satisfy off-shell the algebra of \susy.  Rather, they anti-commute to
give 
\eqn\susyno{\{q_a^\a,q_b^\b\}= \e^{\a\b} \oint  e^{-\phi} \psi_{ab}.}
$\oint e^{-\phi}\psi_m$ is related to the translation
generator $- i\oint  \p x_m$ by multiplication with
the picture-raising operator (since $Z e^{-\phi} \psi^m
= - i\p x^m$), but picture-changing is only
a valid operation when all states are on-shell. For this reason,
it is convenient to define $q_a^-$  in the $-\half$ picture
but to define $q_a^+$ in the $+\half$ picture.  This gives 
\eqn\RNSsusy{q_a^-= \oint e^{-\half\phi} S_a e^{-{i\over 2}H_C^{RNS}},}
$$q_a^+ = \oint (e^{{3\over 2}\phi} b \eta S_a e^{{i\over 2}H_C^{RNS}}
- e^{\half\phi} ({i\over 2}\e_{abcd}S^b\p x^{cd} +i S_a G^{-RNS}_{-3/2 ~C}) 
e^{{i\over 2}H_C^{RNS}}
).$$
As discussed in  \fms, $q_a^+$ is computed by multiplying the 
susy generator in the $-\half$ picture by the 
picture-raising operator $Z$. 
The terms in $q_a^+$ come from the terms
$\p\phi b\eta e^{2\phi}$, 
$-i e^\phi\p x^m\psi_m$ and $-i e^\phi G^{-~RNS}_C$ in $Z$.  Also,
$G^{- RNS}_{-3/2~C} e^{{i\over 2}H_C^{RNS}}$ means the square root pole
in the OPE of $G^{-RNS}_C$ with
$e^{{i\over 2}H_C^{RNS}}$.

These operators now commute to give the spacetime supersymmetry algebra
even off-shell since  
$ \{q_a^\a,q_b^\b\}=- {i\over 2}\e^{\a\b} \oint \e_{abcd}\p x^{cd}$.
Having made a definite choice of pictures for all
supersymmetry charges, it is also possible to make a definite choice
of pictures for all vertex operators in a manifestly supersymmetric fashion.
But doing this in a convenient manner
 requires information about the construction
of the vertex operators that we will develop later.

For Type II strings, one can construct $N=2$ $D=6$ susy generators from the
left and right-moving worldsheet fields. The additional generators
coming from the right-moving worldsheet fields can be defined, just as in the
the above lines,  to be 
\eqn\RNSsusybar
{\bar q_{\bar a}^-= \oint  e^{-\half\bar \phi} 
\bar S_{\bar a} e^{-{i\over 2}\bar H_C^{RNS}},}
$$\bar q_{\bar a}^+ = \oint (e^{{3\over 2}\bar\phi} \bar b \bar\eta 
\bar S_{\bar a} e^{{i\over 2}\bar H_C^{RNS}}
- e^{\half\bar\phi} ({i\over 2}\e_{\bar a\bar b\bar c\bar d}
 \bar S^{\bar b}
\bar\p x^{\bar c\bar d} 
+i\bar S_{\bar a} \bar G^{-RNS}_{-3/2 ~C}) 
 e^{{i\over 2}\bar H_C^{RNS}}.$$
We are here using an unusual notation with the symbol $\bar a$
meant to cover the Type IIA and Type IIB cases simultaneously.
For the Type IIA superstring, a lowered $\bar a$ index denotes a $4$
representation of SU(4)
and a raised $\bar a$ index denotes a $\bar 4$ representation of SU(4)
(which is the opposite convention from that of the $a$ index).
For the Type IIB superstring, a raised $\bar a$ index denotes a $4$
representation of SU(4)
and a lowered $\bar a$ index denotes a $\bar 4$ representation of SU(4)
(which is the same convention as that of the $a$ index).
 
These generators satisfy the $N=2$ $D=6$ susy algebra 
\eqn\susytwo{\{q_a^\a,q_b^\b\}= \half\e^{\a\b}\e_{abcd} P^{cd},\quad 
\{\bar q_{\bar a}^\a,\bar q_{\bar b}^\b\}= \half\e^{\a\b}
\e_{\bar a\bar b\bar c\bar d} P^{\bar c\bar d}.}
The left and right moving supersymmetry algebras each admit
$SU(2)$ groups of outer automorphisms, with generators
$R^{\a\b}$ and $\bar R^{\a\b}$ satisfying
the algebra 
$$[R^{\a\b}, q_a^\g] = \half(\e^{\b\g} q_a^\a +\e^{\a\g} q_a^\b), \quad
[\bar R^{\a\b},q_a^\g]=0,$$
$$[R^{\a\b},\bar q_{\bar a}^\g]=0,\quad
[\bar R^{\a\b}, \bar q_{\bar a}^\g] = \half(\e^{\b\g} \bar q_{\bar a}^\a +
\e^{\a\g} \bar q_{\bar a}^\b).$$

Although the supersymmetry algebra now closes off-shell, spacetime 
supersymmetry is
not yet manifest in the RNS formalism
since the worldsheet variables transform in a complicated 
manner under commutation with the susy generators defined in \RNSsusy.
In order to make susy manifest, one should therefore find a field
redefinition of the worldsheet variables such that they transform in
a simpler manner. 

\subsec{Field redefinition to Green-Schwarz-like variables}

Supersymmetry generates translations of the odd coordinates of superspace.
So the first step in defining the new variables is to look for
a $\t^a_\a$ variable which transforms as $\{q_a^\a,\t^b_\b\}=
\delta_a^b \delta^\a_\b.$ This is easily done by 
defining 
$$\t^a_-= e^{\half(\phi+i H_{C}^{RNS})}S^a,\quad
\t^a_+= c \xi e^{-\half(3\phi+iH_{C}^{RNS})}S^a.$$
However, it is easy to show that $\t^a_-$ and $\t^a_+$ are
not independent fields since they satisfy
$\t^a_-\t^b_+ = \t^a_+\t^b_-$. In fact, one can write 
$\t^a_+$ in terms of $\t^a_-$ as $\t^a_+ =c\xi e^{-2\phi-iH_C^{RNS}}\t^a_-$, 
so one cannot choose all eight
of these variables to be free fields. However, one can choose
half of them, e.g. $\t^a_-$, to be free fields. This breaks half
of the eight manifest spacetime supersymmetries, but leaves four of them
manifest. It might be possible to restore all eight supersymmetries
by treating $\t^a_-$ as a harmonic-like
variable, but this has not yet been done. Since we will no longer
refer to $\t^a_+$, we shall rename $\t^a_-=\t^a$ for the rest of this paper.

The conjugate momentum to $\t^a$ is easily seen to be
$$p_a =  
 e^{-\half(\phi+iH_{C}^{RNS})}S_a,$$
whose zero mode is just the susy generator $q_a^-$. Note that
$\t^a$ and $p_a$ carry conformal weight 0 and 1 and satisfy the OPE
$$p_a(y) \theta^b (z)\to {\delta_a^b\over{y -z}}. $$ 

Because of the $J_C^{RNS}$ dependence in $\t^a$, $\t^a$ has non-trivial
OPE's with any RNS compactification variable $\chi_C^{RNS}$ carrying non-zero 
U(1) charge. For this reason, it is convenient to redefine the
compactification
variables to have no OPE singularities with the six-dimensional
variables.  To do so, we bosonize the ($\xi$, $\eta$) variables
as ($\xi=e^{-i\kappa}$, $\eta=e^{i\kappa}$).  Then we make a chiral
$U(1)$ rotation of the compactification variables generated by the 
parameter $\phi+i\kappa -{i\over 2}\pi$.  
A field $\chi_C^{GS}$ that is a $U(1)$ primary of charge
$n$ (so its dependence on $H_C^{RNS}$ is precisely $e^{inH_C^{RNS}}$) 
is transformed to
\eqn\charge{\chi_C^{GS} =  e^{n(i\kappa+\phi -{i\over 2}\pi)} \chi_C^{RNS}.}
It is easy to check that 
the fields transformed in this way have no singular OPE's with
$\t^a$ or $p_a$ (and likewise with fields $\rho $ and $\sigma$
introduced below).  

This chiral U(1) rotation of the compactification variables  transforms
the $N=2$ superconformal generators which describe the compactifaction. 
If $[T_C^{RNS}$,$G^{+~RNS}_C$,
$G^{-~RNS}$,$J_C^{RNS}$] are these generators,
then the chiral rotation transforms these generators as 
$$[T_C^{GS},G^{+~GS}_C,G^{-~GS},J_C^{GS}]\to
[ e^R T_C^{RNS} e^{-R}, e^R G^{+~RNS}_C e^{-R}, e^R G^{-~RNS}_C e^{-R}, 
e^R J_C^{RNS} e^{-R}]$$
where $R=\oint (i\kappa+\phi) J^{RNS}_C$.
It is straightforward to show this implies that 
$$T^{GS}_{C}=T_{C}^{RNS}+\p^2(\phi+i\kappa)-
\p(\phi+i\kappa)J_{C}^{RNS},$$ 
$$ G^{+~GS}_{C}
=-i e^{\phi}\eta  G_{C}^{+~RNS},\quad 
G_{C}^{-~GS} 
=i e^{-\phi}\xi  G^{-~RNS}_{C}, $$
\eqn\defjc{J^{GS}_{C}=J^{RNS}_C+2\p(\phi+i\kappa). }
For example, for compactification on $T^4$,
$$G^{+~RNS}=\psi^{J ~RNS} \p x^{\bar J},\quad J^{RNS}_C=\psi^{J~RNS} 
\psi^{\bar J ~RNS}$$
is replaced with 
$$G^{+~GS}=\psi^{J ~GS} \p x^{\bar J}
=-i e^{\phi+i\kappa}\psi^{J ~RNS} \p x^{\bar J},$$ 
$$J^{GS}_C=\psi^{J~GS} \psi^{\bar J~GS}=
\psi^{J~RNS} \psi^{\bar J~RNS}
+2\p(\phi+i\kappa), $$
where $x^J,x^{\bar J},\psi^J,\psi^{\bar J}$ are the compactification
variables and $J=1$ to 2.

Not counting the compactification variables and the six $x^m$'s, 
the original RNS
theory contained two chiral bosons ($\beta,\gamma$) and 
eight fermions ($\psi^m,b,c$). Since the $\t^a$ and $p_a$ variables
describe eight fermions, one still needs to define 
two chiral bosons in the supersymmetric variables to recover
the original degrees of freedom. These two chiral bosons should
be defined such that they have no singular OPE's with the other
Green-Schwarz-like variables. One of these chiral bosons, $\sigma$, is
easily defined by 
bosonizing the $b,c$ ghosts as 
\eqn\bosb{b=e^{-i\s},\quad c=e^{i\s}.}
The remaining chiral boson, $\rho$,
is defined by 
\eqn\defrho{\rho= -2\phi-i\kappa-i H_C^{RNS},}
which is the unique combination which has no singular OPE's with 
the other supersymmetric variables. These chiral bosons satisfy
the standard OPE's 
$$ \rho(y) \rho(z) \to -\log(y -z),\quad
\quad \sigma(y) \sigma(z) \to -\log(y -z).$$  The completeness
of the new variables to the old ones (and the equivalence of the two
descriptions) can be checked by bosonization of all variables.
Using the conformal weights of the RNS fields, one can compute that
the operator $e^{m\rho+in\s}$ has conformal weight equal to
$\half(n^2-3n -m^2 +3m)$. This operator has conformal weight
zero when $n=m$. Note that 
typical $\rho$-dependent operators in the formalism are the real
exponentials $e^{n\rho}$ with integer $n$; in this sense,
$\rho$ is like the $\phi$ boson of the RNS formalism.

For the Type II superstring, all of the above definitions
can be repeated for the right-moving sector by simply placing bars
over all fields. 
Ignoring the left and right-moving chiral bosons, 
the worldsheet action for the Green-Schwarz-like
variables is given by 
\eqn\action{S=\int d^2 z (\half
 \p x^m \bar\p x_m + p_a\bar\p\t^a + \bar p_{\bar a}
\p\bar \t^{\bar a} ) + S_C}
where $S_C$ is the action for the compactification variables. 
Note that $S_C\to S_C + \int d^2 z \bar\p (i\kappa+\phi) J_C^{RNS}$ when
one redefines the charged compactification variables as in \charge. But
since $\bar\p(i\kappa+\phi)=0$, this redefinition does not affect $S_C$.

\subsec{ $N=2$ superconformal generators and
$D=6$ spacetime-supersymmetry generators}

It
is straightforward to write the RNS versions of the
$N=2$ superconformal generators and 
$D=6$ spacetime-supersymmetry generators in terms of the
Green-Schwarz-like variables defined in
the previous subsection.
As will be explained below,
the $N=2$ superconformal generators of \embed\ get mapped
after twisting  (which in this case just
removes the $\half \partial(bc+\xi\eta)$ term from $T$) to
$$T=   
\half\p x^m \p x_m +
p_a\p \t^a +\half\p\rho\p\rho
+\half\p\sigma\p\sigma
-{3\over 2}\p^2 (\rho+i\sigma)+T^{GS}_{C}$$
\eqn\gsgen{ G^{+} = 
- e^{-2\rho-i\sigma} (p)^4 ~ +{i\over 2}
e^{-\rho} p_a p_b \p x^{ab} +}
$$
e^{+i\sigma}(\half\p x^m \p x_m +
p_a\p \t^a  +\half(\p\rho\p\rho -\p\s\p\s) -\half\p^2(3\rho+i\s)
+T^{GS}_C)$$
$$
+\p^2(e^{i\s})+\p(e^{i\s}(\p\rho+J_C^{GS})) + 
G^{+~GS}_{C} + e^{-2\rho} (p)^4 G^{-~GS}_C , $$
$$G^{-}  =
e^{-i\sigma} ,$$
$$J=\p(\rho+i\sigma)~+J_{C}^{GS} $$
where $(p)^4={1\over 24}\e^{abcd}p_a p_b p_c p_d$.
The mapping of $T$ is easily found by noting that the
supersymmetric variables are all free fields and by
computing their conformal weights. The mapping of $G^+$ contains
various terms which come from the different terms in $j_{BRST}$. 
For example, using the definitions of $p_a$, $\rho$ and $\s$ of the
previous subsection, 
$(p)^4 = e^{-2\phi -2i H_C^{RNS}}$
and $e^{2\rho-i\s}= e^{4\phi+2i\kappa +2i H_C^{RNS}} b$, so 
the first term in $G^+$ is 
$$-(p)^4 e^{2\rho-i\s} =- e^{2\phi+2i\kappa}b=- \eta\p\eta e^{2\phi} b =
-\gamma^2 b.$$
Similarly, the second term in $G^+$
comes from the $-i\eta e^{\phi} \psi_m\p x^m$ term in
$\gamma G$,
the third  term comes from the $c T -bc\p c$ term, the fourth and
fifth terms  come from the $\p^2 c +\p(c\xi\eta)$ terms, and the
sixth and seventh terms come from
the $-i\eta e^{\phi}(G_C^{+~RNS}+G_C^{-~RNS})$ term in $\gamma G$.
The mapping of $G^-$ comes from the bosonization of the $b$ ghost in
\bosb. Finally, the mapping of $J$ comes from using \defrho\ and
\defjc\ to show that $i\kappa=\p\rho+ J_C^{GS}$, so
$$cb+\eta\xi=\p(i\s+ i\kappa) =\p(i\s+\rho)+J_C^{GS}.$$ 

The generators of \gsgen\ can be put in a more elegant form by performing
the following
similarity transformation on all Green-Schwarz-like variables:
$y^{GS}\to e^R y^{GS} e^{-R}$
where 
\eqn\simil{R=\oint e^{i\sigma} G^{-~GS}_C.} 
For the rest of this paper, we will only
refer to 
Green-Schwarz-like
variables which have been transformed as in \simil, and
to simplify the notation, we will not change the symbol for these variables.

The similarity 
transformation of \simil\ preserves the free-field action of \action\
since $e^{i\s}G_C^{-GS}$ is holomorphic. Furthermore, $e^{i\s}G^{-GS}_C$ is
a $U(1)$-neutral primary field of
conformal weight one, so $R$ commutes with $T$ and $J$. It is straightforward
to compute that 
$$[R,G^+]= - e^{-2\rho} (p)^4 G_C^{- GS}+e^{2i\s}G_C^{-GS} - e^{i\s} T_C^{GS}-
\p(e^{i\s} J_C^{GS}) -\p^2(e^{i\s}),\quad
[R,G^-]= G_C^{-GS},$$
\eqn\simcomp{
[R,[R,G^+]]= -2 e^{2i\s} G_C^{-GS},\quad [R,[R,[R,G^+]]]=[R,[R,G^-]]=0.}
So after performing this
similarity transformation, 
the above $N=2$ superconformal generators decompose into a $\hat c=0$
six-dimensional part
and a $\hat c=2$ compactification-dependent piece as 
$$T=   
\half\p x^m \p x_m +
p_a\p \t^a +\half\p\rho\p\rho
+\half\p\sigma\p\sigma
+{3\over 2}\p^2 (\rho+i\sigma)+T^{GS}_{C}$$
\eqn\gensusy{ G^{+} = 
- e^{-2\rho-i\sigma} (p)^4 ~ +{i\over 2}
e^{-\rho} p_a p_b \p x^{ab} +}
$$
e^{+i\sigma}( \half\p x^m \p x_m +
p_a\p \t^a  +\half\p(\rho+i\sigma)\p(\rho+i\sigma)-\half\p^2(\rho+i\sigma))
+ G^{+~GS}_{C} , $$
$$G^{-}  =
e^{-i\sigma}+ G^{-~GS}_C ,$$
$$J^{GS}=\p(\rho+i\sigma)~+J_{C}^{GS}. $$

Finally, using these $N=2$ $\hat c=2$ generators, one can construct 
N=4 generators in the way described in subsection (2.3). The additional $N=4$
generators are given by 
\eqn\gentildesusy{\tilde G^{+}  =
e^{iH_{C}^{GS}+\rho} +e^{\rho+i\sigma}\tilde G^{+~GS}_{C},}
$$\tilde G^{-}=
e^{-iH_{C}^{GS}}(-  e^{-3\rho-2i\sigma} (p)^4 ~ +{i\over 2}
e^{-2\rho-i\sigma} p_a p_b \p x^{ab} +
$$
$$e^{-\rho}(
\half\p x^m \p x_m +
p_a\p \t^a  +\half\p(\rho+i\sigma)\p(\rho+i\sigma)-\half\p^2(\rho+i\sigma))) +
e^{-\rho-i\sigma}\tilde G_{C}^{-~GS},$$
$$J^{++}=e^{\rho+i\sigma}~J_{C}^{++~GS}, $$
$$J^{--}=e^{-\rho-i\sigma}~J_{C}^{--~GS}.  $$

In terms of the Green-Schwarz-like variables, $[x^m,\t^a,p_a,\rho,\sigma]$,
the $D=6$ spacetime-supersymmetry generators of \RNSsusy\
drastically simplify to 
\eqn\susy{q_a^-=\oint p_a,\quad 
q_a^+=\oint (e^{-\rho-i\sigma} p_a -{i\over 2}\e_{abcd} \theta^b\p x^{cd}).} 
Note that the term containing $G^{-RNS}_C$ in $q_a^+$ was eliminated by the
similarity transformation defined in \simil.
Since the compactification variables no longer appear in these
generators, they are inert under susy transformations. 

\newsec {Massless Six-Dimensional Vertex Operators}

Given the formulas in \defp, \gensusy, and \gentildesusy, the  
only missing element for computing manifestly spacetime-supersymmetric
six-dimensional
superstring amplitudes is explicit expressions for the vertex operators.
After giving the definition of physical closed string
vertex operators, we shall give explicit expressions for
the massless vertex operators, first for those which are independent
of the compactification fields, and then for those which depend on
the compactification fields.

\subsec{ Definition of Physical Vertex Operators}

As discussed in \nefp, physical vertex operators $\Phi^+$ are
defined for the open $N=4$ topological string by the conditions
$$G^+_0 \Phi^+ = \tilde G^+_0 \Phi^+ = (J_0 -1)\Phi^+ =0,
\quad \Phi^+ \sim\Phi^+ + G^+_0  \Gtp_0 \Lambda^-.$$ 
Since the $\Gtp_0$ cohomology is trivial in this case (this is
easiest to check in RNS language where $\tilde G^+=\eta$), 
it is always possible to define a $V$ satisfying 
$$\Phi^+ = \Gtp_0 V, \quad G^+_0 \Gtp_0 V = J_0 V =0
,\quad V \sim V + G^+_0 \Lambda + \Gtp_0 \tilde\Lambda.$$ 
It will be 
more convenient to describe the massless compactification-independent
states in terms of the U(1)-neutral vertex operator $V$, while it
will be more convenient to describe the massless 
compactification-dependent states in terms of the U(1)-charged vertex
operator $\Phi^+$. 
Using 
the relationship described above, it is straightforward to go from one type
of vertex operator to the other. 

The gauge-invariance of $V$, 
$$\delta V = G^+_0 \Lambda + \Gtp_0 \tilde\Lambda$$ implies that
one can choose the gauge-fixing conditions
\eqn\opengauge{G^-_0 V = \tilde G^-_0 V =0}
in a manner similar to the gauge-fixing of $b_0=0$ in the
RNS formalism. Applying $G^-_0 \tilde G^-_0$ to the equation of motion
$G^+_0 \tilde G^+_0 V=0$ in this gauge
implies that $(T_0)^2 V=0$, which implies
that $T_0 V=0$ since $T_0$ is just the conformal weight of $V$.
Similarly, since $\Phi^+=\tilde G^+ V$, 
\eqn\openphi{G^-_0\Phi^+=\tilde G^-\Phi^+=T_0\Phi^+=0}
in this gauge. 

As discussed in \picturedis,
the integrated form of the $\Phi^+$ vertex 
operator is given by $\int G^-_{-1}\Phi^+$ or $\int \tilde G^-_{-1}\Phi^+$,
where the two different choices are related by picture-changing. 
For the integrated form of the $V$ vertex operator, one has four
different choices:
\eqn\fourdif{\int G^-_{-1} G^+_0 V,\quad \int \tilde G^-_{-1}G^+_0 V,\quad
\int G^-_{-1} \tilde G^+_0 V,\quad \int \tilde G^-_{-1}\tilde G^+_0 V,}
but these four choices are also all related by picture-changing. The 
first two and last two choices are related for the same reason as
in the $\Phi^+$ case. The first and third choices (or second
and fourth choices) are related by
picture-changing for the following reason:
If $\Phi^+=\tilde G^+_0 V$ is physical, then up to a 
gauge transformation 
$\delta V=\tilde G^+_0\Lambda$
one has   $V=\xi_0\Phi^+$ where $\xi_0$ is the $\xi$ zero mode
and $G^+_0\Phi^+=\tilde G^+_0\Phi^+=0$. 
Then
$G^+_0 V= Q\xi_0\Phi^+= Z \Phi^+$ is the picture-raised
version of $\Phi^+$ where $Z=\{Q,\xi\}$ is the picture-raising
operator. So the integrated vertex operator
$\int dz G^-_{-1}  G^+_0 V =
\int dz G^-_{-1}  Z\Phi^+$ 
is related to the integrated vertex operator
$\int dz G^-_{-1}  \tilde G^+_0 V=
\int dz G^-_{-1} \Phi^+$ by replacing $\Phi^+$ with its picture-raised
$Z \Phi^+$. 
It will later be convenient to choose the picture of the integrated
vertex operator to be the same as that of $V$, so we shall define the 
integrated vertex operator for $V$ to be the first choice of \fourdif, i.e.
\eqn\opendef{U = \int dz G^-_{-1} G^+_0 V.}

Physical gauge-fixed vertex operators for closed $N=4$    
topological strings will be defined to be operators which satisfy
the conditions of open $N=4$   string physical vertex operators
in terms of both the left and right-moving 
constraints independently.
\foot{This definition is naive since, in
the usual closed string field theory,
the off-shell closed string field is 
annihilated by $b_0-\bar b_0$. Furthermore,
the on-shell closed string field is annihilated by $Q+\bar Q$,
but not necessarily by $Q$ and $\bar Q$ independently. In practice, however,
all physical vertex operators except for those describing
special zero-momentum states are killed by
$Q$ and $\bar Q$ independently. Also, 
one can use the $Q+\bar Q$ gauge-invariance to fix
$b_0+\bar b_0 =0$, which then combines with the off-shell
$b_0-\bar b_0=0$ constraint to imply that $b_0=\bar b_0=0$.
This is just the open string gauge-fixing condition imposed
on both the left and right-moving sectors. 

Although
a field theory
action for the the open $N=4$ topological string has been constructed
\ref\me{N. Berkovits, ``Super-Poincar\'e Invariant Superstring Field
Theory'', Nucl. Phys. B450 (1995) 90.},  
a field theory action for the closed $N=4$ topological string has
not yet been constructed.
So we cannot
justify our closed string vertex operator conditions 
on the basis of an action principle,
but are assuming that one can choose them
to be the left-right product of the open string gauge-fixed 
conditions of \opengauge\ and \openphi. It is possible that our
conditions are too strong, which might imply that we are missing
some special states. } 
So the physical U(1)-charged vertex operators $\Phi^{++}$ of the closed $N=4$  
topological string must satisfy
\eqn\gfc{G^-_0\Phi^{++}=\Gtm_0\Phi^{++}= T_0\Phi^{++}=0,}
$$\bar G^-_0\Phi^{++}=
\bar{\tilde G}^-_0 \Phi^{++} =\bar T_0\Phi^{++}=0,$$
$$G^+_0\Phi^{++}=\Gtp_0\Phi^{++}=\bar G^+_0\Phi^{++}=
\bar{\tilde G}^+_0 \Phi^{++} =0.$$
Note that the last line of \gfc\ is implied by the other lines
if $\Phi^{++}$ happens to satisfy the condition $J^{++}_0\Phi^{++}=0$.
This condition will be satisfied for the massless compactification-dependent
vertex operator considered later in this section.       

Similarly to the open superstring, one can construct a 
U(1)-neutral closed superstring vertex operator $V$ defined by 
$\Phi^{++}= \tilde G^+_0 \bar {\tilde G}^+_0 V$. Imposing the left-right
product of the open string conditions of \opengauge,
$V$ must satisfy the conditions  
\eqn\gfcv{G^-_0 V=\Gtm_0 V=\bar G^-_0 V=
\bar{\tilde G}^-_0 V =T_0 V=\bar T_0 V= 0,}
$$G^+_0\tilde G^+_0 V=
\bar G^+_0\bar{\tilde G}^+_0 V=0.$$
Note that the last line of \gfc\ is implied by the other lines
if $V$ happens to satisfy the condition $J^{++}_0\tilde G^+_0 V=
\bar J^{++}_0\bar{\tilde G^+_0}=0$.
This condition will be satisfied for the massless compactification-independent
vertex operator considered later in this section.       
The integrated closed superstring vertex operator 
in terms of $V$ will be defined similar to the open superstring
vertex operator of \opendef\ as 
$$U = \int d^2 z G^-_{-1} \bar G^-_{-1} G^+_0 \bar G^+_0 V. $$

\subsec{ Massless Compactification-Independent Vertex Operator} 

For the closed superstring, the massless vertex operators which
are independent of the compactification fields are 
most conveniently described by
a $U(1)$-neutral vertex operator
$V$ which must have conformal weight zero at zero momentum.
The most general such vertex operator which is independent of the
compactification fields is 
$$V=\sum_{m,n =-\infty}^{+\infty} e^{m(i\sigma+\rho)+n(i\bar\sigma+\bar\rho)}
V_{m,n} (x,\t,\bar\t) $$
where $V_{(m,n)}$ is any function of the zero modes of $x,\t^a,$ and
$\bar\t^{\bar a}$. It will now be shown that the physical conditions
imply that $V$ describes the $N=(1,1)$ $D=6$ supergravity multiplet for the
Type IIA superstring, and describes the $N=(2,0)$
 $D=6$ supergravity multiplet plus
an anti-self-dual tensor multiplet for the Type IIB superstring. 

It is easy to check that $T_0 V=0$ implies that $\p^p\p_p V_{m,n}=0$
and
$G^-_0 V =\bar G^-_0 V=0$
implies that 
$V_{m,n}=0$ for $m>1$ or $n>1$. 
(For instance, $G^-_0V$ is the double pole in the $G^-\cdot V$ OPE,
and as $G^-=e^{-i\sigma}$, it vanishes if $V_{m,n}=0$ for $m>1$.)
This implies
that  $V$ satisfies $J^{++}_0\tilde G_0^+ V=0$
since $\tilde G_0^+ V=
e^{iH_C^{GS}+2\rho+i\s } V_{1,1}$. So the only remaining condition
comes from 
$\tilde G^-_0 V=\bar{\tilde G}^-_0 V =0$.  
For convenience, let us define 
\eqn\abcd{\tilde G^- = A + B + C + D}
where $$A=
- e^{-3\rho-2i\sigma -iH_C^{GS}} (p)^4,\quad
B= {i\over 2}
e^{-2\rho-i\sigma -iH_C^{GS}} p_a p_b \p x^{ab} ,$$
$$C=
e^{-\rho- i H_C^{GS}}(\half
\p x^m \p x_m +
p_a\p \t^a+\half\p(\rho+i\sigma)\p(\rho+i\sigma)-\half\p^2(\rho+i\sigma)) ,$$
$$D=e^{-\rho-i\sigma}\tilde G_{C}^{-~GS}.$$

When acting on $V_{1,1}$, $\tilde G^-$ has a 
double pole proportional to 
\eqn\vone{[e^{-2\rho-i\s-iH_C^{GS}}(
\p(-3\rho-2i\s-iH_C^{(GS)})(\N)^4  -{1\over 6}\e^{abcd} p_a\N_b\N_c\N_d }
$$ +{i\over 2}e^{-\rho-iH_C^{GS}} \N_a\N_b\p^{ab} )]V_{1,1}$$
where the first two terms come from $A$ and the third term comes
from $B$. We will use the notation that
$\N_a =d/d\t^a$ and $\bar\N_{\bar a} = d/d\bar \t^{\bar a}$. 
The first and third terms of \vone\ imply that 
\eqn\vonee{(\N)^4 V_{1,1}= \N_a\N_b \p^{ab} V_{1,1} = 0,}
however the second term can be cancelled by the double
pole of $B$ on $V_{0,1}$ which is proportional to 
\eqn\vzero{{i\over 2}e^{-2\rho-i\s-iH_C^{GS}}(
\p(-2\rho-i\s-iH_C^{GS})\N_a\N_b\p^{ab}  +
2 p_a \N_b \p^{ab} + \p x^{ab} \N_a\N_b]V_{0,1}.}
This implies that 
$$\e^{abcd}\N_b\N_c\N_d V_{1,1} = 
 6i\N_b \p^{ab} V_{0,1},\quad \N_a\N_b V_{0,1}=0.$$
Elimination of double poles of $\tilde G^-$ on $V_{m,n}$ for $m<0$
implies that $V_{m,n}=0$ for $m<0$.

Similarly, elimination of double poles of $\bar{\tilde G}^-$
with $V$ implies that 
$$(\bar\N)^4 V_{1,1}= \bar\N_{\bar a}\bar\N_{\bar b} \p^{\bar a\bar b} 
V_{1,0} = 0,$$
$$\e^{\bar a\bar b\bar c\bar d}\bar\N_{\bar b}\bar\N_{\bar c}\bar\N_{\bar d}
 V_{1,1} = 6i \bar\N_{\bar b} \p^{\bar a\bar b} V_{1,0},
\quad \bar\N_{\bar a}\bar \N_{\bar b} V_{1,0}=0$$
and $V_{m,n}=0$ for $n<0$. 

Furthermore, there is a residual gauge-invariance of $V$
coming from the fact that the physical conditions
are preserved by
the gauge transformation 
$$\delta V = G^+_0\Lambda +\bar{G}^+_0\bar\Lambda
 +\tilde G^+_0\tilde\Lambda +\bar{\tilde G}^+_0\bar{\tilde\Lambda}$$
if the gauge parameters
$\Lambda$ and ${\tilde\Lambda}$ are annhilated by 
$\bar G^+_0\bar{\tilde G}^+_0$, the gauge parameters
$\bar\Lambda$ and $\bar{\tilde\Lambda}$ are annihilated by 
$G^+_0\tilde G^+_0$, and all gauge parameters are annihilated 
 by $T_0$, $\bar T_0$,
$G^-_0$, $\tilde G^-_0$, $\bar G^-_0$, and  $\bar {\tilde G}^-_0$, 
Using this gauge invariance, one can 
gauge-fix the 
components of $V_{1,1}$ with no
$\t$'s or no $\bar\t$'s by choosing the gauge parameters to be 
$$\Lambda= e^{-\rho-iH_C^{GS}+\bar\rho+i\bar\s} \lambda(x,\t,\tb),\quad
\bar{\Lambda}= e^{2\bar\rho+i\bar\s} \bar{\lambda}(x,\t,\tb),$$
which transforms 
$$\delta V_{1,1}={i\over 2}
\N_a\N_b \p^{ab}\lambda +{i\over 2}\bar\N_{\bar a}\bar \N_{\bar b}
\p^{\bar a \bar b}\bar\lambda,$$
$$\delta V_{0,1}=(\N)^4\lambda,\quad
\delta V_{1,0}=(\bar\N)^4 \bar\lambda.$$
Note that the usual residual gauge transformations of the graviton and 
NS-NS two-form, 
$$\delta (g_{mn}+b_{mn})=\p_m \xi_n +\p_n \eta_m,$$
come from the gauge parameters
$$\lambda=(\t)^4 \tb^{\bar a}\tb^{\bar b}\xi_{\bar a\bar b},\quad
\bar\lambda= \t^{ a}\t^{b}(\tb)^4\eta_{ a b},$$ since the graviton
and NS-NS two-form 
appear as
$$ (g_{mn}+b_{mn})\s^m_{ab}\s^n_{\bar c\bar d}\t^a\t^b\tb^c\tb^d$$ 
in $V_{1,1}$.

Also, one can eliminate the remaining components of $V_{0,n}$ and $V_{n,0}$
by
using the gauge parameters
$$\tilde
\Lambda= e^{-\rho-iH_C^{GS}+n(\bar\rho+i\bar\s)}\tilde 
\lambda_n(x,\t,\tb),\quad
\bar{\tilde \Lambda}= e^{-\bar\rho-i\bar H_C^{GS}+n(\rho+i\s)} \bar
{\tilde\lambda}(x,\t,\tb),$$
which transforms $\delta V_{0,n}=\tilde\lambda$ and 
$\delta V_{n,0}=\bar{\tilde\lambda}$. 

The remaining degrees of freedom are described by the superfield
$$V_{1,1}=\t^a\bar\t^{\bar a} V^{--}_{a\bar a}+
\t^a\t^b\t^{\bar a}\s^m_{ab} \bar\xi^-_{m~\bar a}+
\t^a\bar\t^{\bar a} \bar\t^{\bar b}\s^m_{\bar a\bar b} \xi^-_{m~ a}$$
$$+
\t^a\t^b\bar\t^{\bar a} \bar\t^{\bar b}\s^m_{ab}\s^n_{\bar a\bar b}
( g_{mn}+b_{mn}+\phi \eta_{mn}) +
\t^a(\bar\t^3)_{\bar a} A^{-+~\bar a}_{ a}
+(\t^3)_a\bar\t^{\bar a} A^{+-~ a}_{\bar a} $$
$$
+\t^a\t^b(\bar\t^3)_{\bar a}\s^m_{ab} \bar\chi_m^{+~\bar a}+
(\t^3)^a\bar\t^{\bar a} \bar\t^{\bar b}\s^m_{\bar a\bar b} \chi_m^{+~ a}+
(\t^3)_a(\bar\t^3)_{\bar a} F^{++~ a\bar a} .$$
This describes either the N=(1,1) supergravity multiplet for the Type IIA
superstring, or the N=(2,0) supergravity multiplet plus an anti-self-dual
tensor multiplet for the Type IIB superstring. The NS-NS fields
are described by the symmetric traceless
$g_{mn}$, the anti-symmetric $b_{mn}$, and the dilaton $\phi$. 
The gravitino field is described by $\chi^{\pm a}_m$ and
$\bar\chi^{\pm\bar a}_m$ where  $\chi^{- a}_m=\p^{ab}\xi^-_b$
and $\bar\chi^{- \bar a}_m=\p^{\bar a\bar b}\xi^-_{\bar b}$.
The four Ramond-Ramond field strengths are described by
$F^{\pm\pm~a\bar a}$ where 
$$F^{+-~a\bar a}=\p^{\bar a\bar b} A^{+-~a}_{\bar b},\quad 
F^{-+~a\bar a}=\p^{a b} A^{-+~ \bar a}_{b},\quad 
F^{--~a\bar a}=\p^{ab}\p^{\bar a\bar b} V^{--}_{b\bar b}.$$
For the Type IIA superstring, the four 
Ramond-Ramond fields have spinor indices of opposite 
chirality and therefore describe the field-strengths of the four vectors in
the N=(1,1) supergravity multiplet. For the Type IIB superstring,
the four Ramond-Ramond fields have two spinor indices of the same chirality
and therefore describe the field-strengths of four self-dual 2-forms
and four scalars. The four Type IIB Ramond-Ramond self-dual 2-forms combine
with the NS-NS self-dual 2-form and the graviton to give the bosonic
fields in the N=(2,0) supergravity multiplet. The remaining four 
Type IIB Ramond-Ramond scalars combine with the 
NS-NS dilaton and anti-self-dual 2-form to give the bosonic fields of 
an anti-self-dual tensor multiplet. 
Note that 
the pictures of the component fields of $V_{1,1}$ 
coincide with their conventional $(R^{+-},\bar R^{+-})$ charge
in supergravity (whose low-energy
action is invariant under
$SU(2)_R\times SU(2)_{\bar R}$).

Besides the mass-shell condition $\p_m \p^m=0$, these component
fields satisfy 
$$\p^m g_{mn} =\p^m b_{mn}=0,\quad
\p^m\chi_m^{\pm b} = 
\p^m\bar\chi_m^{\pm \bar b} = 0,$$
$$
\p_{cb} \chi_m^{\pm b} =  
\p_{\bar c\bar b} \bar\chi_m^{\pm \bar b} = 0,\quad 
\p_{cb} F^{\pm\pm~b\bar a} =  
\p_{\bar c \bar b} F^{\pm\pm~a\bar b} = 0 $$ 
which come from component analysis of the superfield constraint
$$
\N_a\N_b \p^{ a b} 
V_{1,1} = 
\bar\N_{\bar a}\bar \N_{\bar b} \p^{\bar a\bar b} 
V_{1,1} = 0.$$

The vertex operator in integrated form is given by 
$$U = \int d^2 z G^-_1 \bar G^-_1 G^+_0 \bar G^+_0 V$$
$$=\int d^2 z
 |e^{-i\sigma-\rho} {-{\e^{abcd}}\over 6} p_a ~(\N_b\N_c\N_d ) +i 
  p_a ~(\N_b \partial^{ab} ) +{i\over 2}
 \p x^{ab} ~(\N_a\N_b)|^2 V_{1,1}.$$
So the vertex operators for $g_{mn}$, $b_{mn}$ and $\phi$ are 
proportional to the symmetric
traceless, antisymmetric, and trace parts of 
$$ \int d^2 z  (\p x^m +  (\sigma^{mp})^a_b p_a\t^b k_p )
(\bar\p x^n + (\sigma^{nq})^c_d  \bar p_c \bar\t^d k_q ) e^{ik\cdot x}$$ 
while the vertex operator for 
$u_{i} \bar u_{j} F^{ij~a\bar a}$ is proportional to\foot
{In the formula below, $u^i$ is simply a notational device
for writing the SU(2) doublets using a single formula. However, it
appears to be related to the $u^i$'s in \rot\ since
$SU(2)_{outer}$ acts in a similar way to the $SU(2)$ of the $R^{jk}$
generators in \rgen. Understanding this relation
could be useful for making
all of the $D=6$ spacetime supersymmetries manifest.}
$$ \int d^2 z  |u^{1} (
- e^{-i\sigma-\rho}  p_a -{i\over 2}\e_{abcd}\t^b \p x^{cd} 
+{i\over 2}\e_{abcd} \t^b \t^c k^{de} p_e)
+u^{2} p_a )|^2  e^{ik\cdot x}.$$

\subsec   {Massless Compactification-Dependent Vertex Operators } 

The vertex operators for the massless moduli are constructed from 
the chiral-chiral $N=2$ worldsheet moduli
of the compactification manifold, $P_I$, which satisfy
$$G^+ P_I = \bar G^+ P_I = \Gtp P_I =\bar{\tilde G}^+ P_I =0,$$
$$J P_I = \bar J P_I = P_I.$$ 
These vertex operators are most conveniently
described by a $U(1)$-charged vertex operator $\Phi^{++}$
whose compactification-independent
part must contribute zero conformal weight at zero momentum. The most general
such vertex operator is  
$$\Phi^{++} = 
\sum_{m,n =-\infty}^{+\infty} e^{m(i\sigma+\rho)+n(i\bar\sigma+\bar\rho)}
\Phi^I_{m,n} (x,\t,\bar\t) P_I  .$$
It is easy to check that 
$ T_0\Phi^{++}=0$ implies that $\p^p\p_p\Phi^{++}=0$ and
that $G^-_0 \Phi^{++}  =\bar G^-_0 \Phi^{++} =0$ 
implies that 
$\Phi_{m,n}=0$ for $m>1$ or $ n>1$. 
 Furthermore, $J^{++}_0\Phi^{++}=
\bar J^{++}_0\Phi^{++}= 0$,
so the only remaining conditions to be analyzed is
$\tilde G^-_0 \Phi^{++}=\bar{\tilde G}^-_0\Phi^{++}=0$.

When acting on $\Phi_{1,1}$, $\tilde G^-$ has a quartic pole proportional
to $(\N)^4 \Phi^I_{1,1}$ and a triple pole with a term proportional to 
$(\N^3)^a \Phi^I_{1,1}$. When these are set to zero, $\tilde G^-$ still has
a double pole coming from $A$ and $B$ of \abcd\ which is proportional to 
$$[e^{-2\rho-i\s}(-{1\over 4}\e^{abcd} p_a p_b \N_c\N_d)
+i e^{-\rho }p_a \N_b \p^{ab}]\Phi^I_{1,1}.$$  
The double pole coming from $B$ cannot be cancelled and implies that  
$$\N_b\p^{ab}\Phi^I_{1,1}=0.$$
But the double pole coming from $A$ can 
be cancelled from the double pole of $B$
with $\Phi^I_{0,1}$, which is proportional to
$${i\over 2}
e^{-2\rho-i\s}(p_a p_b \p^{ab} + 2p_a \p x^{ab}\N_b)\Phi^I_{0,1},$$
implying that 
$$\e^{abcd}
\N_c\N_d\Phi^I_{1,1}=2i\p^{ab}\Phi^I_{0,1},\quad \N_a\Phi^I_{0,1}=0.$$
Using similar arguments, one can show that $\tilde G^-_0 \Phi^{++}=0$ 
implies that 
$$\e^{abcd}\N_c\N_d\Phi^I_{1,0}=2i\p^{ab}\Phi^I_{0,0},
\quad \N_a\Phi^I_{0,0}=0$$
and that $\Phi^I_{m,n}=0$ for $m<0$.
Analogously, $\bar{\tilde G}^-_0\Phi^{++}=0$ implies that 
$$\bar\N_{\bar b}\p^{\bar a\bar b}\Phi^I_{1,1}=
(\bar\N^3)^{\bar a} \Phi^I_{1,1}=0,$$
$$\e^{\bar a\bar b\bar c\bar d}
\bar\N_{\bar c}\bar\N_{\bar d}\Phi^I_{1,1}=2i
\p^{\bar a\bar b}\Phi^I_{1,0},\quad 
\bar\N_{\bar a}\Phi^I_{1,0}=0,$$
$$\e^{\bar a\bar b\bar c\bar d}
\bar\N_{\bar c}\bar\N_{\bar d}\Phi^I_{0,1}=2i
\p^{\bar a\bar b}\Phi^I_{0,0},\quad 
\bar\N_{\bar a}\Phi^I_{0,0}=0.$$

There are no residual gauge invariances in this case, 
so the physical degrees of freedom are described by the 
superfields
$$\Phi^I_{1,1}= t^I_{++} +\t^a \xi^{+ I}_a +
\bar\t^{\bar a} \bar \xi^{+~I}_{\bar a} +\t^a\bar\t^{\bar a} F^I_{a\bar a}
+\t^a\t^b \p_{ab}\Phi^I_{0,1} +
+\bar\t^{\bar a}\bar\t^{\bar b} \p_{\bar a\bar b}\Phi^I_{1,0},$$
$$\Phi^I_{0,1}= t^I_{-+} +\bar\t^{\bar a} 
\bar\xi^{- I}_{\bar a} +\bar\t^{\bar a}
\bar\t^{\bar b}\p_{\bar a \bar b} t^I_{--},$$
$$\Phi^I_{1,0}= t^I_{+-} +\t^a \xi^{- I}_a +\t^a\t^b\p_{ab} t^I_{--},$$
$$\Phi^I_{0,0}= t^I_{--}.$$
These describe the vector (or tensor)
multiplets coming from the Type IIA (or Type IIB) compactification. 
For example, the four NS-NS scalar moduli are described by
$t^I_{\pm\pm}$, the modulinos are described by $\xi^{\pm I}_a$ and
$\bar\xi^{\pm I}_{\bar a}$,  
while the Ramond-Ramond field-strength is described
by
$F^I_{a\bar a}$. 
Besides the mass-shell condition $\p_m \p^m =0$, these 
fields satisfy $\p^{ab}\xi^{\pm I}_a=
\p^{\bar a\bar b}\bar\xi^{\pm I}_{\bar a}=
\epsilon^{abcd}\p_{bc} 
F^I_{a\bar a} =
\epsilon^{\bar a\bar b\bar c\bar d}\p_{\bar b\bar c} 
F^I_{a\bar a} = 0$, which comes from the
superfield constraint 
$\N_{ b}\p^{ab}\Phi^I_{1,1}=
\bar\N_{\bar b}\p^{\bar a\bar b}\Phi^I_{1,1}= 0$.
For Type IIA (or Type IIB) compactifications,
$F^I_{a\bar a}$ describes the field-strength of a vector (or a 
self-dual tensor plus a scalar).  

The vertex operator in integrated form is given by 
$$U = \int d^2 z G^-_1 \bar G^-_1 \Phi^{++}$$
$$=\int d^2 z [(|e^{i\sigma +\rho}|^2 \Phi^I_{1,1} +
e^{i\sigma +\rho}\Phi^I_{1,0} + 
e^{i\bar\sigma +\bar\rho}\Phi^I_{0,1} + \Phi^I_{0,0}) (G^-_1 \bar G^-_1 P_I) 
$$
$$+
(e^{\rho+i\bar\sigma +\bar\rho} \Phi^I_{1,1} + e^\rho \Phi^I_{1,0})(\bar
G^-_1 P_I)$$
$$+
(e^{\rho+i\sigma +\bar\rho} \Phi^I_{1,1} + e^{\bar\rho} \Phi^I_{0,1})(
G^-_1 P_I)$$
$$+
e^{\rho+\bar\rho} \Phi^I_{1,1} P_I].$$
The vertex operator for the component fields are easily computed from
the $\t$ expansion of $U$. For example, the vertex operator for
$t^I_{--}$ is
$$\int d^2 z |e^{i\sigma +\rho}\t^a\t^b k_{ab}  + 1)G^-_1 
+
e^\rho\t^a\t^b k_{ab} |^2 P_I e^{ik\cdot x}$$
and for $t^I_{++}$ is 
$$\int d^2 z |e^{i\sigma +\rho}G^-_1 
+
e^\rho|^2 P_I e^{i k\cdot x}.$$ The vertex operator for the Ramond-Ramond
field-strength $F^I_{a\bar a}$ is 
$$\int d^2 z \t^a \bar\t^{\bar a}  |e^{i\sigma +\rho}G^-_1 
+
e^\rho|^2 P_I e^{i k\cdot x}.$$ 

Note that replacing $G^-$ with $\Gtm$ in the definition of $U$
only changes the `picture' and therefore does not provide new
physical states. One therefore
does not need to consider vertex operators constructed from
chiral/anti-chiral or anti-chiral/anti-chiral moduli of the
compactification manifold.  
\def\e{{\epsilon}}
\def\half{{1\over 2}}
\def\p{{\partial}}
\def\pb{{\bar\partial}}
\def\t{{\theta}}
\def\tb{{\bar\theta}}

\newsec{Almost Flat Approximation to the ${\rm AdS}_3\times {\bf S}^3$ 
Background}

\def\S{{\bf S}}
\def\T{{\bf T}}
\def\R{{\bf R}}
\subsec{Linearized Perturbation}

As discussed in \action, the six-dimensional part of the superstring
action in a Minkowski background  is proportional to 
$$S_0= \int d^2 \sigma \left[
{1\over 8} \epsilon_{abcd} \p x^{ab} \pb x^{cd} + p_a \pb\t^a +
\bar p_a \p\tb^a\right]. $$ 
where $\p={\p\over{\p\sigma_1}} -
i{\p\over{\p\sigma_2}}$ and
$\bar\p={\p\over{\p\sigma_1}} +
i{\p\over{\p\sigma_2}}$.
Here we have substituted $x^{ab}=x^m (\s_m)^{ab}$
and have omitted the chiral boson action for
$\rho$ and $\sigma$, which are free fields.

We now want to make a small deformation of this free-field
action, adding the vertex operators of a suitable self-dual RR three-form
and graviton, so as to describe an ${\rm AdS}_3\times \S^3$ background that is 
almost
flat.  This is appropriate for describing 
an ${\rm AdS}_3 \times \S^3$ background obtained by compactification
on K3 or $\T^4$ with large values of the onebrane and fivebrane charges
$Q_1$ and $Q_5$. By studying the
almost flat case, we will get clues about how ${\rm AdS}_3\times \S^3$
should be described in general in the framework we are using here,
and this will make possible a more complete description in section 7.

We begin by describing the explicit perturbation that is needed to get
from $\R^6$ to ${\rm AdS}_3\times \S^3$ in the almost flat case.  The symmetry
group of $\R^6$ is, of course, the six-dimensional Poincar\'e group,
and the subgroup that leaves fixed a point is the six-dimensional rotation
group $SO(6)$ (or $SO(5,1)$ if we use Lorentz signature).  The subgroup
of the symmetry group of ${\rm AdS}_3\times \S^3$ that leaves fixed a point
is $SO(3)\times SO(3)$ (or $SO(2,1) \times SO(3)$ if we use Lorentz signature),
since one can make $SO(3)$ rotations about any given point in the
homogeneous, isotropic spaces ${\rm AdS}_3$ and $\S^3$, but there are not
rotations that mix the two.

The reason to focus on the symmetries that fix a given point is the 
following.  No matter how small the cosmological constant may be,
${\rm AdS}_3\times \S^3$, if one looks at it ``in the large,'' is not a small
perturbation of $\R^6$.  The $\S^3$ is compact, for example, and the
deviation of ${\rm AdS}_3$ from flat $\R^3$ grows as one goes to infinity.
Viewing ${\rm AdS}_3\times \S^3$ as a perturbation of $\R^6$ is a local 
process;
for small cosmological constant, a portion of $\R^6$ near a given point,
say the origin $P$, can be regarded as an approximation to a portion of
${\rm AdS}_3\times \S^3$.  In view of what was said in the last paragraph,
the perturbation that goes from $\R^6$ to ${\rm AdS}_3\times \S^3$ must
break the $SO(6)$ group of rotations around $P$ to $SO(3)\times SO(3)$.

In the flat case, our notation was adapted to the local isomorphism
$SO(6)\cong SU(4)$.  The $SO(3)\times SO(3)$ subgroup of $SO(6)$ is instead
locally isomorphic to an $SO(4)$ subgroup of $SU(4)$, under which
the ${\bf 4}$ and $\bar {\bf 4}$ of $SU(4)$ become isomorphic and
transform as the ${\bf 4}$ of $SO(4)$.  We  can usefully proceed with
much the same notation as before, with one crucial difference that
in $SO(4)$ there is an invariant metric tensor $\delta_{ab}$, which
can be used to raise and lower indices.  

A self-dual two-form $H$ that preserves an $SO(3)\times SO(3)$ group of
rotations must at $P$ be of the form
\eqn\kson{-iH_{012}=H_{345}=2f,}
for some real constant $f$.  
Here we identify $x^0,x^1,x^2$ as ${\rm AdS}_3$ coordinates, and $x^4,x^5,x^6$
as $\S^3$ coordinates.
Note that in Euclidean signature the self-duality
condition reads $H=i*H$, which is the reason for the $i$ in \kson.
A Wick rotation on the $x^0$ coordinate (taking us to Lorentz signature
for ${\rm AdS}_3$) will make $H$ real.
\foot{$H$ is also real if the signature
is $(3,3)$, a case that is natural in group theory though not in physics,
as we will see in the next section.}
${\rm AdS}_3\times \S^3$ also has ``broken'' symmetries that do not leave
$P$ fixed.  In the almost flat approximation, these correspond to the
translations of $\R^6$, and invariance under them says that $H_{012}$
and $H_{345}$ should be constant.

We want to embed this $H$ field in the six-dimensional Type IIB
supergravity multiplet for which vertex operators were  constructed in the 
last section.  As we have seen, this multiplet contains the four RR 
tensor fields
$H^{\pm \pm}$.  There is also an NS tensor field.  However, we will begin
by considering the case of a perturbation by an RR tensor field,
and will extend the discussion to the NS perturbation only at the end
of the present section.

The low energy supergravity in six dimensions has
an $SO(5)$ R symmetry group.  The subgroup of $SO(5)$ that
does not mix RR and NS worldsheet vertex operators is
$SO(4)\cong SU(2)\times SU(2)$, where one $SU(2)$ acts on the first
$\pm$ index of $H^{\pm\pm}$, and the other acts on the second.
The four fields $H^{\pm \pm}$ 
thus transform as a vector of $SU(2)\times SU(2)
\cong SO(4)$, so if $H$ is real, it can be rotated into any desired
``direction'' by an $SU(2)\times SU(2)$ transformation.  (By $H$ being
real we mean of course that it is real if the spactime signature is taken
to be Lorentzian.)  If we denote
$H^{\pm\pm}$ as $H^{\alpha\alpha'}$, where $\alpha,\alpha'=1,2$ label
the two-dimensional representations of the two $SU(2)$'s, then the
reality condition on $H$ is $H^{\alpha\alpha'}=\epsilon^{\alpha\beta}
\epsilon^{\alpha'\beta'}\overline H_{\beta\beta'}$.  So a representative
example of a real field is a field with $H^{++}$ and $H^{--}$ equal
and real.  Hence with no essential loss of generality, we take
$H^{++}=H^{--}=H/2$, with $H$ as above.\foot{Since the $SU(2)\times SU(2)$ 
mentioned in this paragraph
is not a symmetry of the string theory, we should explain more
accurately in what sense we may rotate $H$ by $SU(2)\times SU(2)$.
The point is that the conformal field theory of the compactification
manifold K3 or ${\rm T}^4$ actually has $(4,4)$ worldsheet supersymmetry,
but we viewed it as a $(2,2)$ theory in the flat space construction
in sections 4 and 5.  By rotating the choice of  $(2,2)$ embedding in $(4,4)$,
we rotate what we mean by $H^{++}$ and $H^{--}$.}

{}From what we have seen in section 4, the vertex operators
for $H^{--}$ and $H^{++}$ are $p^a \bar p^{a'}$
and $(p^a e^\phi-{i\over 2} \epsilon_{abcd}\theta^b\partial x^{cd})
(\bar p^{a'}e^\phi-{i\over 2} \epsilon_{a'b'c'd'}\bar\theta^{b'}\bar
\partial x^{c'd'})$.  To get the precise $H$ field of eqn. \kson,
we must combine the left and right-moving parts of the vertex operators
in a way that is invariant under an $SO(4)$ subgroup in which the $p$'s
and $\theta$'s transform as a vector.  This is done just by contracting
with $\delta^{aa'}$, 
so the desired vertex operator is
\eqn\hv{V^{RR}_{H}=
f\int d^2 \sigma\left[
 p^a \bar p^a  + (p^a e^\phi - {i\over 2}\e_{abcd} \t^b \p x^{cd})
(\bar p^a  e^{\bar\phi} - {i\over 2}\e_{ab'c'd'} \tb^{b'} \pb x^{c'd'})\right] 
}
where $\phi = -\rho-i\sigma$ and $\bar\phi=-\bar\rho-i\bar\sigma.$
Here $f$ is a small parameter measuring the strength of the perturbation.
This vertex operator is just the sum of the ``squares'' of the two spacetime 
susy generators defined in \susy.  

Note that even after adding such an interaction
term to the Lagrangian, $e^\phi=e^{-\rho-i\sigma}$ is still a chiral
field of dimension zero, as its operator product singularity with
itself (and hence with $V^{RR}_{H}$) vanishes.  Other combinations of $\rho$
and $\sigma$ such as $e^{\rho-i\sigma}$ are no longer chiral.

To order $f$, the deformation to ${\rm AdS}_3\times \S^3$ is simply made
by adding $V_H$ to the Lagrangian.  In order $f^2$, we must expect
additional corrections.
For example, the Einstein equations read roughly $R_{ab}=(H^2)_{ab}$, so the
departure of ${\rm AdS}_3\times \S^3$ from flatness is in order $f^2$.
Near $P$, which we identify as
the point $x=0$, the ${\rm AdS}_3\times {\bf S}^3$ 
background is described by the
metric 
$$ g_{mn}= \eta_{mn} + {1\over 6} f^2 (\sigma_m)^{ae}
(\sigma_n)^{bf}\e_{abcd} x^{ce} x^{df} +O(f^4),$$
where $f$ is the same constant as in the three-form.
The vertex operator for the order $f^2$ deformation cannot be constructed
just using the results of section 5; in that section, we considered
metric perturbations of flat empty space
 obeying $R_{ab}=0$, while now the  relevant equation is $R_{ab}=(H^2)_{ab}$.
Nonetheless, it is reasonable to expect the $O(f^2)$ terms in the action
to have the form
\eqn\grav{\eqalign{ V_g=  f^2\int d^2 \sigma\biggl[&
{1\over {24}}\e_{abcd} x^{ae} x^{bf} \p x^{ce} \pb x^{df} 
- \dots \biggr]\cr }}
with the $\dots$ terms involving $\theta$-dependent
couplings such as $p\partial \theta x\partial
x$.   We will determine the details of the additional couplings in section
7 using spacetime supersymmetry, but for now we leave them undetermined
and analyze some general properties that will give us important clues
for later.

Already at this level, the problems of principle of adding RR fields
to the vacuum, in the RNS formalism, have been circumvented.
There are no spin fields in the Lagrangian that might make
computations difficult, and no ambiguity coming
from the different pictures in which an RR field can be written, since
a consistent choice of pictures has been made.
The price we pay for these simplifications is that --
in contrast to more familiar worldsheet actions in the RNS formalism,
or the flat case that we reviewed in the present formalism in section 4 --
 there are couplings of the ``matter'' fields
$x,\theta,$ and $p$ to the ``ghost'' fields $\phi$, $\bar\phi$.  
What saves the day is that the ghost couplings
have a very simple structure.  Because the Lagrangian
only has positive powers of $e^\phi$ and $e^{\bar \phi}$ (a fact that
reflects the fact that the vertex operators have this property, and
hence will persist in higher orders), the ghost
couplings are rather like ``screening charges''; in any given
computation of spectrum or perturbative scattering amplitudes, they
can be treated as infinitesimal perturbations, and taken into account
only up to some finite order.

If one performs the rescaling\foot{Note 
that the rescaling of $e^\phi$ and $e^{\bar \phi}$
in the following formula
amounts to the addition of a constant to the fields $\phi,$
$\bar\phi$, and so is a symmetry of the free action of these fields. The
factor of $-{1\over 4}$ is included to simplify comparison with the
exact non-linear action.}
$$p_a \to f^{-\half} p_a,\quad
\bar p_a \to f^{-\half} \bar p_a,
\quad \t^a \to f^{+\half} \t^a,\quad
\bar \t^a \to f^{+\half} \bar \t^a , $$
$$e^\phi \to -{f\over 4} e^\phi,\quad 
e^{\bar\phi} \to -{f\over 4} e^{\bar\phi},$$
then the action $S_{AdS}=S_0+V_H+V_g$ becomes
\eqn\nxjns{\eqalign{ S_{AdS}
=& \int d^2 \s \biggl[{1\over 8}
\epsilon_{abcd} \p x^{ab} \pb x^{cd} + p_a \pb\t^a +
\bar p_a \p\tb^a + p^a\bar p^a \cr &
+ f^2 \biggl((-{1\over 4}p^a e^\phi - {i\over 2}\e^{abcd} \t^b \p x^{cd})
(-{1\over 4}
\bar p^a  e^{\bar\phi} - {i\over 2}\e^{ab'c'd'} \tb^{b'} \pb x^{c'd'}) \cr
&
+ {1\over {24}}\e_{abcd} x^{ae} x^{bf} \p x^{ce} \pb x^{df}
+\dots \biggr) \biggr].\cr}}
This exhibits $f$ as a ``coupling constant''; the kinetic terms
are independent of $f$, and the interaction are all of order $f^2$.
Since the interaction terms are homogeneous and quartic
in the variables
$x,p,\theta,\,e^{\phi},\dots$ 
 a further rescaling 
\eqn\pxol{(x,p,\theta,e^{\phi},\bar p,\bar \theta, e^{\bar \phi})
\to f^{-1}(x,p,\theta,e^{\phi},\bar p,\bar \theta, e^{\bar\phi})}
puts the action in a form in which $f$ appears only
as an overall multiplicative constant:
\eqn\nxjns{\eqalign{ S_{AdS}'
=& {1\over f^2}\int d^2 \s 
\biggl[{1\over 8} \epsilon^{abcd} \p x_{ab} \pb x_{cd} + p_a \pb\t^a +
\bar p_a \p\tb^a + p^a\bar p^a \cr &
+  \left(-{1\over 4}p^a e^\phi - {i\over 2}\e^{abcd} \t^b \p x^{cd}\right)
\left(-{1\over 4}
\bar p^a  e^{\bar\phi} - {i\over 2}\e^{ab'c'd'} \tb^{b'} \pb x^{c'd'}\right) 
\cr &
+ {1\over {24}} \e_{abcd} x^{ae} x^{bf} \p x^{ce} \pb x^{df}
+\dots \biggr].\cr}}
{}From this point of view, the expansion we are making around the flat
case is an expansion in powers of the fields, which we consider
small; $S'_{{{AdS}}}$ is the ${\rm AdS}_3\times \S^3$ action up to fourth order
in the fields, except that some of the back reaction terms have not
yet been determined.

\subsec{Sigma Model With Supermanifold As Target}

Now we meet a crucial fact which substantially changes the character of
the problem from what we have seen up to this point,
and leads to one of the main insights of the present
paper.  In the flat case, the Lagrangian is linear in $p$ and
has the typical structure  ``second order in bosons, first order in 
fermions,'' of most physical supersymmetric actions.  But now, in \nxjns,
we  see that there is a $\bar p p$ term, and moreover that (at least
perturbatively in the fields) its coefficient is everywhere nonzero.  Hence
$p $ and $\bar p$ can be integrated out to give a Lagrangian for
$x, \theta$, and $\bar \theta$ only (plus couplings to the ``ghosts''
$\phi$, $\bar \phi$).

Solving for the equation of motion for $p^a$ and $\bar p^a$, one gets
the action  
\eqn\hxono{\eqalign{ S'_{AdS}
= {1\over f^2}
\int d^2 \s \biggl[&
{1\over 8}\epsilon^{abcd} \p x_{ab} \pb x_{cd} - \pb\t^a \p\tb^a \cr
& +{i\over 8} \e^{abcd}( e^\phi \p\tb^a \tb^b \pb  x^{cd}
+ e^{\bar\phi} \pb\t^a \t^b \p x^{cd}) \cr
& +{1\over {24}} \e_{abcd} x^{ae} x^{bf} \p x^{ce} \pb x^{df}
+\dots \biggr].\cr}}

Now let us try to interpret this Lagrangian.  The spacetime supersymmetry
generators, constructed in \susy, depend on $\phi$ and $\bar \phi$,
but only via positive powers of $e^\phi $ and $e^{\bar \phi}$.
There likewise are only positive powers of $e^{\phi}$ and $e^{\bar \phi}$
in $S_{AdS}$.
Hence, the part of $S'_{AdS}$ that is independent of $\phi$ and $\bar\phi$
must be spacetime supersymmetric
by itself.  Let us call this action $\widehat S_{AdS}$.

$\widehat S_{AdS}$ has some properties that are not obvious from its
origin via a small perturbation of the flat model.  
First of all, there is a continuous $SU(2)$ symmetry under 
which $\theta^{a \alpha}=(\theta^a,\bar \theta^a)$ transforms as a doublet
(for each fixed value of the $SO(4)$ index $a$).
To write the action in a manifestly $SU(2)$-invariant way, we introduce
the antisymmetric tensor $\epsilon_{\alpha\beta}$ with
$\epsilon_{\alpha\beta}\theta^{a \alpha}\theta^{b \beta}$=$\theta^a
\bar\theta^b +\theta^b\bar\theta^a$, 
$\epsilon^{\alpha\beta}\epsilon_{\beta\gamma}=
\delta^\alpha_\gamma$.  We furthermore note that $S'_{AdS}$ (and all
previous actions written in this section) has a manifest symmetry
under $z\leftrightarrow \bar z$, $\theta\to\bar\theta$, $\bar\theta\to 
 -\theta$.
Since $\theta\to\bar\theta,$ $\bar\theta\to -\theta$ is an $SU(2)$
transformation, the fact that the action has $SU(2)$ symmetry means
that this operation itself (without $z\leftrightarrow\bar z$) is a symmetry,
and hence that there is also symmetry under $z\leftrightarrow \bar z$,
with no action on the fields.  Hence we make a further change of notation
to exhibit this symmetry, which we will call worldsheet parity.  
To do so, we introduce real coordinates
$\sigma^1$, $\sigma^2$, with $z=\half(\sigma^1+i\sigma^2)$.
With these choices, we can write $\widehat S_{AdS}$ in the form
\eqn\osno{\eqalign{\widehat S_{AdS}= &{1\over f^2}\int d^2\sigma\biggl(
{1\over {8}}\epsilon_{abcd}\partial_ix^{ab}\partial_ix^{cd}-{1\over 2}
\partial_i\theta^{a\alpha}\partial^i\theta^{a\beta}\epsilon_{\alpha\beta}
\cr
& +{1\over {24}}\epsilon_{abcd}x^{ae}x^{bf}\partial_ix^{ce}\partial_i
x^{df}+\dots \biggr),\cr}}
which is manifestly invariant under $SU(2)$ and also under worldsheet parity.
The $SU(2)$ symmetry is not yet so striking at this point, because
the dependence on $\theta $ and $\bar\theta$ is so simple and we have
not determined the $\dots $ terms.  However, we will see in section 7
that the full sigma model action (modulo ghost couplings) does have the
$SU(2)$ symmetry.  

What is $\widehat S_{AdS}$?  The kinetic energy is quadratic for all bose
and fermi fields.  Thus, $\widehat
S_{AdS}$ is a ``sigma model with a supermanifold
as target space.'' 
The action, in other words, is of the general form $\int d^2\sigma
g_{IJ}\partial_i\Phi^I\partial^i\Phi^J$, with bosonic and fermionic
coordinates $\Phi^I$, $\Phi^J$ and metric $g_{IJ}$ on a target supermanifold
$M$.  The coordinates
are in fact  $x^{mn}$ and $\theta^{a\alpha}$, so the bosonic
and fermionic dimension of the
target space is $6|8$.  The worldsheet parity symmetry of $\widehat S_{AdS}$
means that there is no ``Wess-Zumino term.'' 
In this context, a Wess-Zumino term would be a term of the general form
$\int d^2\sigma \epsilon^{ij}B_{IJ}
\partial_i\Phi^I\partial_j\Phi^J$, with $B_{IJ}$ a two-form on $M$.

Moreover, spacetime supersymmetry acts on the sigma model in the most
elementary way: by  geometrical transformations of $M$.  
Recall that in the flat Minkowski
background, the spacetime supersymmetry generators
were given by
$$q^-_a=\oint p_a,\quad q^+_a=\oint (e^\phi p_a -{i\over 2}
\e_{abcd}\t^b\p x^{cd}),$$
$$\bar q^-_a=\oint \bar p_a,\quad q^+_a=\oint (e^{\bar\phi}\bar p_a -
{i\over 2}\e_{abcd}\tb^b
\bar\p x^{cd}).$$
Replacing $p^a$ with $\p\tb^a$ and $\bar p^a$ with $-\bar\p\t^a$, and
using the OPE's 
$$x^{ab}(y)x^{cd}(z)\to -\e^{abcd} \log|y-z|,\quad 
\t^a(y)\bar\t^b(z)\to \eta^{ab} \log|y-z|,$$ 
it is easy to check that 
under the susy transformation generated by
$v_\pm^a q^\pm_a +\bar v_\pm^a \bar q^\pm_a,$
the worldsheet fields transform as
\eqn\transf{
\delta\t^a = v_-^a + v_+^a e^\phi 
-{i\over 2}\epsilon_{abcd}\bar v_+^b x^{cd},}
$$\delta\tb^a = \bar v_-^a + \bar v_+^a e^{\bar\phi}  
-{i\over 2}\epsilon_{abcd} v_+^b x^{cd},$$
$$\delta x^{ab}=i( v_+^a \t^b -v_+^b \t^a
+\bar v_+^a \tb^b -\bar v_+^b \tb^a )$$
$$\delta \phi=\delta \bar\phi = 0.$$
This has the following very important property: if we drop the
ghosts $e^\phi$ and $e^{\bar\phi}$, then under supersymmetry,
$\theta$, $\bar \theta, $ and $x$ just transform 
under spacetime supersymmetry into functions of themselves.  
This means that spacetime supersymmetry acts (modulo ghost couplings)
by an automorphism of the target space $M$ of the sigma model,
and is a symmetry of the sigma model action (again modulo ghosts)
if and only if the metric of $M$ is invariant under that automorphism.
Thus, spacetime supersymmetry is just a supergroup of isometries of $M$.
Knowledge of this fact will enable us to determine $M$ and the sigma
model action in section 7.    Even if
the $e^\phi$ and $e^{\bar \phi}$ terms are included, spacetime supersymmetry
still acts by automorphisms of $M$, but now these are $\phi$ and 
$\bar\phi$-dependent automorphisms.

It is also illuminating to examine, in the present ``almost flat'' 
approximation
to ${\rm AdS}_3\times \S^3$, the structure of the spacetime 
supersymmetry currents.
Recall that in flat Minkowski space, the susy currents for 
$q^\pm_a$ were holomorphic (i.e.
$j_{\bar za}^\pm=0$ and $\bar\p j_{z a}^\pm=0$), 
while the susy currents
for $\bar q^\pm_a$ were similarly anti-holomorphic,
(i.e.
$\bar j_{za}^\pm=0$ and $\p \bar j_{\bar z a}^\pm=0$). 
Is this true after the perturbation to ${\rm AdS}_3\times \S^3$?
The susy transformations of \transf\ 
are symmetries of the quadratic part of
the $\widehat S_{AdS}$.  (They are not symmetries of the quartic
terms as we have not determined the $\dots$ terms; also, as we will
see in the next section, there are higher order corrections to the
supersymmetry transformations.  But the  quadratic approximation
to the Lagrangian is sufficient for our present purposes.)
However, as symmetries of the free kinetic energy of $x,\theta,\bar\theta$,
these transformations are not generated by purely holomorphic or
antiholomorphic currents.
Using the Noether method to find the currents that generate the
symmetry \transf\ of the free action, 
one finds for example that $j_{\bar z a}^+$ no longer vanishes 
in the almost flat approximation to ${\rm AdS}$, but is equal to 
$j_{\bar z a}^+={i\over 2} \e_{abcd}
(\bar\p\t^b) x^{cd}$. There is thus a conserved supercurrent
$\bar\p j_{z a}^+ +\p j_{\bar z a}^+ =0$, but it has both $(1,0)$ and
$(0,1)$ pieces.  In the flat case, the rotation symmetries have this
property, but the supersymmetries are purely holomorphic or purely
antiholomorphic.  After deformation to ${\rm AdS}_3\times \S^3$ (with
RR background), the spacetime supersymmetries are carried by mixtures
of left and right-movers.

\subsec{Structure Of Unbroken Supersymmetries}

The following further remark will help in determining the sigma model
action $\widehat S_{AdS}$ which describes the system modulo ghost
terms.  Though there will be corrections to \transf\ of higher order in the
fields, these formulas suffice for determining which supersymmetries
are broken and which are unbroken in the classical vacuum of the
sigma model at $x=0$.  Setting $e^\phi$ and $e^{\bar \phi}$ to zero,
we see from \transf\ that the unbroken supersymmetries
(which act trivially on $\theta$, $\bar\theta$ if $x=0$) are precisely
those with $v_-^a=\bar v_-^a=0$.
In particular, in the sigma model, half of the supersymmetries are 
broken and half are unbroken.

 It is instructive
to calculate the algebra generated by the unbroken supersymmetries.
The anticommutator of two unbroken supersymmetries, if not zero,
must of course be an unbroken  bosonic symmetry.  The unbroken 
bosonic symmetries are the $SO(4)$ rotations around $P$.
A straightforward computation shows that the anticommutator  of two unbroken
supersymmetries is indeed the generator of an infinitesimal
rotation of the $x$'s and $\theta$'s.  
The unbroken supersymmetries and rotations generate a supergroup that
we will call $SU'(2|2)$; its properties are described more fully in the
next section.  We can think of the unbroken symmetry group $H\cong SU'(2|2)$
 as a supergroup of
``rotations'' of $x,\theta,\bar\theta$.  The broken bosonic
symmetries are the translations $\delta x={\rm constant}$.  Similarly
the broken supersymmetries (with $v_-^a$, $\bar v_-^a$ nonzero) are
translations of $\theta$, $\bar\theta$.  
There is a spontaneously broken translation symmetry for every
coordinate $x,\theta,\bar\theta$, so the target space $M$ of the sigma
model is a homogeneous space for the spacetime supersymmetry.
The coordinates transform in the adjoint representation of $SU'(2|2)$
(this will become more obvious in section 7 when we describe this
supergroup in detail),
so the broken symmetries, that is the translation generators of
$x,\theta,\bar\theta$, do likewise.

Even though our analysis has been based on knowing the action only
to very low order in the perturbation parameter $f$, it is highly
plausible that the identification of the target space as a homogeneous
space is general.
If so, we can determine what $M$ must be.
In fact, the spacetime supersymmetry of Type IIB string theory
on ${\rm AdS}_3\times \S^3\times {\rm K3}$ is $SU'(2|2)^2$, where $G=SU'(2|2)$
is the supergroup encountered in the last paragraph.
For $M$ to be a homogeneous space for $G\times G$, it must be of the
form $(G\times G)/H$ for some subgroup $H$ of $G\times G$; to get
the right dimension for $M$, $H$ must be of dimension $6|8$.
$H$ is in fact the supergroup generated by the unbroken supersymmetries
at any given classical vacuum, which we may as well take to be the vacuum
at $P$.  From the analysis in the last paragraph, $H$ is therefore
isomorphic to $G$.  Moreover, to get the unbroken supersymmetries to
transform in the adjoint representation of $H$, $H$ must be a diagonal
subgroup of $G\times G$.  (This embedding of $H\cong G$ in $G\times G$ is
practically unique even without the knowledge of how the broken symmetries
transform.) If $H$ is such a diagonal subgroup, then $M=(G\times G)/H$
is just a copy of $G$, with $G\times G$ acting on $M=G$ by left and right
multiplication.

This identification of $M$ can further be confirmed by the following simple
consideration.  
Ignoring fermionic variables, the ${\rm AdS}_3\times \S^3$ manifold
can be identified as the group manifold $SU(1,1)\times SU(2)$.
This is the bosonic reduction of the $SU'(2|2)$ manifold.\foot{To be more
precise, it is the bosonic reduction of the alternative
real form $SU'(1,1|2)$ of $SU'(2|2)$.}  
So on this
grounds alone, the bosonic reduction of $H$ must be the bosonic reduction
of a diagonal subgroup of $G\times G$. 
%% Combining this knowledge
%% with the fact that $H$ has dimension $6|8$ uniquely determines $H$ to
%% be a diagonal subgroup of $G\times G$.

What about the terms in $S'_{AdS}$ with ghosts? 
They cannot be written in a manifestly $G\times G$-invariant way,
but must be $G\times G$-invariant anyway, with a suitable action of
$G\times G$.  We postpone the analysis of how this comes about to section 8.

\subsec{NS Perturbation And Wess-Zumino Term}

We will see in the next section that, considering only operators
with two derivatives, the sigma model with target space $G$ can be
generalized to include one more $G\times G$-invariant interaction,
namely a Wess-Zumino term.
 In fact, usually, in a sigma  model
with target space a group manifold, conformal invariance can be achieved
only if such a  term is present with the right coefficient.  
In the next section, we will
see that the sigma model with $SU'(2|2)$ target has an exceptional
property: it is conformally invariant
for any value of the Wess-Zumino coefficient (including zero). 
Thus, omitting the Wess-Zumino term, we get a conformal field theory
description of the ${\rm AdS}_3\times \S^3$ model with RR background.
But the Wess-Zumino term should also have a physical interpretation
(as it admits a deformation of the $N=4$ symmetry of the flat model
described in section 4).

In fact, its interpretation is rather simple. 
So far in this section, we have focused on giving an expectation value to
the RR three-form fields on ${\rm AdS}_3\times \S^3$.  
One can also turn on an NS three form field, with the same ansatz \kson.  
(In the interpretation of the model in terms of strings on $\R^6\times {\rm 
K3}$
or $\R^6\times \T^4$, this corresponds to having NS as well as Dirichlet 
fivebranes wrapped on K3 or $\T^4$.)  To first order in $f$, this
perturbation is made by adding a vertex operator $V_H^{NS}$ of the
$NS$ $B$-field.  As shown in section 5, that 
vertex operator contains no ghost couplings, so
if one perturbs the flat model by the NS three-form only, matter
and ghosts are decoupled as in the flat case, while any generic
combination of the RR and NS fields gives matter-ghost couplings.

In fact, any generic linear combination of RR and NS perturbations leads
to the general structure we have found: the action contains 
a term $\bar p p$ (which appears in the RR vertex operator) as a result
of which $p$ and $\bar p$ can be integrated out, to give a second
order sigma model action for the group $SU'(2|2)$.  The NS perturbation
appears as a Wess-Zumino term in this action.  We will describe
the spacetime supersymmetric Wess-Zumino term in section 7 and see
its relation to $V_H^{NS}$.

If instead of a generic combination of RR and NS fields, one makes
the NS perturbation only, then one has two options: 

(1) There is
no $p \bar p$ term in the action,
so it is natural to use a description with first order kinetic
energy for fermions.  This description will be given in section 10 and
turns out to be rather simple because the fermions can be treated as
free fields.
 
 (2) On the other hand, one can regard the NS model as a model
obtained with a perturbation by $V_H^{NS}+\epsilon V_H^{RR}$, in
the limit as $\epsilon\to 0$.  For every nonzero $\epsilon$,
there is a $\bar p p$ term, making it possible to integrate out $p$.
Moreover, after rescaling the $\theta$'s by a factor of $\epsilon^{-1/2}$, 
the resulting Lagrangian has a limit
as $\epsilon\to 0$.  This gives a different description of the
NS model, as a WZW model of $SU'(2|2)$.  

It will be shown in section 10 that these two descriptions
are equivalent by rewriting the currents of WZW $SU'(2|2)$ model
in terms of free fermions and an additional bosonic WZW model
on $SU(2)\times SL(2)$.
Whether one follows route (1) or route (2), the model with the NS perturbation
only has left and right-moving $SU'(2|2)$ current algebra, while
any model that includes also (or only) an RR perturbation
has spacetime supersymmetries that are carried by a mixture of
left-movers and right-movers.
\def\S{{\bf S}}
\def\R{{\bf R}}
\def\2{{\bf 2}}
\def\T{{\bf T}}
\def\Sdet{{\rm Sdet\,}}
\newsec{Sigma Model Of $SU'(2|2)$}

\subsec{Construction Of The Model}

In this section, we will investigate the model that was suggested
by the discussion in section 6, namely the two-dimensional sigma
model with target the group manifold of $G=SU'(2|2)$.\foot{
Aspects of sigma
models on supergroup manifolds have also been recently considered
independently in \ref\betl{M. Bershadsky, A. Vaintrob and S. Zhukov,
to appear.}.}

First of all, in general the group (or rather supergroup)
$U(n|m)$ is the group of unitary
transformations of a complex vector space of dimension $n|m$, that is,
bosonic dimension $n$ and fermionic dimension $m$.  An element $X$ of $U(n|m)$
can be represented by a matrix
\eqn\huis{X=\left(\matrix{ A & B \cr C & D\cr}\right),}
where $A$ and $D$ are $n\times n$ and $m\times m$ bosonic matrices,
and $B$ and $C$ are $n\times m$ and $m\times n$ fermionic matrices.
Such a matrix has a Berezinian or superdeterminant, characterized
by the fact that for $X,Y\in U(n|m)$, $\Sdet XY=\Sdet YX$ and that
if $B=C=0$, $\Sdet X= \det A \det^{-1}B$.

\def\Str{{\rm Str\,}}
The Lie algebra (or to be more precise, superalgebra)
of $U(n|m)$ consists of matrices of the
form
\eqn\xnsn{ x=\left(\matrix{ a & b \cr c & d\cr}\right),}
where $a$ and $d$ are 
bosonic and $b$ and $c$ are fermionic; $a$ and $d$ are hermitian
and $b$ and $c$ are hermitian conjugates.  One defines the supertrace
of $x$ as $\Str x = \Tr \,a - \Tr\, d$.

There are two similar ways to relate $U(n|m)$ to a group of dimension
one less.  We can require that $\Str x=0$ or equivalently that $\Sdet X=1$.
This gives a supergroup $SU(n|m)$.
Or we can take the quotient by scalars, considering $x$ trivial if it
is a multiple of the identity and identifying two $X$'s that are scalar
multiples of each other.

For $n\not= m$, the two operations are equivalent to each other locally.
If $x$ is a nonzero 
multiple of the identity, its supertrace is nonzero if $n\not= m$,
so requiring the supertrace to be zero is equivalent at the Lie
algebra level to removing the constants.

For $n = m$, the story is different.  One has $\Str 1=0$, so 
asking that the supertrace of $x$ vanishes does not remove the scalars.
One can form a smaller supergroup, of dimension two less than that
of $U(n|n)$, by requiring that $\Str x=0$ and also working modulo the
constants.  We will call this group $SU'(n|n)$ (mathematically
the name ${\cal A}_{n-1,n-1}$ has been used).  
The Lie algebra of $SU'(n|n)$ consists of matrices as
in \xnsn\ with $\Tr \,a = \Tr \,d = 0$.

\def\n{{\bf n}}
The bosonic part of $SU'(n|n)$ is $SU(n)\times SU(n)$, generated by
$a$ and $d$.  The fermionic generators of $SU'(n|n)$ -- that is, the
matrices $b$ and $c$ -- transform as $\n\otimes \bar\n \oplus \bar\n
\otimes \n$.  At this stage we notice a further coincidence that arises
only for $n=2$.  In this case, the representations $\n$ and $\bar\n$
are isomorphic, so the odd generators of $SU'(2|2)$ consist of two
copies of $\n\otimes \n$.  As a result, it turns out that the group
$SU'(2|2)$ has a group $R= SL(2,\R)$ of outer automorphisms, commuting with
the bosonic generators and rotating the two fermionic copies of ${\bf 2}\otimes
{\bf 2}$.\foot{ After constructing the action, we will make a Wick
rotation to a group $SU'(1,1|2)$ that is more directly relevant to
the physics in Lorentz signature.  For this group, the outer automorphism
group is $SU(2)$ rather than $SL(2,\R)$.  We will also consider
later the case of Euclidean signature.}

To describe explicitly the Lie algebra of $SU'(2|2)$ with its outer 
automorphism
group, we proceed as follows.  The bosonic part of $SU'(2|2)$ is
$SU(2)\times SU(2)$, which we identify at the Lie algebra level with
$SO(4)$.  The ${\bf 2}\otimes {\bf 2}$ of $SU(2)\times SU(2)$ is the
vector of $SO(4)$.  So the odd or fermionic generators of $SU'(2|2)$
consist of a pair of $SO(4)$ vectors; we write them as $S_{a\,\alpha}$,
where $a=1,\dots, 4$ is a vector index of $SO(4)$, and $\alpha=1,2$
labels the two vectors. The outer automorphism group will act on the $\alpha$
index.  The bosonic generators of $SU'(2|2)$ transform in the adjoint
or antisymmetric tensor representation of $SO(4)$; we write them
as $K_{ab}$, where $a,b=1,\dots, 4$ and $K_{ab}=-K_{ba}$.  The Lie
algebra of $SU'(2|2)$ can be described by the following formulas:
\eqn\intu{\eqalign{ [K_{ab},K_{cd}] & = \delta_{ac}K_{bd}-\delta_{ad}K_{bc}
              -\delta_{bc}K_{ad}+\delta_{bd}K_{ac}  \cr
                    [K_{ab},S_{c\alpha}] & = \delta_{ac}S_{b\alpha}
                                   -\delta_{bc}S_{a\alpha}\cr
                    \{S_{a\alpha},S_{b\beta}\}& = \half \epsilon_{\alpha\beta}
                         \epsilon_{abcd}K^{cd}.\cr}}
Here $\epsilon_{\alpha\beta}$ and $\epsilon_{abcd}$ are the antisymmetric
tensors of $SL(2,\R)$ and $SO(4)$ and we shall define
$\e_{12}=\e^{21} = \e_{1234}=1$.  The Jacobi identity is readily verified;
the $SL(2,\R)$ invariance is manifest.

Now we would like to describe a two-dimensional
sigma model, with target space the
group manifold of $G=SU'(2|2)$, which is invariant under the left
and right action of $G$ on itself.
The basic field of the sigma model will be a field $g$ that takes values
in  $G$; the model should be invariant under $G\times G$ acting by
\eqn\nxonx{g\to AgB^{-1}, ~{\rm with}~ A,B\in G.}
By a sigma model, we mean a model with an action of the general form
\eqn\ansin{S={1\over 2}\int d^2\sigma g_{IJ}\partial_i  
\Phi^I\partial^i\Phi^J,}
where $\Phi^I$ are coordinates on the group manifold, and $g_{IJ}$
is a metric on $G$.
The condition for the sigma model to be left and right invariant
is that the metric $g$ should be invariant.   Given such a metric,
its restriction to the identity element $1\in G$ is certainly
invariant under the subgroup of $G\times G$ that leaves the identity
element fixed (this is the diagonal subgroup defined by $A=B$).  Conversely,
any metric at the identity that is invariant under the stabilizer
of the identity
can be transported over the whole manifold using the $G\times G$
symmetry to give an invariant metric on the whole group manifold.

\def\ll{\langle}
\def\rr{\rangle}
The tangent space to $G$ at the identity is naturally isomorphic with
the Lie algebra of $G$; the diagonal subgroup $A=B$ acts on the tangent
space by conjugation.
So sigma models of this form are classified by $G$-invariant inner products
on the Lie algebra of $G$.  If we denote the inner product of two
elements $x,y$ of the Lie algebra as  $\ll x,y\rr$, then
the condition of invariance is that for any $x,y,z$ one has
\eqn\hsin{\ll [x,y\},z\rr+(-1)^{xy}\ll y,[x,z\}\rr=0.}
(Here $[~,~\}$ denotes the bracket in the Lie superalgebra;
we will henceforth write just $[~,~]$.)
Up to a scalar multiple, the most general invariant inner product on the
$SU'(2|2)$ Lie algebra reads
\eqn\sinsin{\eqalign{ \ll K_{ab}, K_{cd}\rr & = \epsilon_{abcd} \cr
                      \ll S_{a\alpha},S_{b\beta}\rr & = \delta_{ab}
                      \epsilon_{\alpha\beta},  \cr}}
with other components vanishing.   

Note that, unlike most familiar
situations encountered in physics, this inner product cannot conveniently
be written as $\ll x,y\rr=\Tr\,xy$ with the trace taken in some representation
of $G$.    Perhaps it is important to point out that the
basic $2|2$-dimensional representation of $SU(2|2)$ cannot be interpreted
as a representation of $SU'(2|2)$ (operators that one would try to define
as $SU'(2|2)$ generators in this representation do not close on $SU'(2|2)$
but on its extension $SU(2|2)$), so we cannot
define a quadratic form on the $SU'(2|2)$ Lie algebra via a supertrace
in this representation.  The smallest representation of $SU'(2|2)$
is the adjoint representation.

The sigma model action for a $G$-valued field $g$ can now be introduced.
Up to a possible multiplicative constant, it is
\eqn\posnon{S={1\over 2}\int d^2\sigma\, \ll g^{-1}\partial_i g,
g^{-1}\partial_i g\rr.}
We would now like to expand this action near $g=1$,
and compare the expansion to the action found from a different
point of view in the last section.
For this, we will pick an explicit parametrization of the group
manifold and compute the action in detail.

\subsec{ Evaluation Of The Action}

We wish to evaluate the action explicitly up to quartic order in the
fields.  Let us first discuss this procedure in general, for any
Lie group $G$ with generators $T_A$ and an invariant quadratic form
$\ll~,~\rr$ on the Lie algebra.
We introduce coordinates $\Phi^A$ and parametrize an element
of the group near the identity as
\eqn\nogho{g=\exp(\Phi^AT_A).}
Now we compute
\eqn\ogho{g^{-1}dg=d\Phi^AT_A+{1\over 2}d\Phi^A\Phi^B[T_A,T_B]
+{1\over 6}d\Phi^A\Phi^B\Phi^C[[T_A,T_B],T_C]+O(\Phi^4).}
Hence we have
\eqn\unnu{\ll g^{-1}\partial_ig,g^{-1}\partial_ig
\rr = \partial_i\Phi^A \partial_i\Phi^B\ll T_A,T_B\rr
-{1\over 12}\partial_i\Phi^A\Phi^B \partial_i\Phi^C\Phi^D\langle
[T_A,T_B],[T_C,T_D]\rangle +O(\Phi^5).}
One can also explicitly describe the transformation of the $\Phi^A$
under the $G\times G$ symmetry.  We introduce infinitesimal parameters
$\epsilon_L^A$, $\epsilon_R^B$ for the left and right group actions,
and abbreviate $\Phi=\Phi^AT_A$, $\epsilon_L=\epsilon_L^AT_A$,
$\epsilon_R=\epsilon_R^AT_A$.
The transformation
\eqn\snonx{\exp(\Phi)\to \exp(\epsilon_L)\exp(\Phi)
\exp(-\epsilon_R)}
 can be expanded in powers of $\Phi$
to give
\eqn\huccup{\delta\Phi=\epsilon_L-\epsilon_R+{1\over 2}[\epsilon_L+\epsilon_R,
\Phi]+{1\over 12}\left[[\epsilon_L-\epsilon_R,\Phi],\Phi\right]+O(\Phi^3).}
  Clearly, the unbroken symmetry,
which leaves  fixed the classical vacuum at $g=1$, is generated  
by $\epsilon_L+\epsilon_R$, and the broken symmetry, which 
shifts the $g=1$ vacuum, is generated by $\epsilon_L-\epsilon_R$.
The conserved currents generating the symmetries are
\eqn\sonxonxx{\eqalign{J_L(\epsilon_L)=& \ll\epsilon_L,dg g^{-1}\rr
=\ll\epsilon_L, d\Phi^AT_A-{1\over 2}d\Phi^A\Phi^B[T_A,T_B]
\cr
&
+{1\over 6}d\Phi^A\Phi^B\Phi^C[[T_A,T_B],T_C]\rr +O(\Phi^4), \cr
 J_R(\epsilon_R)=& -\ll\epsilon_R, g^{-1}dg\rr
=-\ll \epsilon_R, d\Phi^AT_A+{1\over 2}d\Phi^A\Phi^B[T_A,T_B] \cr
& +{1\over 6}d\Phi^A\Phi^B\Phi^C[[T_A,T_B],T_C]\rr +O(\Phi^4). \cr}}

For the $SU'(2|2)$ case, we further write
\eqn\groupexp{
\Phi={1\over 4}\epsilon_{abcd} x^{ab}K^{cd}+\theta^{a\alpha}S_{a\alpha},}
with bosonic
and fermionic coordinates $x^{ab}$, $\theta^{a\alpha}$.
(The coefficient of $K^{cd}$ has been written as a multiple
of $\epsilon_{abcd}x^{ab}$ rather
than $x^{cd}$, since this leads to formulas that agree better with the
$SU(4)$-invariant formulas of the flat case.)
In this case we can compute
\eqn\umbox{d \Phi^A\Phi^B[T_A,T_B]= \left(dx^{ae}x^{be}
      -\half \epsilon^{abcd}\epsilon_{\alpha\beta}
        \theta^{c\alpha}\theta^{d\beta}\right)K_{ab}
      +\half\epsilon_{abcd}\left(dx^{ab}\theta^{c\alpha}-x^{ab}
      d\theta^{c\alpha}\right)S^d_{\alpha}.}
With the aid of this formula, one finds that to this order, the action is
\eqn\gumbox{\eqalign{
S=&{1\over 2}\int d^2\sigma\left({1\over 4}\epsilon_{abcd}d
x^{ab}*dx^{cd} -d\theta^{a\alpha}*d
\theta^{a\beta}\epsilon_{\alpha\beta}\right.  \cr &
-{1\over 12} \epsilon_{aba'b'}\left(d x^{ae}x^{be}-\half\epsilon_{abcd}
d\theta^{c\alpha}\theta^{d\beta}\epsilon_{\alpha\beta}\right)*
\left(d x^{a'e'}x^{b'e'}-\half\epsilon_{a'b'c'd'}
d\theta^{c'\alpha}\theta^{d'\beta}\epsilon_{\alpha\beta}\right)
\cr
& \left.+{1\over 48}\epsilon_{\alpha\beta}\epsilon_{abcd}
\left(dx^{ab}\theta^{c\alpha}-x^{ab}d\theta^{c\alpha}\right)*\epsilon_{a'b'c'd}
\left(dx^{a'b'}\theta^{c'\beta }-x^{a'b'}d\theta^{c'\beta}\right)\right).\cr}}
Here to avoid clutter, we have adopted the convention that if $A$ and $B$
are fields, $dA * dB$ is short for $\partial_i A \partial^iB$.
Using \sonxonxx, we can also now compute that
the supersymmetry currents  are
\eqn\nonson{\eqalign{S_{L\alpha}^a & = d\theta^a_\alpha-{1\over 4}
\e^{abcd} (d\t^b_\alpha x^{cd}
-\t^b_\alpha d x^{cd})
\cr
S_{R\alpha}^a & =-d\theta^a_\alpha -{1\over 4} 
\e^{abcd} (d\t^b_\alpha x^{cd}
-\t^b_\alpha d x^{cd} )
 \cr}}
up to second order in the fields.
The rotation currents  $K_{L\,ab}$ (generating the left
action of $K_{ab}$) can be
similarly computed:
\eqn\honco{K_L^{ab}= dx^{ab}-\half\epsilon^{abcd}dx^{ce}x^{de}
+\half\epsilon_{\alpha\beta}(d\theta^{a\alpha}\theta^{b\beta}-d\theta^{b\alpha}
\theta^{a\beta}).} 
When it is likely to cause no confusion, we use the same name $K$ or $S$
for an element of the Lie algebra and the corresponding conserved current
or charge.

To analyze the ghost couplings in the next section, we will also
need the Maurer-Cartan equations.  For a general group $G$ with Lie algebra
basis $T_A$, obeying $[T_A,T_B]=f_{AB}^CT_C$, 
define the right-invariant currents by expanding $dg g^{-1} =\sum_AT_AJ^A$.
The $J^A$ are related to $J_A=\ll T_A,dg g^{-1}\rr$ by ``raising an index''
with the quadratic form $\ll ~,~\rr$.
  Since $d(dg\,g^{-1})=-dgg^{-1}\wedge
dg g^{-1}$, the $J^A$ obey the Maurer-Cartan equations
\eqn\theyobey{dJ^A=-{1\over 2}f_{BC}^AJ^B\wedge J^C.}
Of course, the $J_A$ obey an equivalent equation.
In the case of $SU'(2|2)$, these equations give 
\eqn\turgog{\eqalign{dS_{a\,\alpha} & 
= -{1\over 2}\epsilon_{abcd} K^{bc}\wedge S^d_\alpha\cr
 dK^{ab} & = 
\half\epsilon^{abcd}K^{ce}\wedge  K^{de}-\epsilon_{\alpha\beta}
 S^{a\alpha}\wedge S^{b\beta},\cr}}                                
where here the currents can be either left or right currents.

Finally, we shall meet in the next section the right-invariant two-forms
\eqn\ploopo{\omega_L^{\alpha\beta}=S_L^{a\alpha}\wedge S_L^{a\beta}.}
Note that $\omega_L^{\alpha\beta}=\omega_L^{\beta\alpha}$; 
the $\omega_L$'s transform
in the spin one representation of the $SU(2)$ group $R$
of outer automorphisms.
A small calculation using the Maurer-Cartan equations shows that
the $\omega_L$'s viewed as two-forms on the group manifold
are closed, $d\omega_L^{\alpha\beta}=\half 
\epsilon_{abcd}S^{a\alpha}\wedge
S^{b\beta}
\wedge K^{cd}+{\alpha\leftrightarrow\beta}=0$, where fermi statistics have been
used.  Since the second Betti number of the $SU'(2|2)$ group manifold
is zero, the two-forms $\omega_L^{\alpha\beta}$ are actually exact,
$\omega_L^{\alpha\beta}=d\lambda_L^{\alpha\beta}$ for some $\lambda_L^{\alpha
\beta}$.  Though the $\omega_L$'s are right-invariant, the $\lambda_L$'s cannot
be chosen to be right-invariant.  The $\omega_L$'s are thus non-trivial
if viewed as elements of the Lie algebra cohomology of $SU'(2|2)$ and are,
in fact, related to the existence of nontrivial central extensions of this 
group.  But the fact that $\omega_L^{\alpha\beta}=d\lambda_L^{\alpha\beta}$
for some $\lambda_L$'s, even though not right-invariant,
means that the interaction terms
\eqn\koopo{\int \sum_{\alpha\beta}d_{\alpha\beta}\omega_L^{\alpha\beta} }
that one might think of adding to a two-dimensional sigma model action
(with ``coupling constants'' $d_{\alpha\beta}$) are actually trivial.
This fact will be important.

It will also be useful to note that $\omega_L^{\a\b}$ is invariant under
all $SU'(2|2)\times SU'(2|2)$ transformations except for the left susy
transformations generated by the charges $\oint S_L^{a\a}$. The commutation
relations of \intu\ together
with the Maurer-Cartan equations of \turgog\ imply that 
\eqn\omlcom{
[\omega_L^{\a\b}, \oint S_L^{a\g}] =\half\e_{abcd}(\e^{\b\g} K_L^{bc}
S^{d\a}_L + \e^{\a\g} K_L^{bc} S_L^{d\b}) = - \e^{\b\g} dS_L^{a\a} - 
\e^{\a\g} dS_L^{a\b}.}
Since $\omega_L^{\a\b}= d\lambda_L^{\a\b}$, \omlcom\ implies that 
\eqn\llcom{
[\lambda_L^{\a\b}, \oint S_L^{a\g}] =
- \e^{\b\g} S_L^{a\a} - 
\e^{\a\g} S_L^{a\b} + dM_L^{a\a\b\g}}
for some $M_L^{a\a\b\g}$.
Similarly, one can show that  
\eqn\omrcom{
[\omega_R^{\a\b}, \oint S_R^{a\g}] = - \e^{\b\g} dS_R^{a\a} - 
\e^{\a\g} dS_R^{a\b},\quad 
[\lambda_R^{\a\b}, \oint S_R^{a\g}] =
- \e^{\b\g} S_R^{a\a} - 
\e^{\a\g} S_R^{a\b} + dM_R^{a\a\b\g}}
where $\omega_R^{\a\b}=$
$S_R^{a\a} \wedge S_R^{a\b}$
$ = d\lambda_R^{\a\b}.$

\subsec{ The Signature Of Spacetime}

Because of the  $\epsilon_{abcd}$ symbol multiplying the term quadratic
in $x$, the action introduced above describes motion of strings in a target
space with signature $---+++$.
The reason for this signature is that we started with the  group $SU'(2|2)$,
whose bosonic part is $SU(2)\times SU(2)$.  Going back to the description
of the Lie algebra in eqn. \xnsn, if $a$ and $d$ are generators of the
two $SU(2)$'s, then our metric on the Lie algebra was a multiple
of $\ll a,a\rr=-\Tr\, a^2$, $\ll d,d\rr =\Tr\, d^2$.  Thus one $SU(2)$
has $---$ signature, and the other has $+++$ signature.  To get
Lorentz signature, we should replace one of the $SU(2)$'s by $SU(1,1)$
(which is locally isomorphic to $SL(2,\R)$).  $SU(1,1)\times SU(2)$
is the bosonic part of a supergroup that we might call $SU'(1,1|2)$.
Since the analytic
continuation from $SU(2)$ to $SU(1,1)$ maps the quadratic form $-\Tr\, a^2$
of signature $---$ to a form of signature $-++$, the $SU'(1,1|2)$ model
has Lorentz signature   $-+++++$.  Indeed, the manifold ${\rm AdS}_3\times
\S^3$ (with Lorentz signature on ${\rm AdS}_3$) can be understood as the group
manifold $SU(1,1)\times SU(2)$.

With Euclidean signature, ${\rm AdS}_3$ is not a group manifold, so there is no
real form of $SU'(2|2)$ that will give a positive signature ${\rm AdS}_3\times
\S^3$ model.  However, at the cost of losing the reality of the Lagrangian,
it is possible, as follows,
 to make an analytic continuation to a model with
Euclidean signature in the target space.    
Let $x^+$ and $x^-$ be the self-dual and anti-self-dual
parts of $x$.  Then the analytic continuation $x^+\to ix^+$ (with
no change in $x^-$)
maps to a section in which the bosonic part of the action
 is real and positive definite.  However, there is apparently no
change of variables that makes the resulting couplings to fermions real.
This should not be a complete surprise.  Irrespective of our formalism,
the Euclidean version of  ${\rm AdS}_3\times \S^3$
cannot be represented with real fields.  For example, the  threeform field $H$
that enters the ${\rm AdS}_3\times \S^3$ supergravity solution obeys
a self-duality condition which with Euclidean signature contains
a factor of $i$, which appeared in subsection (6.1).

\subsec{ Adding A Wess-Zumino Term}

To generalize the sigma model while preserving its symmetries -- including
two-dimensional conformal invariance -- the remaining option is to
add a term 
\eqn\juccx{\int  B_{IJ}d\Phi^I\wedge d\Phi^J}
with $B$ a two-form on the $SU'(2|2)$ manifold. (We already analyzed in \koopo\
some examples of such terms that happen to be trivial.)   There are in fact no 
$SU'(2|2)\times SU'(2|2)$-invariant two-forms on the group manifold.  However, 
for invariance of the interaction \juccx, it suffices that the
three-form $H=dB$ should be invariant.  $H$ is of course automatically closed.
Conversely, given a left and right-invariant and closed three-form $H$,
one can construct a new invariant interaction -- the Wess-Zumino term --
that is defined locally as \juccx.

A three-form $H$ on the $SU'(2|2)$ manifold
with the necessary properties is determined by
a third order antisymmetric function on the Lie algebra.
Up to a scalar multiple, there is a unique such function.  It can
be defined by
\eqn\sininx{\Psi(x,y,z)=\langle x, [y,z]\rangle.}
The resulting interaction can be written most invariantly as
\eqn\soop{S'= 
\int_Xd^3y\, \epsilon^{ijk}\ll g^{-1}\partial_ig,
     [g^{-1}\partial_jg,g^{-1}\partial_kg]\rangle.}
The integration is carried out over a three-manifold $X$ whose
boundary is spacetime.

Up to third order in the fields, the integrand is simply
\eqn\xnxin{\epsilon^{ijk}\langle \partial_i\Phi, 
[\partial_j\Phi,\partial_k\Phi]\rangle
=\partial_i\left(\epsilon^{ijk}\langle \Phi,[\partial_j\Phi,\partial_k\Phi]
\rangle\right).}
An integration by parts thus reduces the Wess-Zumino action in cubic
order to an ordinary integral over spacetime:
\eqn\uxxin{S'=\int d^2\sigma \epsilon^{ij}\langle \Phi,[\partial_i\Phi,
\partial_j\Phi]\rangle.}  
For $G=SU'(2|2)$, we get, to this order,
\eqn\snoop{S'=\int d^2\sigma \epsilon^{ij}\left(\partial_ix^{ab}\partial_j
x^{cb}x_{ac}-{3\over 2}\epsilon_{abcd}x^{cd}
\epsilon_{\alpha\beta}\partial_i
\theta^{a\alpha}\partial_j\theta^{b\beta}\right).}
But \snoop\ describes the first order deformation of flat $\R^6$ by
an $H$ field from the NS sector;
it is in fact the integrated vertex operator of section 5 for
an antisymmetric NS-NS field with $B_{mn}=(\sigma_m)^{ab} 
(\sigma_n)^{cb} x^{ac}.$ Note that this $B$ field satisfies $\p^m B_{mn}
=\p_p \p_p B_{mn}=0$, so its vertex operator can be evaluated using the
methods of section 5.

The general $G\times G$-invariant invariant action that is conformally
invariant at the classical level is
\eqn\goopo{I={S\over f^2}+ikS'.}
$f$ is a constant that one might interpret as the inverse radius of
${\rm AdS}_3\times \S^3$; $k$ determines the ``level'' of the Wess-Zumino
coupling.  

\subsec{ Interpretation Of Parameters}

In the application to string theory,  ${\rm AdS}_3\times \S^3$
arises by compactification on $\T^4$ or K3 with onebranes and fivebranes.
In general, let $Q_5^{NS}$ and $Q_5^{RR}$ be the numbers of NS and
RR fivebranes, and let $Q_1^{NS}$ and $Q_3^{RR}$ be the analogous
onebrane numbers. 

One parameter is easy to identify:
$k$ (if topologically normalized by including a constant
that we have omitted) simply equals $Q_5^{NS}$.  In fact, NS fivebranes
determine, in the usual way, an NS $H$-field with a topologically
quantized flux.  Elementary strings are electric-magnetic dual to
NS fivebranes, and $Q_5^{NS}$ determines the ``level''
of the Wess-Zumino interaction.  ($Q_5^{RR}$ would similarly determine
the level of a Wess-Zumino interaction for a $D$-string probe, but in the
present paper we are considering elementary string actions only.)

On the other hand, $f^{-1}$ 
is the radius of the ${\rm AdS}_3\times \S^3$ spacetime.
This radius, when it is large, can be measured by probing 
${\rm AdS}_3\times \S^3$
by massless particles.  The long distance action for massless particles
is governed by supergravity, so -- when the radius is large -- it can
be determined in terms of the brane charges by solving supergravity 
equations.  In the special cases $Q_5^{NS}=0$ and $Q_5^{RR}=0$, the
solution is given in \mava .
In general, the solution of the supergravity equations gives 
\eqn\lxoc{{1\over f^2}=\sqrt{(Q_5^{NS})^2+\lambda^2(Q_5^{RR})^2},}
with $\lambda$ the ten-dimensional string coupling constant.
When $Q_5^{RR}=0$ , this relation
becomes  $f^{-2}=|k|$.  This is the familiar relation between
metric and $H$-field on the group manifold
which determines the model to be a WZW model of
$SU'(2|2)$.  It thus has left and right-moving $SU'(2|2)$ current
algebra, and will be studied from that point of view in section 10.
In general, if the charges are real, one has $1/f^2\geq |k|$.

Supersymmetry requires that the onebrane charges $(Q_1^{NS},Q_1^{RR})$
be a multiple of the fivebrane charges $(Q_5^{NS},Q_5^{RR})$.
So we can write 
$$(Q_5^{NS},Q_5^{RR})=Q_5(p,q), ~~~~(Q_1^{NS},Q_1^{RR})
=Q_1(p,q),$$ with relatively prime integers $p,q$ and integers
$Q_1,Q_5$.  The parameter $Q_1$
does not enter the $SU'(2|2)$ sigma model, but it determines a certain
relation between  the string coupling constant and the volume $v$ of the
K3 or $\T^4$.  In fact, $v$ is determined
by equating the tensions of the strings derived from 
onebranes and fivebranes\foot{This condition is equivalent
to minimizing the tension in the Einstein frame
of the bound string made of wrapped fivebranes and onebranes.},
via $vT_5Q_5=T_1Q_1$.  With $T_1= \sqrt{p^2+q^2/\lambda^2}$,
$T_5=\sqrt{(p/\lambda^2)^2+q^2/\lambda^2}$, we get 
\eqn\icoc{v={Q_1\lambda\over Q_5}\sqrt{p^2\lambda^2+q^2\over 
p^2+\lambda^2q^2}.}
Our treatment in the present paper is valid in the limit
$\lambda\to 0$, with  $\lambda q$, $p$, $Q_1$, and $Q_5$ fixed.   
In this  limit (assuming $\lambda q\not= 0$), the formula simplifies to
\eqn\bicoc{v={Q_1\over Q_5}{\lambda q\over\sqrt{p^2+(\lambda q)^2}}.}
As $p,q$, $Q_1,$ and $Q_5$ are integers,
 $v$ is not a continuous parameter for given $\lambda$, 
but with $\lambda$ very
small and $q\sim 1/\lambda$, the possible values of $v$ 
are very closely spaced.  In our perturbative treatment, $v$ appears
to be a continuous parameter.
The fundamental reason for this is that, as we are not considering
$D$-brane probes in the present paper, but only fundamental strings,
we do not see the quantization of RR flux and hence $Q_5^{RR}$
and $Q_1^{RR}$ appear to be real-valued parameters.  Quantization
of RR flux is -- as always -- a nonperturbative phenomenon from the
point of view of weakly coupled perturbative string theory.

\subsec{\it Conformal Field Theory}

The sigma model Lagrangian \goopo\ can be constructed for any Lie group
or supergroup $G$ with invariant quadratic form $\ll ~,~\rr$.
Generically, this Lagrangian has a nonzero beta function unless
$f$ and $k$ are suitably related.  However, we will now argue that
for $SU'(2|2)$, the beta function vanishes for any $f$, $k$.  
This means that in this particular case, we have a conformal field
theory description of a Ramond-Ramond background in string theory.

The first step is to look at the one-loop beta function.  For any
$G$, it is proportional to the quadratic form on the Lie algebra
that is defined by 
\eqn\uxnon{\Str_{adj}\, xy}
where $\Str_{adj}$ 
is the supertrace in the adjoint representation.  For conventional
simple Lie groups, one defines an invariant $c_2(G)$ by stating that
\uxnon\ equals $c_2(G)$ times a basic quadratic form $\ll~,~\rr$.
For $SU'(2|2)$, or in general $SU'(n|n)$, \uxnon\ vanishes, and hence
the one-loop beta function is zero.
This may be shown by explicit computation.  Since any invariant
quadratic form on the Lie algebra is a multiple of $\ll~,~\rr$,
it is sufficient to consider
the case that 
\eqn\snon{x=y=\left(\matrix{ 1 & 0 \cr 0 & -1 \cr}\right)}
equal diagonal generators of one of the $SU(2)$'s in 
$SU'(2|2)$.  Under this generator, the bosonic part of the adjoint
representation has eigenvalues $2,-2,0,0,0,0$, while the fermionic
part has eight states of eigenvalue $\pm 1$. So the vanishing comes
from $2^2+(-2)^2-8\cdot 1^2=0$.  A more conceptual proof
can be given by comparing to $U(n|n)$.  We will omit this
argument (which is similar in spirit to arguments we give below
for the all orders beta function).
 
We will give two arguments to 
show that the $SU'(n|n)$ beta function vanishes to all orders; one of the
arguments shows that the beta function vanishes exactly and not just
to all orders of perturbation theory.  Since one of the arguments is
based on showing that the $SU'(n|n)$ beta function is independent of $n$,
we pause here to show that this beta function is zero for $SU'(1|1)$.
Indeed, $SU'(1|1)$ has no bosonic generators at all and is an abelian
supergroup, so the $SU'(1|1)$ sigma model is a free conformal field theory.

We will show that the beta function of  the $SU'(n|n)$ sigma model vanishes
by comparing it to properties of
a $U(n|n)$ sigma model.  In $U(n|n)$, one can define an invariant
quadratic form by a supertrace in the $n|n$-dimensional
fundamental representation:
\eqn\hmoerfree{\ll x,y\rr =\Str xy.}
There is also an invariant but degenerate quadratic form $\Str x\cdot \Str y$.
The sigma model action thus has two possible couplings
\eqn\bomerfree{ S_{U(n|n)}=\int d^2\sigma\left(
{1\over 2f^2}\Str (g^{-1}dg)^2 + h(\Str g^{-1}dg)^2\right),}
with constants $f$, $h$.  One can also have a Wess-Zumino coupling, which
we do not write explicitly.  Its inclusion does not affect the argument.

It is useful to single out two generators of the Lie algebra,
\eqn\toggof{I=\left(\matrix{ 1 & 0 \cr 0 & 1 \cr}\right),~~~~~
L=\left(\matrix{1 & 0 \cr 0 & -1\cr}\right).}
Let $T_a$ be a basis of generators that are orthogonal to both
$I$ and $L$.
In the basis $I,L$, and $T_a$, the nonzero brackets of  $U(n|n)$ take the form
\eqn\xnxo{\eqalign{
[L,T_a]& =c_{ab}T_b\cr
[T_a,T_b]&=f_{ab}^cT_c +d_{ab}I.\cr}}
Here $f_{ab}^c$ are the structure constants of $SU'(n|n)$,
and $c_{ab}$, $d_{ab}$ are constants, whose details will not concern
us, that reflect the fact that $L$ generates an outer automorphism
of $SU'(n|n)$ and $I$  appears as a central extension.

Because of this structure of the Lie algebra, the Lagrangian takes a
very simple form.
We parametrize the group manifold by
\eqn\xxonp{g=\exp(uI+vL+\Phi^aT_a).}
The quantities $g^{-1}dg$ via which the Lagrangian is expressed
can be expanded in terms of repeated commutators (the first few terms
are in \ogho).  $L$ never appears on the right hand side except
in the linear term $d\Phi$, because it never appears on the right hand
side of the commutation relations \xnxo.  
Likewise, since all commutators $[I,\cdot]$ vanish,
$u$ does not appear in $g^{-1}dg$ except in the linear term.
  Because of these
facts, $u$ only appears in the action in a quadratic $du\,dv$ term.
The terms that do not involve
$I$ or $L$ at all just give the $SU'(n|n)$ sigma model action.
So we get the simple structure:
\eqn\xcpo{S_{U(n|n)}=S_{SU'(n|n)}+\int d^2\sigma\left({2n \,du\,dv\over f^2}
+ n^2 h (dv)^2+{1\over f^2}\Delta L(v,\Phi^a)\right).}
Because the kinetic energy is of the form $du\,dv + dv^2$,
there is a $\ll u\,u\rr$ and a $\ll u\,v\rr$ propagator, but no
$\ll v\,v\rr$ propagator.  The field $v$ is coupled to the $SU'(n|n)$
fields $\Phi^a$ by couplings $\Delta L$, whose details need not concern
us.  The important thing is that there are no such couplings involving $u$.

Since only $v$ and not
$u$ appears
in the interaction vertices and the $\ll v\, v\rr$ propagator
vanishes, it follows that all $U(n|n)$ Feynman diagrams all of whose
external lines
are in $SU'(n|n)$  are the same as $SU'(n|n)$ Feynman diagrams.
Hence the beta function of the $SU'(n|n)$ theory is determined by the
renormalization of the $\Str (g^{-1}dg)^2$ term for $U(n|n)$.
But here are two reasons that this renormalization vanishes:

(1) The operator ${\cal O}=\Str (g^{-1}dg)^2$
  contains a $du\, dv$ term.
But the $s$-loop effective action, for all $s\geq 1$, is a function
of only $v$ and not $u$, since $u$ nowhere appears in the interaction
vertices.  So there can be no infinity proportional to ${\cal O}$.
Note that this argument is not limited to perturbation theory.

(2) Consider the $n$-dependence of the $U(n|n)$ beta function.
To probe it, we parametrize the group by $g=\exp(\sum_A\Phi^A T_A)$
where now $T_A$ runs over all generators of the Lie algebra and $\Phi^A$
is a set of coordinates on the group manifold.  To probe the beta
function, we look at the two point function $\langle \Str C\Phi(x)
\,\Str D\Phi(y)\rangle$ with $c$-number-wave functions $C$ and $D$
taking values in the Lie algebra.  The general  form
of the two point function allowed by $U(n|n)$ invariance is
\eqn\xnono{\langle \Str C\Phi(x)\,\Str D\Phi(y)\rangle
= E(x-y) \Str CD +F(x-y) \Str C\, \Str D,}
with functions $E, F$.\foot{To avoid infrared divergences in this discussion,
one could, for example, add to the Lagrangian
a mass term $\Str g$ that is invariant under the
diagonal subgroup of the $U(n|n)\times U(n|n)$ symmetry of the sigma
model.  The argument uses only  invariance under this diagonal subgroup.}  
Now let us consider the $n$ dependence of this two point function.
Feynman diagrams in $U(n|n)$ theory can be constructed
using `t Hooft's ``double line'' notation for fields in the adjoint
representation of a unitary group or supergroup \ref\thooft{G. 't Hooft,
``A Planar Diagram Model For Strong Interactions,'' Nucl. Phys. B (1974).}.
The  $n$-dependence comes entirely from supertraces that arise
from an ``index loop'' in a Feynman diagram.  For $U(n|m)$, the supertrace
of the identity in the fundamental representation equals $n-m$, and
one gets this factor for each index loop.  For $U(n|n)$, this factor
vanishes, and hence all diagrams that contain index loops vanish.
Hence the two point function \xnono\ is independent of $n$.  So in
particular $E$, which determines the $SU'(n|n)$ beta function, is finite
for all $n$ if it is finite for some $n$.  We have already noted
that the $SU'(1|1)$ theory
is free and finite, so the            $E$ function in \xnono\ must have
no infinity for $n=1$ and hence for all $n$.  Hence, the $SU'(n|n)$ theory
is finite for all $n$.

To avoid confusion, we should note that we are not claiming that
in $U(n|n)$, there is no infinite renormalization of the $(\Str g^{-1}dg)^2$
term in the Lagrangian.  On the contrary, for $U(n|n)$,
 $\Str_{adj} L^2\not= 0$, so there is such an infinity at one-loop order.   
(For $U(n|n)$, $\Str_{adj}xy$ is a multiple of $\Str \,x\,Str \,y$, so
the one-loop infinity is a multiple of $(\Str \,g^{-1}dg)^2$.)
Likewise, we make
no claim that the $F$ term in the correlator is finite, only the $E$
term which is related to the $SU'(n|n)$ beta function.

\def\e{{\epsilon}}
\def\s{{\sigma}}
\def\N{{\nabla}}
\def\half{{1\over 2}}
\def\p{{\partial}}
\def\pb{{\bar\partial}}
\def\t{{\theta}}
\def\Gtp{{\tilde G^+}}
\def\Gtm{{\tilde G^-}}
\def\tb{{\bar\theta}}

\newsec{Structure Of The Ghost Couplings}

\subsec{First Order Treatment}

\def\S{{\bf S}}
In section 7, we argued that the ${\rm AdS}_3\times \S^3$ model
with Ramond-Ramond background can be described by a sigma
model with  couplings to ghost fields $e^\phi$ and $e^{\bar \phi}$.  
The simplifying feature that makes the theory manageable is that the
ghost couplings involve only positive powers of
$e^\phi$ and $e^{\bar \phi}$.  Hence, there is a consistent
truncation in which they are neglected in first approximation.
We have described this truncation in the last section in terms of
a sigma model with manifest spacetime supersymmetry.
But looking back to the formulas of section 7, it is clear
that the ghost couplings do not have manifest supersymmetry.
Instead, they have a much more complex structure that we will analyze
here.

The ghost couplings that come directly from the vertex operators are
as we saw in  equation \hv,
\eqn\uncuc{S_\phi=
{i\over 8}\int d^2\s
 e^\phi p^a\epsilon_{abcd}\bar\theta^b\bar \partial x^{cd},}
with a similar term linear in $e^{\bar\phi}$ and a higher order
$e^{\phi}e^{\bar\phi}$ term.  Let us analyze the structure of $S_\phi$.  
Just as in section 6, we will work up to quartic order in fields
(that is, in $p,\theta, e^\phi$, and their barred counterparts)
and try to guess the general structure. 
Now, at least in the flat space case considered in section 4,
$p^a$ is one of the supercurrents and $\epsilon_{abcd}
\bar\theta^b\partial x^{cd}$ is the ghost-independent
part of another supercurrent.
Let us see if in the sigma model, the ghost coupling has a similar
current-current form.
In going to the sigma model, it is convenient
to solve for $p^a$ by the equations of
motion and write the above interaction as
\eqn\ixxo{S_\phi={i\over 8}
\int d^2\s  e^{\phi}\partial\bar\theta^a \epsilon_{abcd}
\bar\theta^b\partial x^{cd}.}
Here we have ignored the cubic corrections to $p^a=\partial\bar\theta^a$
because they would give sixth order terms in \ixxo, which we are neglecting.
Now, though this is not immediately apparent,
\ixxo\ can actually be rewritten as
\eqn\pixxo{S_\phi=
-{i\over 2}
\int d^2\s  e^{\phi}\left(\partial\bar\theta^a-{1\over 4}\epsilon^{abcd}
(\bar\theta^b\partial x^{cd}-\partial\bar\theta^b x^{cd})\right)
\left(\bar\partial\bar\theta^a-{1\over 4}
\epsilon^{aefg}(\bar\theta^e\bar\partial x^{fg}
-\bar\partial\bar\theta^e x^{fg})\right).}
We have added a variety of terms which vanish modulo fifth order
expressions for various reasons.
Some terms are explicitly fifth order in the fields.  Also,
the $e^{\phi}\partial\bar\theta^a\bar\partial\bar\theta^a$ term is
equivalent to a fifth order expression,
after integrating by parts and using the fact that $\bar\partial e^{\phi}=0$
and that $\bar\partial\partial\theta^a=0$ modulo cubic terms.
A term $e^{\phi}\partial\bar\theta \bar\partial(\bar\theta x)$ is similarly
fifth order after integration by parts.   

Let us recall from equation \nonson\ the form, up to quadratic order in fields,
of the currents that generate the left supersymmetries of the sigma model:
\eqn\finducr{S^a_{L\alpha}=d\theta^a_\alpha -{1\over 4} \e^{abcd}
(d\theta^b_\alpha x^{cd} -\theta^b_\alpha d x^{cd}).}
This has been written in a form that is manifestly invariant under the
outer automorphism group $R=SU(2)$.  For the time being, however,
it will be more useful to relax this manifest invariance.  We recall
that $(\theta,\bar\theta)$ are the upper and lower components of an
$R$ doublet.  We denote the upper and lower components of $S_L$ as
$E,F$.  We have up to second order
\eqn\ginducr{\eqalign{E^a&=d\theta^a 
-{1\over 4} \e^{abcd}
(d\theta^b x^{cd} -\theta^b d x^{cd})\cr
F^a&=d\bar\theta^a -{1\over 4}
 \e^{abcd}
(d\bar\theta^b x^{cd} -\bar\theta^b d x^{cd}).\cr}}
Evidently then, we can write the coupling in \pixxo\ as
\eqn\xoppop{S_\phi=-{i\over 2} \int d^2\s 
 e^{\phi}F^a_z F^a_{\bar z}=\half\int d^2\sigma 
\epsilon^{ij}e^\phi
F^a_i F^a_j=\half\int  e^\phi F^a\wedge F^a.}
We have introduced real coordinates by $z=\half(\sigma^1+i\sigma^2)$, and
used fermi statistics to express the coupling in terms of a wedge
product of one-forms $F^a$.
Where it will cause no confusion, we also write $E^a$, $F^a$ for
the conserved charges obtained by integrating those currents;
they are of course the upper and lower components of the supersymmetry
doublet $S^a_\alpha$. We write $\eta^a_+,$ $ \eta^a_-$ for infinitesimal
parameters generating left-moving supersymmetry.  The left supersymmetry
of the $SU'(2|2)$ sigma model action 
\eqn\sonxonnc{S_0=\half\int d^2\sigma \langle g^{-1}dg,g^{-1}dg\rangle}
studied in section 7 is thus
\eqn\boxson{\delta g=(\eta_+^aE_a+\eta_-^a F_a)g.}
 We will also denote as $K_{ab}$ the
current generating the left action of the bosonic generators $K_{ab}$
of $SU'(2|2)$ on $g$.

The ghost coupling in \pixxo\
is invariant under the right action of $SU'(2|2)$
(the symmetry
$g\to g B^{-1}$) since the left supercurrents that appear
there are all right-invariant.
It is also invariant under the left action of rotations (the bosonic
symmetries in $SU'(2|2)$), and under the left action of $F^a$.  However,
under the left action of $E^a$, $S_{\phi}$ is not invariant. Because
of the commutation 
relation $\{E_a,F_b\}=\half \epsilon_{abcd}K^{cd}$, the variation
of $S_\phi$ under a left supersymmetry is 
\eqn\cobbob{\delta S_\phi = \half\int  e^{\phi}\eta_{+}^a\epsilon_{abcd}
K^{cd}\wedge F^b.}

Now, recall from section 7 the Maurer-Cartan equation for the
right-invariant one-forms on the $SU'(2|2)$ manifold:
\eqn\xnonnc{dF_a=-\half\epsilon_{abcd} K^{bc}\wedge F^d,}
and
\eqn\xnonnd{dE_a=-\half\epsilon_{abcd} K^{bc}\wedge E^d.}
Using the formula for $dF_a$, we can write
\eqn\vobbo{\delta S_{\phi}=-\int e^{\phi}\wedge \eta_{+a} dF^a.}

The only hope of canceling this variation is to add ghost-dependent
terms to the symmetry transformation \boxson.  A clue comes from
the flat case, where the supercharges have ghost dependent terms:
$q_a^-=\oint p_a$,  $q_a^+=\oint(e^\phi p_a-
{i\over 2}\e_{abcd} \t^b \p x^{cd}).$
This suggests
that in the flat case, to take the ghosts into account, the left supercurrents
must be subjected to a $\phi$-dependent rotation which mixes $E$ and $F$. 
As will
be shown below, the appropriate field-dependent rotation
is $(E,F)\to (E+ie^\phi F,F).$ 
This is an $R$ transformation by the matrix
\eqn\tutugo{W=\left( \matrix{1 & ie^{\phi }\cr 0 & 1\cr }\right) .}
This strongly suggests that we should modify
the transformation law \boxson\ under the left supersymmetries
to be
\eqn\newboson{\delta g=\left(\eta_+^a(E_a+ie^\phi F_a)+\eta_-^aF_a\right)g.}
Now in general, under $\delta g=\epsilon g$, the sigma model action
$S=\int d^2\sigma \half \langle g^{-1}\partial_i g, g^{-1}\partial^ig\rangle$
changes by $\delta S=\int d^2\sigma \langle \partial_i\epsilon, \partial^ig
g^{-1}\rangle$.  For the transformation in \newboson, the variation 
is 
\eqn\trutuog{\delta S=i\int d^2\sigma \partial_ie^{\phi}\eta_+^a F_i^a.}
Because $e^{\phi}$ is holomorphic, we have $(\partial_i+i\epsilon_{ij}
\partial_j)e^{\phi}=0$.  Using this fact and integrating by parts,
we   get
\eqn\butog{\delta S=\int d^2\sigma e^{\phi}\eta_+^a\epsilon^{ij}\partial_i
F_j^a.} 
Comparing this with \vobbo, we see that to this order, the total
action $S+S_\phi$ is invariant under the modified spacetime supersymmetry
transformation \newboson.

\subsec{Left And Right}

One strange thing about this result is that we have modified
the left action of $SU'(2|2)$ on itself, but not the right action.
However, by a change of variables we could reverse this result,
modifying the right action but not the left action.
Indeed, suppose that we make the $R$-transformation  that is inverse to $W$:
\eqn\uttu{W^{-1}=\left(\matrix{ 1 & -ie^\phi \cr 0 & 1}\right). }
This transforms the modified supersymmetries of \newboson\ into the
standard ones -- showing in particular that the modified supersymmetries
do possess the standard commutation relations.
However, since there is only one $R$ symmetry group that acts on
all degrees of freedom including the left and right supersymmetries,
this transformation will turn the right action of supersymmetry into
\eqn\donno{\delta g = g (\eta_+^a(E^a-ie^\phi F^a)+\eta_-^aF^a).}
Thus, we can put the left action or the right action of $SU'(2|2)$ in the
standard form, but not both. 

The left currents in $S_\phi$ are dictated by the exotic action of the
left supersymmetries in \donno.
The similarity 
transformation that ``straightens out'' the left supersymmetries
and makes the right supersymmetries exotic will therefore
replace the left currents in $S_\phi$ by right currents.

This situation is quite unfamiliar.  The model possesses $G\times G$ symmetry.
One can pick variables to make the action of either copy of $G$ standard.
But one cannot simultaneously put the action of both copies of $G$ in
a standard form.

We will show explicitly how to change variables so that the ghost
couplings are written in terms of right currents rather than left currents.
Under the $R$-transformation of \uttu, the sigma model action of \sonxonnc\
transforms to 
\eqn\rtran{\delta S = i \int d^2 \sigma (\p_i e^\phi ) r^{22}_i}
where 
$r^{\a\b}_i$ is the conserved current 
associated to the $R$-transformation. Instead
of computing $r^{\a\b}_i$
directly using the Noether method, we shall compute it
indirectly from its commutation relations with the $SU'(2|2)$ charges.
If $\oint r^{\a\b}$ is the $R$-charge, then
the only non-zero commutation relations are with the left and
right susy charges which gives
\eqn\chargec{[
\oint r^{\a\b}, \oint S_{La}^\g ] = \half(\e^{\beta\gamma} \oint S_{La}^\alpha 
+\e^{\a\g} \oint S_{La}^\b),}
$$
[\oint 
r^{\a\b}, \oint S_{Ra}^\g ] = \half(\e^{\beta\gamma} \oint S_{Ra}^\alpha 
+\e^{\a\g} \oint S_{Ra}^\b). $$
Using \llcom\ and \omrcom, one sees that $\oint r^{\a\b}$ has the same
commutation relations as $-\half\oint (\lambda_L^{\a\b}+\lambda_R^{\a\b})$.
Similarly, $dr^{\a\b}$ has the same commutation relations as 
$-\half(\omega_L^{\a\b}+\omega_R^{\a\b})$.
Since any two-form which commutes with all the $SU'(2|2)\times SU'(2|2)$
transformations 
must vanish, $dr^{\a\b}+\half(\omega_L^{\a\b}+\omega_R^{\a\b})$ must
vanish, i.e.
\eqn\drdef{dr^{\a\b}=-\half(\omega_L^{\a\b}+\omega_R^{\a\b}).}

Since $\bar\partial e^\phi=0$, \rtran\ is equal to
\eqn\rtwotran{\delta S = 
  \int d^2 \sigma  e^\phi \e^{ij} \p_i r^{22}_j = 
  \int  e^\phi \wedge d r^{22}_j}
$$= 
 -\half \int  e^\phi (F_{L}^a \wedge F_{L}^a +
 F_{R}^a \wedge F_{R}^a).$$
So the $R$-transformation of \sonxonnc\ changes the ghost-coupling from
$S_\phi =\half \int e^\phi F_L^a \wedge F_L^a $ to
$S_\phi = -\half \int e^\phi F_R^a \wedge F_R^a $,
which now depends only on left-invariant currents as expected. 

\subsec{ Incorporation Of $e^{\bar\phi}$}

Having understood the basic structure, it is now not so difficult
to give a full description of the ghost couplings.

First, consider the term linear in $e^{\bar\phi}$ (with no copies of
$e^\phi$).  By manipulations quite similar to what we have seen,
this coupling can be put in the form
\eqn\ximmo{S_{\bar\phi} =-\half\int d^2 \s e^{\bar \phi}E^a \wedge E^a.}
Its variation, furthermore, can be canceled 
by modifying \newboson\ so that the left supersymmetry transformations are
\eqn\newwboson{\delta g=\left(\eta_+^a(E_a+i
e^\phi F_a)+\eta_-^a(F_a-i e^{\bar\phi }
E_a)\right)g.}
Of course, just as in the discussion of $S_\phi$, we could make an
$R$-symmetry transformation to remove this correction to the left
supersymmetry transformations, at the cost of adding a similar correction
to the right supersymmetry transformations.  $S_{\bar\phi}$ would
then be expressed in terms of right currents.

The action $S+S_{\phi}+S_{\bar\phi}$ is invariant under the
transformation \newwboson, up to terms of order $e^{\phi}e^{\bar\phi}$.
To cancel those terms, it is necessary to add to the action further
terms proportional to $e^\phi e^{\bar \phi}$.  This should come
as no surprise, because as we saw in section 6, the vertex
operator for the leading order deformation from flat space to 
${\rm AdS}_3\times 
\S^3$  contains a term proportional to $e^\phi e^{\bar\phi}$.
As we will see presently, there actually are further terms
$e^{n\phi}e^{m\bar\phi}$ with all positive integers $n,m$ satisfying
$|n-m|\leq 1$.
Luckily, with the experience we have gained so far, it is possible
to guess the general structure.

\subsec{ Exact Non-Linear Ghost Couplings}

Up to this order in $e^\phi$ and $e^{\bar\phi}$, the 
action
\eqn\firsto{S=S_0 +\half\int e^\phi F^a \wedge F^a -\half\int e^{\bar\phi}
E^a \wedge E^a}
is invariant under the ghost-modified left supersymmetries
of \newwboson,
where $S_0$ is the $SU'(2|2)$ sigma model action of 
\sonxonnc. We will now generalize \firsto\ and \newwboson\
in a manner that leaves the action invariant to all orders in
$e^\phi$ and $e^{\bar\phi}$.

\def\ll{\langle}
\def\rr{\rangle}
The generalization of the ghost-modified left supersymmetries will
be defined by
\eqn\leftsusy{
\delta g=\big(\eta^a{}_I\,v^I{}_\alpha\, S^{\alpha a}\big)g,}
where $\eta^a_I$ are constant anticommuting parameters, $v^I_\alpha$
is a rotation matrix that depends on $e^\phi$ and $e^{\bar \phi}$ in 
a fashion which will be determined, and we recall
that $S^{\alpha a}=(E^a,F^a)$
is the doublet of supersymmetry generators.
To lowest order in $e^\phi$ and $e^{\bar\phi}$, the matrix elements of
$v$ are given 
by 
\eqn\lowest{v_1^+ =1,\quad v_2^+= i e^\phi,\quad 
v_1^- =-i e^{\bar\phi},\quad v_2^-=1,}
where to avoid
confusing the different types of index
we let $\alpha$ take the values $1,2$ and $I$ takes the   values $+,-$.
Under the transformation of \leftsusy, 
the sigma model action $S_0$ transforms as 
\eqn\sigmatr{\delta S_0= \int d^2 \sigma \eta_I^a (S^{\alpha a}\partial_i
v^I_\alpha. )}

So one would like to construct a generalization
of \firsto\ such that the transformation
of the $\phi$-dependent terms in the action cancels
\sigmatr. This generalization can be found by considering the action
$S=S_0+S_1$ where $S_0$ is the sigma model action of 
\sonxonnc\ and
\eqn\exacto{S_1= \int \sum_{\alpha,\beta =1,2} c_{\alpha\beta}\,\,S^{a\alpha}
\wedge S^{a\beta}.}
Here $c_{\alpha\beta}=c_{\beta\alpha}$; the $c$'s are functions of $e^\phi$
and $e^{\bar\phi}$ that will be determined.  
We recall also that
$S^{a\alpha}$ denotes either the supersymmetry generators or the
corresponding left currents, which are defined as
\eqn\mxonc{\langle S^{a\alpha},dg\,g^{-1}\rangle.}
Note that the constant $\phi$-independent parts of $c^{AB}$ can
be chosen arbitrarily since, as shown in section 7,
$$\int S^{a\alpha}\wedge S^{a\beta} =0.$$
Ignoring this constant part, the lowest order $e^\phi$ contributions
were computed earlier to be  $ c_{11}= e^{\bar\phi}$, $c_{22}= -
e^{\phi}$ and $c_{12}=0$.

Under the modified left susy transformation of \leftsusy, 
the currents  transform as
\eqn\gentransf{\delta S^{a\alpha}=-  \eta^a_I dv^{I\alpha}
-\half\epsilon^{abcd} \eta_{I\,b}K_{cd}v^{I\alpha}.}
This formula is obtained directly from the definition \mxonc\
of the
supercurrents.   Under $\delta g=\epsilon g$, the
variation of the supercurrent has two terms, one proportional to
a derivative of $\epsilon$ and one not; these give the two terms in \gentransf.

So under \leftsusy, $S_1$  transforms as 
\eqn\hcno{\eqalign{
\delta S_1 &=2\int c_{\alpha\beta}\left(-\eta^a_Idv^{I\alpha}S^{a\beta}
-\half\epsilon_{abcd}\eta^b_{I}v^{I\alpha}K^{cd}S^{a\beta}\right)   \cr
& =-2\int c_{\alpha\beta} \eta^a_I\left(dv^{I\alpha}S^{\beta}+v^{I\alpha}
dS^{a\beta}\right)=2\int dc_{\alpha\beta}\,\eta^a_Iv^{I\alpha}S^{a\beta}\cr}}
where $v^{I\alpha}=\e^{\alpha\gamma} v^I_\gamma$.
Here we have used the Maurer-Cartan equations  and  integrated by parts. 

So to cancel the transformation of 
\sigmatr\ for all $\eta^a_I$, we require the relation
\eqn\matrixeq{\partial_i v^I_\alpha=- 2\epsilon_{ij}v^I_\beta \e^{\beta\gamma}
\partial_j c_{\gamma\alpha}.}
We define a $2\times 2$ matrix
\eqn\yuco{C^\alpha{}_\beta=\epsilon^{\alpha\gamma}c_{\gamma\beta}.}
Likewise, we regard $v^I_\alpha$ as the matrix elements of a $2\times 2$
matrix $V$.  The equation then reads
\eqn\coxxo{V^{-1}\partial_iV= -2\epsilon_{ij}\partial_j C.}
As $C$ is traceless, the determinant of $V$ is constant.
In terms of a complex coordinate $z=\half(\sigma^1+i\sigma^2)$,
the equation for $V$ and $C$ becomes 
\eqn\oboxxo{\eqalign{V^{-1}\partial V & = -2i\partial C\cr
            V^{-1}\bar\partial V & =  2i\bar\partial C,\cr}}
with  $\partial=\partial_z$, $\bar\partial=\partial_{\bar z}$.
This implies that 
\eqn\squarec{\p\bar\p C =\half(\bar\p\p C+\p\bar\p C) =- {i\over 4}
[V^{-1}\bar\p V~,~ V^{-1}\p V]
=-i [\bar\p C, \p C].}

So we are looking for solutions to $\p\bar\p C=-i[\bar\p C,\p C]$
such that, ignoring contant $\phi$-independent terms, 
$c_{11}= -\half e^{\bar\phi}$,
$c_{22}= \half e^{\phi}$, and $c_{12}=0$
when $e^\phi$ is small. One can solve these conditions
explicitly to find that 
\eqn\cexact{C =
{1\over {1 - {1\over 4} e^{\phi+\bar\phi}}}\left(\matrix{ i
&  -\half e^{\phi} \cr -\half e^{\bar\phi} & - i}\right).}
So the exact expression for the action is $S=S_0 + S_1$   where 
\eqn\exactex{S_1= \int (1-{1\over 4}e^{\phi+\bar\phi})^{-1}(\half e^\phi
 F^a \wedge F^a
-\half e^{\bar\phi} E^a \wedge E^a -2i E^a \wedge F^a).}

One can now look for a matrix $V$ satisfying 
\oboxxo\ (where $C$ is given by \cexact) such that
\lowest\ is satisfied when $e^\phi$ is small.
Once again, one can explicitly solve these conditions to find that
\eqn\Vexact{V={1\over {1-{1\over 4}e^{\phi+\bar\phi}}}
\left(\matrix{
 1 + {1\over 4} e^{\phi+\bar\phi}& i e^\phi \cr 
-i e^{\bar\phi}
 & 1+{1\over 4} e^{\phi+\bar\phi}}\right). }

\subsec{Exact Ghost Couplings in Presence of Wess-Zumino Term}

One can also ask what happens to the ghost couplings in $S_1$ 
after adding a Wess-Zumino term to $S_0$ with coupling parameter $N$.
For this, we restore the coupling parameter $f$ (which we have suppressed
so far in this section) and introduce also the Wess-Zumino coupling
$k$.  So the action is
\eqn\wzterm{S = {1\over {f^2}} (S_0 +S_1) + k S_{WZ}= 
 {1\over {f^2}} (S_0 +S_1 + N  S_{WZ})}
where $N=kf^2$ and 
$S_{WZ}={i\over 2} \int d^3 \sigma \langle dg g^{-1},  [dg g^{-1} ,
dg g^{-1}]\rangle.$\foot{We take Euclidean signature for the worldsheet
and hence include a factor of $i$ in $S_{WZ}$.}
In this case, 
the transformation of $S_0 + N S_{WZ}$ under the left supersymmetry
transformation of \leftsusy\ is given by 
\eqn\sigmatrwz{\delta S_0 + N\delta S_{WZ}
=\int d^2 \sigma \eta_I^a ( \partial_i v^I_\alpha S_i^{a \alpha } 
 + iN \epsilon^{ij}\partial_i v^I_\alpha S_j^{a\alpha }).}
 Here we have kept the definition $S^{a\alpha}=\ll S^{a\alpha},dgg^{-1}\rr$,
 but with $N\not=0$, the left supersymmetry currents are actually
 the combinations in \sigmatrwz:
 $\tilde S^{a\alpha}_i=S_i^{a\alpha}+iN\epsilon_{ij}S_j^{a\alpha}$.
Defining $c_{AB}$, $V$ and $C$ as above, and requiring
that $\delta S_1$ cancels the transformation of \sigmatrwz, 
one finds that the equation satisfied by $C$ is modified to
\eqn\difmwz{\p C = {i\over 2}(1+N)V^{-1} \p V,\quad \bar\p C= 
-{i\over 2}(1-N)V^{-1}\bar\p V.}
This implies that
$$\p\bar\p C = \half[(1-N)\bar\p\p C +(1+N)\p\bar\p C]=
{i\over 4}(N-1)(N+1)[V^{-1}\bar\p V~, ~ V^{-1} \p V]$$
$$=-i [\bar\p C, \p C].$$
So the equation for $C$ is independent of $N$.

The initial conditions for this equation, however, depend on $N$.
To determine them, we should be a bit more precise
about how the derivation in section 6 generalizes in the presence
of both RR and NS flux.  We start with the flat model as presented
in section 4, and then perturb the Lagrangian by RR and NS vertex
operators with coefficients $f_{RR}$ and $f_{NS}$:
\eqn\yuppo{L\to L+f_{RR}\int V_{RR}+f_{NS}\int V_{NS}.}
In second order, both the RR and NS fluxes give back reaction on the metric,
and the Lagrangian will require a further correction
\eqn\zuppo{(f_{RR}^2+f_{NS}^2)\int d^2\s  
 {1\over {24}}\epsilon_{abcd}x^{ae}x^{bf}\p x^{ce}
\bar\p x^{df}+\dots.}
The coefficient of this term  determines, after some rescalings of
fields,  the sigma model
radius $1/f$.  In other words,
if we let $f=\sqrt{f_{RR}^2+f_{NS}^2}$, and rescale $x$ and the $\theta$'s
by a factor of $f^{-1}$, then the ghost-free part of the action
gets an overall scale $1/f^2$, as in section 6.  Since the $e^\phi$
and $e^{\bar \phi}$ couplings
arise in linear order only from $f_{RR}$, they are proportional
after the rescaling to $f_{RR}/f$ times the expression at $f_{NS}=0$.  Thus,
the term in $S_1$ linear in $e^\phi$ and $e^{\bar\phi}$ is multiplied
by $f_{RR}/f$ from what it is in the pure RR case.  Also, note that
a Wess-Zumino coupling comes only from $V_{NS}$, and hence after 
the rescaling, $N=f_{NS}/f$.  

The net effect is that turning on NS flux multiplies the
linear contribution to $C$ by a factor $f_{RR}/f=\sqrt{1-N^2}$.  Since
this factor appears with $e^\phi$, $e^{\bar\phi}$ only
in the combinations $\sqrt{1-N^2} e^\phi$ and $\sqrt{1-N^2}e^{\bar\phi}$,
it could be removed by adding a constant to $\phi$ and $\bar\phi$,
an operation under which the equation for $C$ is invariant.  Up to
such a shift, the result for $C$ is independent of $N$.  If we prefer not
to shift $\phi$ and $\bar\phi$, then the general solution for $C$
is
\eqn\moxie{C(e^\phi,e^{\bar\phi},N)=C_0(\sqrt{1-N^2}e^\phi,\sqrt{1-N^2}
e^{\bar\phi}),}
where $C_0$ is the solution found above at $N=0$.

There is an
 important benefit of  {\it not} shifting $\phi$ and $\bar\phi$ 
 to eliminate the $N$ dependence.
Because of the background charge in the worldsheet Lagrangian, 
such a shift would change the string
coupling constant by an amount that would diverge if we go
to the points $N=\pm 1$.  If we want to vary $N$ keeping the string coupling
constant fixed, we should use the form of the $C$ matrix in \moxie.

Equation \moxie\ shows that going beyond the WZW point, that is to $|N|>1$,
would change the reality properties of the solution.  (For $|N|<1$,
$C$ is hermitian if we consider $\bar\phi$ the complex conjugate of $\phi$,
but that fails for $|N|>1$.)   In terms of our discussion in section 7,
this is related to the fact that the fluxes are no longer real if
$|k|>1/f^2$.

To study the limit as $N\to 1$, a subtle choice of shifting
$\phi$ and $\bar\phi$ is useful.
We counter-rotate $\phi$ and $\bar\phi$:
\eqn\ugh{\eqalign{ e^\phi & \to -{2\over {\sqrt{1-N^2}}}e^\phi  \cr
                  e^{\bar\phi} & \to -{{\sqrt{1-N^2}}\over 2}e^{\bar\phi}.\cr}}
Factors of $-2$ have been included to put the resulting formulas
in the nicest form.
 This rescaling has no effect
on the string coupling constant, since the background charge effects
cancel between $\phi$ and $\bar\phi$.
For $N$ near 1, we now get 
\eqn\ompo{C=\left(1+{{1-N^2}\over 4}e^{\phi+\bar\phi}\right)
\left(\matrix{ i &  e^\phi \cr {{1-N^2}\over 4}e^{\bar\phi} & -i\cr}\right)
+O((1-N)^2).}
We have included terms of order $(1-N)$ since these are needed
to determine the solution for $V$.
We now compute that to leading order near $N=1$, 
\eqn\otubo{\eqalign{\partial C & = \partial\phi\left(\matrix{ 0 & e^\phi\cr
                      0 & 0 \cr}\right)  \cr
                     \bar\partial C & = \half(1-N)
                     \bar\partial\,
                     \bar\phi\left(\matrix{ie^{\phi+\bar\phi} &
                                e^{2\phi+\bar\phi} \cr
                            e^{\bar\phi} & -ie^{\phi+\bar\phi}\cr}\right).  }}
Inserting these formulas in \difmwz, we find that the limit of $V$ for
$N\to 1$ is
\eqn\jotubo{V=\left(\matrix{1 & 0 \cr ie^{\bar\phi} & 1 \cr}\right)
\left(\matrix{1 & - ie^\phi \cr 0 & 1 \cr}\right).    }

In the formalism that we have used  up to this point,
the right supercurrents are the standard ones, and the left ones
are rotated by $V$.  $N=1$ is in fact the WZW point, at which the left
currents are anti-holomorphic and the right currents are holomorphic.
A more natural description at the WZW point is one in which the left
currents are rotated by an anti-holomorphic factor, and the right currents
by a holomorphic factor. To get to such a description,  we make
an $R$ transformation by the inverse of the second factor in $V$,
to go to a description in which the anti-holomorphic currents are rotated
from the standard ones by a factor
\eqn\gotubo{\left(\matrix{ 1 & 0 \cr i e^{\bar\phi} & 1\cr}\right),}
and the holomorphic currents are instead rotated by
\eqn\mutto{\left(\matrix{ 1 & i e^\phi \cr 0 & 1\cr}\right).}            
 
It will now be shown that this 
$R$-transformation removes the ghost couplings derived from \ompo, 
so we get a pure WZW model of $SU'(2|2)$, with the currents
rotated as just indicated.
This description, which matches nicely with the treatment of the flat
case in section 4, will be our starting point in section 10, where we
examine $SU'(2|2)$ current algebra.

So we need to show that 
\eqn\want{\delta(S_0 + N S_{WZ}) + \delta (S_\phi) = - S_\phi}
under the $R$-transformation of \mutto\ where
$S_\phi = -\int e^\phi F^a\wedge F^a$. 
We will organize the proof as follows.  We will show that
the first order variation of $S_0+NS_{WZ}$ -- that is, the term
of order $e^\phi$ -- is
\eqn\qant{\delta_1(S_0+N S_{WZ})=-S_\phi.}
This is the result we want to first order in $e^\phi$.  Moreover,
we will prove that there are no higher order terms (proportional
to $e^{n\phi}$ for $n>1$) in $\delta(S_0+N S_{WZ})$.  Combining
this last fact with \qant, which identifies $S_\phi$ as the first
order variation of $S_0+NS_{WZ}$, it follows that $\delta S_\phi=0$.
Hence the left hand side of \want\ is linear in $e^\phi$, and
\qant\ is equivalent to \want.

To show \qant, note first that 
\eqn\dfirst{\delta_1 (S_0 + N S_{WZ}) = 
- i \int d^2 \sigma (\p_i e^\phi ) \tilde r^{22}_i 
=  i\int d^2 \sigma  e^\phi \p_z \tilde r_{\bar z}^{22}}
where 
$\tilde r^{\a\b}_i$ is the conserved current of the WZW model
associated to the $R$-transformation. 
At this stage, we can prove that there can be no higher order variation
of $S_0+NS_{WZ}$.  Indeed, since 
two strictly upper triangular $2\times 2$ matrices commute, the 
$R$-transformation we are making, if the coefficient $e^\phi$ is
constant, would commute with the current $\tilde r_{\bar z}^{22}$ that
generates an upper triangular $R$ transformation.  A variation of
$\tilde r_{\bar z}^{22}$ proportional to derivatives of $e^\phi$ would,
on symmetry and dimensional grounds, be proportional to $\bar\partial 
e^\phi=0$.
So $\delta(S_0+NS_{WZ})=\delta_1(S_0+NS_{WZ})$.  As we discussed in the
last paragraph, it follows that \qant\ and \want\ are equivalent.

As in subsection (8.2), we
shall compute 
$\tilde r^{\a\b}_i$ indirectly from 
its commutation relations with the $SU'(2|2)$ charges 
\eqn\charget{[
\oint \tilde r^{\a\b}, \oint\tilde  S_{La}^\g ] 
= \half(\e^{\beta\gamma} \oint\tilde  S_{La}^\alpha 
+\e^{\a\g} \oint\tilde  S_{La}^\b),}
$$[\oint\tilde  
r^{\a\b}, \oint\tilde  S_{Ra}^\g ] 
= \half(\e^{\beta\gamma} \oint\tilde  S_{Ra}^\alpha 
+\e^{\a\g} \oint\tilde  S_{Ra}^\b), $$
which implies that
\eqn\rcommt{[
\tilde r^{\a\b}, \oint\tilde  S_{La}^\g ] 
= \half(\e^{\beta\gamma} \tilde  S_{La}^\alpha 
+\e^{\a\g} \tilde  S_{La}^\b) +dM_a^{\a\b\g},}
$$[\tilde  
r^{\a\b}, \oint\tilde  S_{Ra}^\g ] 
= \half(\e^{\beta\gamma} \tilde  S_{Ra}^\alpha 
+\e^{\a\g} \tilde  S_{Ra}^\b) +dN_a^{\a\b\g}, $$
for some $M_a^{\a\b\g}$ and $N_{\a\b\g}$.
Note that the susy charges and currents contain tilde's 
since they include
the contribution from the Wess-Zumino term. 

Since $\tilde S_{La}^\alpha$ has only a $\bar z$ component and
$\tilde S_{Ra}^\alpha$ has only a $z$ component, \rcommt\ implies that
\eqn\rcot{[
\tilde r_{\bar z}^{\a\b}, \oint\tilde  S_{La}^\g ] 
= \half(\e^{\beta\gamma} \tilde  S_{La\bar z}^\alpha 
+\e^{\a\g} \tilde  S_{La\bar z}^\b) +\bar\p M_a^{\a\b\g},}
$$
[\tilde
r_{\bar z}^{\a\b}, \oint\tilde  S_{Ra}^\g ] =\bar\p N_a^{\a\b\g}.$$
So, ignoring the $\bar\p $ terms,
$\tilde r_{\bar z}^{\a\b}$ has
the same commutation relation as $-\lambda_{L\bar z}^{\a\b}$
of \llcom\ since $\tilde S_{La\bar z}^\a = 2 S_{La\bar z}^\a$ when $N=1$.
Therefore, $\tilde r_{\bar z}^{\a\b} + \lambda_{L\bar z}^{\a\b}$ is
$SU'(2|2)\times SU'(2|2)$-invariant up to $\bar\p $ terms, which
implies that
\eqn\rtis{\tilde r_{\bar z}^{\a\b}= -\lambda_{L\bar z}^{\a\b} + 
\bar\p  f^{\a\b}}
for some $f^{\a\b}$.
Using \rtis\ and $\bar\p  e^\phi =0$, \dfirst\ implies that
$$\delta(S_0 + N S_{WZ})= -i\int d^2\sigma e^\phi \p_z \lambda_{L\bar z}^{22}
= \int e^\phi \wedge \omega^{22}_L =
\int e^\phi F_L^a \wedge F_L^a = -S_\phi.$$
This completes the proof.

\subsec{Quantum Treatment And Conformal Invariance}

The above analysis of the supersymmetry of the ghost couplings
was really based on a classical manipulation.  One may question
whether there are quantum anomalies that spoil supersymmetry.

First of all, there is no problem in maintaining invariance under
the right action of $SU'(2|2)$, since this is preserved, for example,
by Pauli-Villars regularization.  Likewise, Pauli-Villars regularization
preserves the symmetry under the bosonic part of $SU'(2|2)$, acting
on the left.  The question is to preserve the left supersymmetries.

Let us suppose that, in some regularization, $c$ and $ v$ have
been chosen so that the supersymmetric variation of the action
vanishes up to $k^{th}$ order in $e^\phi$ and $e^{\bar \phi}$,
that is including all terms $e^{a\phi+b\bar\phi}$ with $a+b\leq k$.
We also assume that $c$ and $v$ to this order are polynomials
of order $k$ in $e^\phi$ and $e^{\bar \phi}$.  We will show that
$c$ and $v$ can be corrected by $k+1^{th}$
order polynomials to  cancel the variation in 
that order.  The analysis could be taken as a substitute for the explicit
solution of the classical equations that we have given above.

An important part of the structure is that since $\phi$ and $\bar\phi$
enter only in the combinations $e^\phi$ and $e^{\bar\phi}$, which
have no short distance singularities, $\phi$ and $\bar\phi$ can be
treated in making the analysis as $c$-number functions that
obey $\bar\partial \phi=\partial\bar\phi=0$.  Also, if $\phi$
and $\bar\phi$ are constant, the whole deformation that we have
constructed collapses.  That is because, as explained in section 7,
the two forms $S^{a\alpha}\wedge S^{a\beta}$ can be written 
\eqn\imxono{S^{a\alpha}\wedge S^{a\beta}=d\lambda^{\alpha \beta}} 
for some $\lambda$.  Hence the interaction $S_1$ is
\eqn\inconn{S_1=-\int dc_{\alpha\beta}\wedge  \lambda^{\alpha\beta},}
where we can think of $\lambda$ as a  quantum field operator
of dimension one.  This shows that the corrections to the sigma
model action $S_0$ vanish if $c$ is constant and in general
are ``soft'' or superrenormalizable interactions.
We will not actually use this way of writing $S_1$ explicitly,
because there is no nice choice of $\lambda$ (for instance, one cannot
pick $\lambda$ to be right-invariant), but nevertheless this fact underlies
the arguments that follow.  Because $S_1=0$ if $\phi$ and $\bar\phi$
are constant, it follows that any anomaly is proportional to $\partial_z
\phi$ or $\partial_{\bar z} \bar\phi$.  Since $\phi$ and $\bar\phi$
appear only via powers of $e^\phi$ or $e^{\bar\phi}$, any
$\partial_z \phi$
term multiplies $e^{a\phi}$ for some $a>0$, and any $\partial_{\bar z}
\bar\phi$
term multiplies $ e^{b\bar\phi}$ for some $b>0$.

Let $\tilde S^{a I}=v^I_\alpha S^{a\alpha}$ be the rotated supercurrents
that generate the left supersymmetry to order $k$. Thus, the
divergence of $\tilde S^{aI}$ is of order $k+1$ in $e^\phi$, $e^{\bar\phi}$.
This divergence is of the form
\eqn\xnxon{\partial_i \tilde S_i^{aI}={\cal O}^{aI},}
where ${\cal O}^{aI}$ has the following properties.  
It is right-invariant and of dimension two since $\tilde S$ is right-invariant
and of dimension one; 
it is proportional to $\partial\phi$ or $\bar\partial\phi$
since there is no anomaly if these vanish; and because of the $a$
index of the current, it transforms as a vector under the ``rotation''
subgroup of $SU'(2|2)$.  Any operator with these properties is a linear
combination of the left supersymmetry
currents times $\partial\phi$ or $\bar\partial\phi$.\foot{Any dimension
one classical operator is $\lambda_A \partial_i \Phi^A$, where $\Phi^A$
are the coordinates on the group manifold and $\lambda= \lambda_A d\Phi^A$ is
a one-form on the group manifold.  
The operator in question is right-invariant if and only
if $\lambda$ is right-invariant; but the right-invariant one-forms
are associated precisely with the left currents.}
The divergence of the currents is therefore of this form, and hence
so is the variation of the action $S_0+S_1$ under a supersymmetry generated
by $\tilde S^{aI}$:
\eqn\mumbog{\delta_{aI}(S_0+S_1) =\int d^2\sigma\left(\partial_z\phi  f_{IJ} 
\tilde S_{\bar z\,a}^J+\partial_{\bar z}\bar\phi g_{IJ}
\tilde S_{ z\,  a}^J\right).}
Here $f_{IJ}$ and $g_{IJ}$ are unknown functions of $e^\phi$ and
$e^{\bar \phi}$.
A ``Wess-Zumino consistency condition''\foot{One applies $\delta_{bJ}$ to
\mumbog\ and uses the fact that $\{\delta_{bJ},\delta_{aI}\}(S_0+S_1)=0$
since $\{\delta_{bJ},\delta_{aI}\}$ is a generator of rotations and
the action is manifestly rotation-invariant.
But in  evaluating $\delta_{bJ}(S_0+S_1)$ from the formula \mumbog,
the action of $\delta_{bJ}$ on the right and side is only via
 the action of supersymmetry on the supercurrents $\tilde S$,
 $\delta_{bJ}\tilde S_{aI}\sim \epsilon_{IJ}\epsilon_{abcd}K^{cd}$.
Because of the $\epsilon_{IJ}$,
 $\{\delta_{bJ},\delta_{aI}\}(S_0+S_1)=0$ holds if and
only if $f$ and $g$ are symmetric.}
 shows that
\eqn\humbog{f_{IJ}=f_{JI},~~~g_{IJ}=g_{JI}.}
Furthermore, $f_{IJ}$ is divisible by $e^\phi$, since, as noted at the
end of the last paragraph $\phi$ only appears in this form, so the
 $\partial\phi$ term in \mumbog\ is really a sum of terms 
$\partial\phi e^{a\phi}$ with $a>0$; similarly, $g_{IJ}$ is divisible
by $e^{\bar\phi}$.  Since $\partial \phi e^{a\phi}=\partial(e^{a\phi}/a)$
and similarly for $\bar\phi$, 
 we can write $f_{IJ}=\partial F_{IJ}$,
$g_{IJ}=\bar\partial G_{IJ}$, where $F$ and $G$ like $f$ and $g$
are polynomials in $e^\phi$, $e^{\bar\phi}$.  The anomaly is therefore
\eqn\trumbog{\delta_{aI}(S_0+S_1)=\int d^2\sigma\left(\partial_z F_{IJ}
\tilde S_{\bar z\,a}^{J}+\partial_{\bar z} G_{IJ} \tilde S_{z\,a}^J\right).}
Now we write $F=H+K$, $G=H-K$, and of course we are interested in
canceling the term of order $k+1 $ (in $e^\phi$ and $e^{\bar\phi}$) in
$H$ and $K$.   The contribution proportional to $K$
can be canceled via a $k+1^{th}$ order correction to $c$, and the
contribution proportional to $H$ can be canceled via a $k+1^{th}$ order
contribution to $v$.

This shows that spacetime supersymmetry is not spoiled by any quantum
anomalies; at most such anomalies modify the classical expressions for
$c$ and $v$.

\subsec{ Conformal Invariance}

The contributions 
 $S_0$ and $S_1$  to the action  are conformally invariant
at the classical level.  Now we wish to show that also the quantum theory
is conformally invariant.  We already know from section 7 that this
is true for $S_0$ alone; we wish to show that incorporation of the ghost
couplings does not spoil conformal invariance.  

The interactions in $S_1$ are actually sigma model fields of dimension
one, multiplying $\partial\phi$ or $\bar\partial\phi$, as we have recalled
in eqn. \imxono.  The trace of the stress tensor vanishes if the ``soft''
interactions in $S_1$ are turned off.  Hence, with $S_1$ included,
this trace -- which we denote $T$ -- is a sum of terms each of which
is a sigma model operator of dimension one or zero multiplying
one or two derivatives of $\phi$ and $\bar\phi$ and 
 powers
of $e^\phi$ and $e^{\bar\phi}$.  In addition,
$T$ is invariant under left and right supersymmetry.  There is no
left and right-invariant sigma model operator of dimension less than two
except the identity, so $T$ is of the form
\eqn\iub{T=\partial_z\phi\partial_{\bar z} \bar\phi N(e^{\phi},e^{\bar\phi}).}
Moreover, by familiar arguments,
$N$ is divisible by $e^\phi$ and $e^{\bar\phi}$.   
It follows that
\eqn\jiub{T=\partial_z\partial_{\bar z}\tilde N(e^\phi,e^{\bar\phi})}
for some $\tilde N$; we have used the fact
that  for any $a,b>0$,  $\partial_z\phi\partial_{\bar z}\bar\phi e^{a\phi+
b\bar\phi}=\partial_z\partial_{\bar z}e^{a\phi+b\bar\phi}/ab$.
But a $c$-number trace of the stress tensor of the
form  in \jiub\
can be canceled
by adding to the Lagrangian a ghost coupling to background curvature,
$\int d^2\sigma \tilde N(e^\phi, e^{\bar\phi})R$, with $R$ the curvature
scalar of the worldsheet.  Upon adding such a coupling, we achieve conformal
invariance.  We do not know whether with a natural
regularization such an addition to the action is  actually needed.
        
\def\p{{\partial}}
\def\S{{\bf S}}
\def\R{{\bf R}}
\def\2{{\bf 2}}
\def\T{{\bf T}}
\def\Sdet{{\rm Sdet\,}}

\def\e{{\epsilon}}
\def\a{{\alpha}}
\def\b{{\beta}}
\def\g{{\gamma}}
\def\d{{\delta}}
\def\s{{\sigma}}
\def\N{{\nabla}}
\def\half{{1\over 2}}
\def\p{{\partial}}
\def\pb{{\bar\partial}}
\def\t{{\theta}}
\def\Gtp{{\tilde G^+}}
\def\Gtm{{\tilde G^-}}
\def\tb{{\bar\theta}}

\newsec{$N=2$ Constraints for ${\bf AdS}_3\times {\bf S}^3$ }

In this section, we shall first define the $N=2$
superconformal constraints
associated with  the ${\rm AdS}_3\times {\bf S}^3$ with ghosts
that was described in section 8. The $N=4$
topological constraints can be easily constructed from these
$N=2$ constraints using the method described in section 2. We shall then
show that these $N=2$ constraints commute with the ghost-dependent
$SU'(2|2)\times SU'(2|2)$
transformations found in section 8. Using the formalism of sections
2-5, this implies that scattering amplitudes
as well as physical state conditions are 
$SU'(2|2)\times SU'(2|2)$ invariant.
Finally, we shall show that the $N=2$ constraints are also holomorphic.
The proofs of
$SU'(2|2)\times SU'(2|2)$ invariance and holomorphicity are not as
precise as the rest of this paper, because we treat some aspects
of the $SU'(2|2)$ current operator products classically, potentially
overlooking terms analogous to normal-ordering contributions in free field
theory.  We expect that a more complete treatment would be somewhat
similar to the proof of $SU'(2|2)$ invariance and conformal invariance
of ghost couplings in the last section.  

We have not explicitly checked that our constraints have the standard
$N=2$ OPE's, which would be a useful thing to check.
Probably the only difficult OPE to check would be that $G^+$ has no
singularity with itself.

\subsec{Review of Constraints in Minkowski Background}

Since the proof of $SU'(2|2)\times SU'(2|2)$ invariance is somewhat
complicated, it will be helpful to first review this invariance for the
flat Minkowski case.
Recall that the $N=2$ superconformal constraints in a flat $\d=6$
Minkowski background are given by
$$T=
{1\over 8}\epsilon_{abcd}\p x^{ab}\p x^{cd} +
p_a\p \t^a +\half\p\rho\p\rho
+\half\p\sigma\p\sigma
+{3\over 2}\p^2 (\rho+i\sigma)+T^{GS}_{C}$$
\eqn\gensusyyy{ G^{+} =
- {1\over {24}}e^{-2\rho-i\sigma}\epsilon^{abcd}p_a p_b p_c p_d ~
+{i\over 2}
e^{-\rho} p_a p_b \p x^{ab} +}
$$
 e^{i\sigma}(
{1\over 8}\epsilon_{abcd}\p x^{ab}\p x^{cd} +
p_a\p \t^a)  +\half e^{i\sigma}(\p(\rho+i\sigma)\p(\rho+i\sigma)-
\p^2(\rho+i\sigma))
+ G^{+~GS}_{C} , $$
$$G^{-}  =
e^{-i\sigma}+ G^{-~GS}_C ,$$
$$J=\p(\rho+i\sigma)~+J_{C}^{GS}. $$

Holomorphicity of these constraints is clear.  The only
non-trivial $SU'(2|2)\times SU'(2|2)$ invariance to check is the supersymmetry
corresponding to
$q_a^+=\oint (e^{\phi} p_a -{i\over 2}\epsilon_{abcd} \theta^b\p x^{cd}),$
which generates the transformation
\eqn\nonts{\delta \t^a =u^a e^\phi, \quad \delta p_a ={i\over 2}
\epsilon_{abcd} u^b \p x^{cd},
\quad \delta x^{ab} = i(u^a \t^b -u^b\t^a),}
$$\delta
\p(\rho-i\sigma)=2 u^a p_ae^\phi,\quad \delta\phi=0,$$
$$ \delta e^{m\rho+i(m+1)\sigma}
= e^{(m-1)\rho +im\sigma} u^a p_a,\quad \delta e^{-i\sigma}=0.$$
Note that these transformations can be obtained by taking the contour
integral of the susy generator around the worldsheet field.  We recall
that $\phi=-\rho-i\sigma$ and that $\rho$ and $\sigma $ have canonical OPE's.

Under the transformation of \nonts,  it is trivial to show that
$\delta J=\delta G^-=0$. To show that $\delta T=0$, note that
the transformation of $x^{ab}$ cancels the transformation of $p_a$
and the transformation of $\t^a$ cancels the transformation of
$\p(\rho-i\sigma)$.  Invariance of $G^+$ is more complicated and
inolves cancellation between various terms.
The variation of $e^{-2\rho-i\sigma}$ in the first term does not
contribute since $(p)^5=0$. The variation of $p_a$ in the first term
contributes
$${{i}\over {12}}e^{-2\rho-i\sigma}\epsilon^{abcd} p_b p_c p_d \epsilon_{aefg}
u^e \p x^{fg}= -{i\over 2} e^{-2\rho-i\sigma} u^a p_a (p_c p_d \p x^{cd})$$
which cancels the variation of $e^{-\rho}$ in the second term.
The variation of $p_a$ and $x^{ab}$ in the second term contributes
$${i\over 2}e^{-\rho}(i\epsilon_{abcd} u^b \p x^{cd} p_e \p x^{ae}
+2i u^a p_a p_b \p\t^b)
=-
e^{-\rho}(u^e p_e)({1\over 8}\epsilon_{abcd} \p x^{ab}\p x^{cd}
+p_b \p\t^b)$$
which is cancelled by the variation of $e^{i\sigma}$ in the third term
of $G^+$. Finally, the variation of $x$, $\t^a$ and $p_a$ in the third term
of $G^+$ contributes
$$ e^{i\sigma}( {i\over 2}\epsilon_{abcd}\p x^{ab} u^c \p\t^d
-{i\over 2}\epsilon_{abcd} u^b \p x^{cd}\p\t^a +p_a u^a \p e^\phi)=
-  e^{i\sigma} u^a p_a \p e^\phi,$$
which is cancelled by the variation of the fourth term in $G^+$, namely
$\half e^{i\sigma}( (\p\phi)^2 +\p^2\phi)$.
The easiest way to compute the variation of this fourth term is to write it
as
$\p e^\phi e^{2i\sigma+\rho}$, which is regularized as
\eqn\nor{\p e^\phi e^{2i\sigma+\rho} =
{1\over{2\pi i}}\oint dy {{\p_y e^{\phi(y)}}\over {y-z}}
e^{2i\sigma(z)+\rho(z)}}
where the contour integration of $y$ goes around the point $z$. To
reproduce $\half e^{i\sigma}( (\p\phi)^2 +\p^2\phi)$ from \nor,
one uses
that
$$\p_y e^{\phi(y)} e^{2i\sigma(z)+\rho(z)}
=  e^{i\sigma(z)}\p_y [(y-z)^{-1} +\p\phi(z) +{{y-z}\over 2}
((\p\phi)^2 +\p^2\phi)]. $$

\subsec{$SU'(2|2)\times SU'(2|2)$ Invariance of Constraints}

In the ${\rm AdS}_3\times {\bf S}^3$ background with Ramond-Ramond coupling, we
learned in the last section that the sixteen spacetime-supersymmetry
transformations are now generated by  the eight fermionic components of
the left-invariant currents $ g^{-1} dg$, and by the eight currents
\eqn\susyads{v_1^+ E^a + v_2^+ F^a,\quad v_1^- E^a + v_2^- F^a}
where $E^a$ and $F^a$ are the fermionic components of the right-invariant
currents $ dg ~g^{-1}$, and
\eqn\uttww{V=\left(\matrix{ v_1^+ & v_2^+ \cr v_1^- & v_2^-}\right)=
(1-{1\over 4}e^{\phi+\bar\phi})^{-1}
\left(\matrix{
 1 +{1\over 4} e^{\phi+\bar\phi}& i e^\phi \cr
-i e^{\bar\phi}
 &1+{1\over 4} e^{\phi+\bar\phi}}\right). }

It will now be argued that the following
$N=2$ generators are invariant under all $SU'(2|2)\times SU'(2|2)$
transformations of the action (including the $\phi$-dependent
transformations of \susyads):
$$T=
{1\over 8}\epsilon^{abcd} K_{ab} K_{cd} -\half \e_{\a\b} S^{a\a} S^{a\b}
 +\half\p\rho\p\rho
+\half\p\sigma\p\sigma
+{3\over 2}\p^2 (\rho+i\sigma)+T^{GS}_{C},$$
\eqn\gensu{ G^{+} =
-{1\over {6}} e^{-2\rho-i\sigma} \epsilon^{abcd} U_{ab} U_{cd}
~ +i e^{-\rho} K^{ab} U_{ab} +}
$$
e^{i\sigma}( {1\over 8}\epsilon^{abcd} K_{ab} K_{cd} -\half \e_{\a\b}
S^{a\a} S^{a\b} )
+ \half e^{i\sigma}
(\p(\rho+i\sigma)\p(\rho+i\sigma)-\p^2(\rho+i\sigma))
+ G^{+~GS}_{C} , $$
$$G^{-}  =
e^{-i\sigma}+ G^{-~GS}_C ,$$
$$J=\p(\rho+i\sigma)~+J_{C}^{GS}, $$
where
$$U_{ab}= S_a^\a S_b^\b (\p c_{\a\b}/\p e^\phi)
= S_a^\a S_b^\b
 \p c_{\a\b} (\p e^\phi)^{-1} $$
$$
=  (1- {1\over 4} e^{\phi+\bar\phi})^{-2}
[\half F_a F_b -{1\over{8}} e^{2\bar\phi}
E_a E_b -{i\over{4}} e^{\bar\phi} (E_a F_b + F_a E_b)],$$
$c_{\a\b}=\epsilon_{\a\g}\epsilon_{\b\d} c^{\g\d}$ where $c_{\g\d}$
is defined in \cexact, $S^1_a=E_a$, $S^2_a=F_a$,
and  [$E_a$,$F_a$,$K_{ab}$] are the fermionic and bosonic
$z$-components of the right-invariant currents $\p_z g ~g^{-1}$.
Note that since the $\phi$-dependent part of the
action is anti-symmetric in $z$ and $\bar z$,
it does not contribute to $T$.
Also note that when $E^a = \p \t^a$, $F^a = p^a$, $K^{ab}=\p x^{ab}$,
and $e^{\bar\phi}$ is set to zero, one recovers the $N=2$ constraints
of \gensusyyy\ in a flat Minkowski background. 
The anti-holomorphic $N=2$ generators,
$[\bar T$,$\bar G^+$,$\bar G^-$,$\bar J]$,
are similar to those of \gensu\ but use the $\bar z$-components
of the right-invariant currents, $\bar\p g ~g^{-1}$.

Under $SU'(2|2)$
transformations acting from the right, \gensu\ is clearly invariant since
it is constructed using right-invariant currents. Also, under
the bosonic $SU'(2|2)$ transformations acting from the left, \gensu\
is invariant since all indices are contracted in an SO(4)-covariant
manner. So one only needs to check that \gensu\ is invariant
under the $\phi$-dependent susy transformation
generated by
\eqn\vsusy{\eta_I^a (v_1^I E^a + v_2^I F^a) =\eta_I^a v_\a^I S^{\a a}}
where $v^I_\a$ are defined in \uttww.
Under \vsusy, the right-invariant currents transform as
\eqn\righttr{\delta S^{\a a} =
\epsilon^{\a\b}(-\half\epsilon^{abcd} \eta_I^b v_\b^I
K^{cd} -\eta_I^a \p v_\b^I),
\quad\delta K^{ab}= -\eta_I^a  v_\a^I  S^{\a b} +
 \eta_I^b  v_\a^I  S^{\a a}}
where the transformations of $S^{\a a}$ were explained in \gentransf.
The compactification variables are all inert under this
transformation.

One also needs to define
how the chiral bosons, $\rho$ and $\sigma$, transform under
the susy transformation of \vsusy.
Since \vsusy\ only involves
these chiral bosons in the linear combination
$\phi=-\rho-i\sigma$, $\rho+i\sigma$ commutes with \vsusy\ and is
therefore invariant under the susy transformation.
However, since
\eqn\ffope{\p(\rho(y)-i\sigma(y))~ e^{n\phi(z)} 
\to {{2n }\over{y-z}} e^{n\phi},}
$\p(\rho(y)-i\sigma(y))$
will not be invariant but will be defined to transform as
\eqn\prhot{\delta(\p(\rho-i\sigma))=
2  \eta_I^a S^\a_a (\p v_\a^I/\p\phi)=
{2\over {\p \phi}}
\eta_I^a S^\a_a \p v_\a^I .}
Note that $\p v_\a^I = \p\phi(\p v_\a^I / \p\phi)$
since $\p \bar\phi=0$ even in the presence of $S_1$.  The notation in
\prhot\ is somewhat symbolic.  We have $\partial v_\a^I=\partial\phi
f_\alpha^I(e^\phi,e^{\bar\phi})$ for some function $f$, and we write
$f_\alpha^I$ as ${1\over {\partial \phi}}\partial v_\alpha^I$.  Other
formulas below must be read likewise. 

In defining the above transformation we need to justify using the free field
OPE \ffope\ even after the deformation of the action by coupling in
the ghost fields.  We can treat the ghost couplings as a perturbation
and bring down terms from the action, and recall that 
the action only involves ghost terms of the form $e^{m\phi +k\bar \phi}$.  We
then consider the OPE of any field $A(y)$ with $e^{n\phi (z)}$.
Noting that $\phi(z)$ has no singularity with the fields $\phi$ or 
${\bar \phi} $ coming from the ghost couplings in the deformed action,
we conclude that the deformation
will not modify the singularity structure of the
OPE of $A(y)$ with $e^{n\phi (z)}$, and so these are the same as the
free field OPE singularity. This justifies \ffope\ and the
OPE's we now consider.  In particular we have,
\eqn\opemrhot{
e^{m\rho(y)+i(m+1)\sigma(y)}~ e^{n\phi(z)}\to {{n}\over{y-z}}
 e^{(m-1)\rho(y) +
im \sigma(y)} e^{(n-1)\phi(z)},}
where we have taken the single pole with one of the
$e^\phi$'s in the term $(e^\phi(z))^n$.
So we shall define
\eqn\pmrhot{
\delta(e^{m\rho+i(m+1)\sigma})
=e^{(m-1)\rho+im\sigma}
{1\over{\p e^\phi}}\eta_I^a S^\a_a \p v_\a^I .}
Note that $e^{(m-1)\rho+im\sigma}$ may have poles with
${1\over{\p e^\phi}} \p v_\a^I $, so this expression may need to
be normal-ordered.
It might be possible to explicitly compute the contributions of
such normal-ordering terms, but
we have not yet done so.
Finally, we will need to know the susy transformation of
$e^{-i\sigma}$. Since $e^{-i\sigma}$ has no poles with
$e^{n\phi}$ for $n$ positive, we shall define
\eqn\gmtr{\delta(e^{-i\sigma})=0.}

Using the transformations defined above, it is now straightforward
to check that the $N=2$ constraints of \gensu\ are invariant under \vsusy.
The easiest constraints to check are $G^-$ and $J$, which are invariant
using \gmtr\ and the fact that $\phi$ is invariant. The next easiest
constraint to check is $T$, where the
variation of the the sigma model term
cancels the variation of the
kinetic term of the chiral bosons.
Note that the sigma model term can be written as
${\rm Str}(\p g ~g^{-1} \p g~g^{-1})$, whose variation under \vsusy\ is
$-\eta_I^a S_a^\a \p v_\a^I$.
The kinetic term for the chiral bosons can be written as
$\half \p(\rho-i\sigma)\p(\rho+i\sigma)$, so using \prhot, its
variation is  $\eta_I^a S^\a_a \p v_\a^I$.

The hardest constraint to check is $G^+$. To check its invariance,
it will be useful to note that
\eqn\identity{\p c^{\a\b} \p c_{\b\g}=\p c_{\a\b} \p c^{\b\g}=0.}
Using
\eqn\preveq{ v_\gamma^I \e^{\g\a}\p c_{\a\b}  = {i\over 2} \p v_\b^I}
from equation \coxxo,
this implies that
\eqn\useful{\p v_\a^I \p c^{\a\b} = 2i \epsilon^{\g\d} v_\g \p c_{\d\a} \p 
c^{\a\b}=0.}
Note that
\eqn\gxyz{G^+= e^{-2\rho-i\sigma} X + e^{-\rho} Y + e^{i\sigma} Z
+ e^{2i\sigma +\rho} \p e^{\phi}}
where $X$, $Y$ and $Z$ only depend on the chiral bosons in the combinations
$\phi$ and $\bar\phi$ and the last term comes from \nor.
Defining the supersymmetry transformation of
$G^+$ by
\eqn\sutram{\d G^+=
\d(e^{-2\rho-i\sigma}) X + e^{-2\rho-i\sigma} \d X
+\d(e^{-\rho}) Y + e^{-\rho}\d Y
+\d( e^{i\sigma}) Z + e^{i\sigma} \d Z
+\d(e^{2i\sigma + \rho})\p e^\phi,}
we shall now show that $\d G^+=0$.
Variation of the first
term in $G^+$ gives two contributions, one coming from the
variation of $e^{-2\rho-i\sigma}$ and the other coming from the
variation of $U_{ab}$. Using \pmrhot,
the variation of $e^{-2\rho-i\sigma}$ gives
$$-{{ e^{-3\rho-2i\sigma}}
\over{6 \p e^\phi}}\eta_I^e S^\a_e \p v^I_\a
 \epsilon^{abcd} U_{ab} U_{cd}=
-{{ e^{-3\rho-2i\sigma}}
\over{6 (\p e^\phi)^3}}\eta_I^e   \p v^I_\a \p c_{\b\g} \p c_{\d\tau}
 \epsilon^{abcd} S_e^\a S_a^\b S_b^\g S_c^\d S_d^\tau.$$
Since $S^\a_a$ is anti-commuting,
$$\epsilon^{abcd} S_e^\a S_a^\b S_b^\g S_c^\d S_d^\tau
=(\epsilon_{\kappa\pi} S_e^\kappa S_e^\pi)(\epsilon^{\a\b}
\epsilon^{abcd} S_b^\g S_c^\d S_d^\tau $$
$$+\epsilon^{\a\g}
\epsilon^{abcd} S_b^\b S_c^\d S_d^\tau
+\epsilon^{\a\d}
\epsilon^{abcd} S_b^\b S_c^\g S_d^\tau
+\epsilon^{\a\tau}
\epsilon^{abcd} S_b^\b S_c^\g S_d^\d).$$
but contracting the $\a$ index of $\p v_I^\a$ with any of the
indices of $\p c^{\b\g}\p c^{\d\tau}$ gives zero because of \useful,
so this contribution vanishes identically.

The second contribution from the first term of $G^+$ comes from the
variation of $S_a^\a$ and is
$${{4 e^{-2\rho-i\sigma}}\over {6\p e^\phi}}\epsilon^{abcd}
(\half\epsilon_{aefg}\eta_I^e K^{fg} v^I_\tau +\eta_I^a \p v^I_\tau)
\epsilon^{\a\tau} (\p c_{\a\b}) S^\b_b U_{cd}$$
$$={{ e^{-2\rho-i\sigma}}\over {3\p e^\phi}}
v^I_\tau (2\eta_I^b K^{cd}+4\eta_I^c K^{db})
\epsilon^{\a\tau} (\p c_{\a\b}) S^\b_b U_{cd}$$
\eqn\fvar{=-{{i e^{-2\rho-i\sigma}}\over {6\p e^\phi}}
\p v^I_\b  (2\eta_I^b K^{cd}+4\eta_I^c K^{db})
S^\b_b U_{cd} }
where we have used \preveq,
\useful\ and that $\epsilon^{abcd}\epsilon_{aefg}$
=$\delta^b_e \delta^c_f \delta ^d_g$ plus cyclic permutations.
The variation \fvar\ can be put in a simpler form by noting that
$$\p v^I_\b  S^\b_b U_{cd} =
\p v^I_\b S^\b_b S^\g_c S^\d_d {{\p c_{\g\d}}\over{\p e^\phi}}$$
$$
=  \p v^I_\b  (S^\g_b S^\b_c- \epsilon^{\b\g} (\half\epsilon_{\tau\kappa}
S^\tau_b S^\kappa_c)) S^\d_d {{\p c_{\g\d}}\over{\p e^\phi}}$$
\eqn\triu{= \p v^I_\b  S^\g_b S^\b_c S^\d_d
 {{\p c_{\g\d}}\over{\p e^\phi}}
= -\p v^I_\b S^\b_c U_{bd},}
where we have used \useful\ to go from the second to the last line.
Using \triu, one can finally write the variation of the first
term of $G^+$ in \fvar\ as
\eqn\finvar{- {{i e^{-2\rho-i\sigma}}\over {\p e^\phi}}
\eta_I^b \p v^I_\b S^\b_b  K^{cd} U_{cd}.}
Using \pmrhot, it is easy to see
that \finvar\ is precisely cancelled by the variation of
$e^{-\rho}$ in the second term of $G^+$.

One also gets contributions from varying $S^\a_a$ and $K_{ab}$ in the
second term of $G^+$. Using \righttr, these contribute
$$-{2i\over {\p e^\phi}}
 e^{-\rho}(\eta_I^a S^{b\b} v^I_\b S^\g_a S^\d_b
\p c_{\g\d} +K^{ab}
(\half\epsilon_{aefg}\eta_I^e K^{fg} v^I_\tau +\eta_I^a \p v^I_\tau)
\epsilon^{\a\tau} \p c_{\a\b} S^\b_b )$$
$$=- {2i\over{\p e^\phi}}
e^{-\rho}(
\eta_I^a \epsilon^{\b\d} v^I_\b S^\g_a (\half\epsilon_{\kappa\pi} S^\kappa_b
S^\pi_b)
\p c_{\g\d} +
\half\epsilon_{aefg}\eta_I^e K^{ab} K^{fg} v^I_\tau
\epsilon^{\a\tau} (\p c_{\a\b}) S^\b_b) $$
\eqn\seccon{= {1\over {\p e^\phi}}
 e^{-\rho} 
 (\eta_I^a (\p v_\g^I) S^\g_a (\half\epsilon_{\kappa\pi}S^\kappa_b S^\pi_b)
- \half\epsilon_{aefg}\eta_I^e K^{ab} K^{fg} S^\b_b \p v^I_\b)  }
where we have used \useful\ to go from the first to the second line
and \preveq\ to go from the second to the third line.
Since $\epsilon_{aefg} K^{ab} K^{fg}$
 =${1\over 4}\delta_e^b \epsilon_{acfg} K^{ac} K^{fg}$,
\seccon\ can be written as
\eqn\seccc{ -{1\over {2\p e^\phi}}
 e^{-\rho} (\eta_I^a \p v_\g^I S^\g_a) (-\epsilon_{\kappa\pi}S^\kappa_b S^\pi_b
+ {1\over 4}\epsilon_{acfg} K^{ac} K^{fg})=
-{{e^{-\rho}}\over {\p e^\phi}}
 \eta_I^a \p v_\g^I S^\g_a~{\rm Str}(\p g g^{-1} \p g g^{-1}) ,   }
which is cancelled by the variation of the $e^{i\sigma}$ in the third
term of $G^+$.

Finally, it will be shown that the remaining contribution
to the variation of the
third term in $G^+$ is cancelled by the variation of the
fourth term in $G^+$.  From varying the sigma model stress-tensor,
the remainining contribution to the third term in $G^+$ is given by
\eqn\varspf{
-\eta_I^a S_a^\a \p v_\a^I
e^{i\sigma} .}
but by writing the fourth term in $G^+$ as in \nor, it is easy to see that
\varspf\ is cancelled by the variation of the $e^{2i\sigma+\rho}$ in \nor.
So we have shown, up to normal ordering, that the constraints of
\gensu\ are $SU'(2|2)$-invariant.

\subsec{Holomorphicity of Constraints}

It will now be shown that, in addition to being invariant
under $SU'(2|2)\times SU'(2|2)$,
the $N=2$ constraints of \gensu\ are also holomorphic.
To prove this, we will need to know some facts about the
chiral bosons. The first fact is that $\bar\p\phi=0$. This
is because the action only involves $\rho$
and $\sigma$ in the combination $\rho+i\sigma$, so
$\p\bar\p(\rho+i\sigma)=0$ even after including the $\phi$-dependent
terms of $S_1$ into the action. However, once those $\phi$-dependent
couplings are included, $\rho-i\sigma$ is no longer holomorphic.  Rather,
in the presence of those couplings the equation of motion for $\phi$ gives
$\p\bar\p(\rho-i\sigma)=2\p S_1/\p\phi$.
Another useful fact is that $\bar\p e^{-i\sigma}=0$.
This can be seen by
unbosonizing the $\sigma$ field back into $b=e^{-i\sigma}$
and $c=e^{i\sigma}$. Since the $\phi$-dependent part of the action
only includes positive powers of $e^\phi=e^{-\rho-i\sigma}$,
it contains $b$ dependence but not $c$ dependence. So
the equation of motion for $c$ is still $\bar\p b=0$.
A final useful fact is that
\eqn\pmrh{\bar\p e^{i(m+1)\sigma+m\rho} =
 - e^{im\sigma+(m-1)\rho} {i\over {\p e^\phi}}\bar S_a^\a \p c_{\a\b} S_a^\b }
where $\bar S_a^\b$ is defined to be the $\bar z$ component of the
left-supersymmetry current $S_a^\a$. Note that when there is no bar over
the current, it will be assumed to be the $z$ component of the current.
Equation \pmrh\ can be heuristically justified in a manner similar to
\pmrhot, but there may be normal-ordering contributions that have not
been included.

To prove holomorphicity of the constraints,
we will also need to know what are $\bar\p K_{ab}$ and $\bar\p S_{a}^\a$.
Since the bosonic $SU'(2|2)\times SU'(2|2)$ transformations
are unmodified in the presence of $S_1$, $K_{ab}$ is conserved on-shell,
i.e. $\bar\p K_{ab}= -\p\bar K_{ab}$.
The Maurer-Cartan equation implies that
$$\p \bar K_{ab} -\bar\p K_{ab}
= -\epsilon_{\a\b} (S_a^\a \bar S_b^\b -\bar S_a^\a S_b^\b)
+\epsilon_{abcd}K^{ce} \bar K^{de},$$
so
\eqn\kbar{\bar \p K_{ab} =\half
\epsilon_{\a\b} (S_a^\a \bar S_b^\b -\bar S_a^\a S_b^\b)
-\half\epsilon_{abcd}K^{ce} \bar K^{de}.}

In the presence of $S_1$, the conserved supersymmetries are given
by $v^I_\a S_a^\a$, so
$\bar\p (v^I_\a S_a^\a)= -\p(v^I_\a \bar S_a^\a)$
where $v^I_\a$ is defined in \uttww.
Using \preveq, this implies that
$$\bar\p S_a^\d +2i \e_{\a\b}(\bar\p c^{\b\d}) S_a^\a =-\p \bar S_a^I
+2i \e_{\a\b}(\p c^{\b\d}) \bar S_a^\a.$$
Using the Maurer-Cartan equations,
$\p\bar S_a^\a -\bar\p S_a^\a$
= $-\half\e_{abcd} (K_{bc} \bar S_d^\a
-\bar K_{bc} S_d^\a)$, one learns that
\eqn\pbar{\bar\p S_a^\a =-i \e_{\g\b}( S_a^\g \bar\p c^{\a\b} -
\bar S_a^\g \p c^{\b\a}) +{1\over 4}\e_{abcd}
(K_{bc} \bar S_d^\a
-\bar K_{bc} S_d^\a).}

Using these equations for the chiral bosons and the right-invariant currents,
one is now ready to prove the holomorphicity of the constraints.
That $\bar\p T=0$ was proved in section 8.   Also, $\bar\partial G^-=
\bar\partial J=0$ using the above facts about chiral bosons.
Once again, $\bar\partial  G^+=0$ is the most difficult equation to prove,
but our task will be simplified by the knowledge that
$G^+$ is invariant under left supersymmetry.
Expanding $G^+$ as in \gxyz, we shall define
\eqn\dbarg{\bar\p G^+=
\bar\p(e^{-2\rho-i\sigma}) X + e^{-2\rho-i\sigma} \bar\p X
+\bar\p(e^{-\rho}) Y + e^{-\rho}\bar\p Y
+\bar\p( e^{i\sigma}) Z + e^{i\sigma} \bar\p Z
+\bar\p(e^{2i\sigma + \rho})\p e^\phi.}

First, note that if one takes the supersymmetry parameter to be
\eqn\susybar{\eta_I^a=- {1\over 2}\e_{\a\b}(v^{-1})_I^\a \bar S^{\b a},}
then the supersymmetric transformation law of $e^{m\rho+i(m+1)\sigma}$
becomes 
$\d(e^{m\rho+i(m+1)\sigma})=\bar\p e^{m\rho+i(m+1)\sigma}$
using \preveq, \pmrhot, and \pmrh.  Here we have defined
$v^{-1}$ by  $(v^{-1})_I^\a v^I_\b =\d^\a_\b$.
Second, note that when $\eta_I^\a$ is chosen as in \susybar,
\righttr\ and \kbar\ imply that
\eqn\difk{\bar\p K_{ab} - \d K_{ab}=
-\half\epsilon_{abcd}K^{ce} \bar K^{de} .}
Finally, note that when $\eta_I^\a$ is chosen as in \susybar,
\righttr\ and \pbar\ imply that
\eqn\difp{\bar \p S_a^\a -\delta S_a^\a =
- i \e_{\g\b} S_a^\g \bar\p c^{\a\b}
-{1\over 4}\e_{abcd}
\bar K_{bc} S_d^\a.}

Using the knowledge that $\d G^+=0$, one can therefore conclude that
$\bar\p G^+$ gets contributions from only three sources.
The first source is from \difk, the second source is
form \difp, and the third source is from when the $\bar\p$ hits
a $\bar\p c_{\a\b}$ in $G^+$. These three sources contribute only to
the terms $\bar\p X$, $\bar\p Y$ and $\bar\p Z$ in the
computation of \dbarg. All other contributions cancel out using
the proof of the previous subsection. It will now be shown
that the contributions from these three sources cancel each other
out in $\bar\p X$, $\bar\p Y$, and $\bar\p Z$, implying that $\bar\p G^+=0$.

First, consider the contribution of \difp\ to $\bar\p X$, which is
$${2\over {3(\p e^{\phi})^2}}
\e^{abcd}(i\e_{\g\b} S_a^\g \bar\p c^{\a\b} +{1\over 4}\e_{aefg}\bar K_{ef}
S^\a_g) \p c_{\a\d} S^\d_b S^\tau_c \p c_{\tau\kappa} S^\kappa_d.$$
The term involving $\bar K_{ef}$ is zero since,
after writing $\e^{abcd}\e_{aefg}$ as $\d^b_e\d^c_f\d^d_g$ plus
cyclic permutations, one can use
$S_a^\a \p c_{\a\b} S_a^\b=0$ since
$c_{\a\b}$ is a symmetric matrix, and
$$S_c^\a \p c_{\a\d} S^\tau_c \p c_{\tau\kappa}
= -\half\e^{\a\tau} (\e_{\pi\sigma} S_c^\pi S_c^\sigma)
\p c_{\a\d} \p c_{\tau\kappa}=0$$
using \identity. The remaining term is
$${2i\over {3(\p e^{\phi})^2}}
\e^{abcd}\e_{\g\b} 
S_a^\g 
\bar\p c^{\b\a} \p c_{\a\d} S^\d_b S^\tau_c \p c_{\tau\kappa} S^\kappa_d$$
\eqn\canone{ =
{1\over {3(\p e^{\phi})^2}}
\e^{abcd} S_a^\g \p\bar\p c_{\g\d} S^\d_b S^\tau_c \p c_{\tau\kappa} 
S^\kappa_d}
since \squarec\ implies that
$$\bar\p\p c_{\g\d}= i(\e_{\g\b} \bar\p c^{\b\a}\p c_{\a\d} +
 \e_{\d\b} \bar\p c^{\b\a}\p c_{\a\g}).$$
But \canone\ is cancelled by the contribution when the $\bar\p$ hits the
$c_{\a\b}$'s, so the sources cancel out in the $\bar\p X$ term.

Next, consider the contribution of \difp\ to $\bar\p Y$, which is
\eqn\termone{-{2i\over {\p e^\phi}}
 (i\e_{\g\b} S_a^\g \bar\p c^{\a\b} +{1\over 4}\e_{aefg}\bar K_{ef}
S^\a_g) \p c_{\a\d} S^\d_b K^{ab}.}
The term involving $\bar K_{ef}$ in \termone\
is cancelled by the contribution of
\difk\ to $\bar\p Y$, which is
$$ -{i\over 2}  U^{ab}\e_{abcd} K^{ce} \bar K^{de} 
= {i\over 2} \e_{aefg}\bar K^{ef} K^{ab} U^{gb}.$$
The remaining term in \termone\ is
$${2\over{\p e^\phi}}\e_{\g\b} S_a^\g \bar\p c^{\b\a} \p c_{\a\d} S^\d_b K^{ab}
= -{i\over {\p e^\phi}} S_a^\g \bar\p\p c_{\g\d} S^\d_b K^{ab},$$
which is cancelled by the contribution to $\bar\p Y$ when the $\bar\p$
hits the $c_{\a\b}$.

Finally,
consider the contribution of \difp\ to $\bar\p Z$, which is
$$ \e_{\a\b}
(i\e_{\g\d} S_a^\g \bar\p c^{\a\d} +{1\over 4}
\e_{aefg}\bar K_{ef} S^\a_g) S_a^\b.$$
But this vanishes since $S_a^\a$ is anti-commuting. Similarly,
the contribution of \difk\ to $\bar\p Z$ vanishes since
$$\e_{abcd} K^{ab} \e_{cdef} K^{eg} \bar K^{fg}
= 4 K^{ef} K^{eg}\bar K^{fg} =0$$
by symmetry in $f$ and $g$ of $K^{ef}K^{eg}$.
Since there are
no $c_{\a\b}$ terms in $Z$, we have
proven (up to normal-ordering) that $\bar\p G^+=0$.

\def\e{{\epsilon}}
\def\a{{\alpha}}
\def\b{{\beta}}
\def\g{{\gamma}}
\def\d{{\delta}}
\def\s{{\sigma}}
\def\N{{\nabla}}
\def\half{{1\over 2}}
\def\p{{\partial}}
\def\pb{{\bar\partial}}
\def\t{{\theta}}
\def\Gtp{{\tilde G^+}}
\def\Gtm{{\tilde G^-}}
\def\tb{{\bar\theta}}

\newsec{WZW on $SU'(2|2)$}

It would be nice to be able to completely
solve the two parameter conformal theory we have defined
above. One of the two parameters, $k$ is quantized and so is
not associated with a marginal deformation of the theory, whereas
the other parameter, $1/f^2$,  controls the radius of the target
and is associated with such a  deformation.  The experience
with conformal theories suggests that finding an exactly solvable
conformal theory along a whole line of marginal deformations is
extremely rare. In fact, the only known examples involve tori
and their orbifolds.  It is not clear to 
us whether one should expect in the case at hand an exactly
solvable theory for all $1/f^2$ and fixed $k$.  However there
is one point at which this must be so and that is the WZW limit where
${1/f^2}=|k|$.  In this limit we should get an affine Kac-Moody
algebra based on the supergroup $SU'(2|2)$ at level $k$, which should
be soluble to a large extent.  The $SU'(2|2)$ current algebra
can be written explicitly as follows:
$$K_{ab}(z)K_{cd}(0)\sim {{k }\epsilon^{abcd}\over z^2}+
{\delta_{ac}K_{bd}(0)-\delta_{ad}K_{bc}(0)-\delta_{bc}K_{ad}(0)+\delta_{bd}
K_{ac}(0)\over z}$$
$$S_{a\alpha}(z)S_{b\beta}(0)\sim {k\delta_{ab}\epsilon_{\alpha \beta}\over
z^2}+{\half\epsilon_{\alpha \beta}\epsilon_{abcd}K^{cd}(0)\over z}$$
$$K_{ab}(z) S_{c\alpha}(0)\sim {\delta_{ac}S_{b\alpha}(0)-\delta_{bc}
S_{a\alpha}(0)\over z}$$
We will see how this current algebra can be realized in terms of
more familiar theories.\foot{It would be interesting
to revisit sigma models on super Calabi-Yau manifolds proposed
in \ref\seth{S. Sethi,``Supermanifolds, Rigid Manifolds
and Mirror Symmetry,'' Mirror Symmetry II, ed. B. Greene and
S.-T. Yau, Int. Press 1997.}\ as mirror to rigid Calabi-Yau manifolds,
in light of the observation here that sigma model on supermanifolds
may in some cases be equivalent to more familiar systems.}  

The flat model on $\R^6$ times K3 or ${\bf T}^4$ was
described  in section 4 in the spacetime supersymmetric formalism
using free fermions $(p_a,\theta^b)$ of spins $(1,0)$.
{}From our discussion in section 6.4, we anticipate that after
deformation to ${\rm AdS}_3\times \S^3$ with only NS flux,
the model can be described by a WZW model of $SU'(2|2)$ that is
obtained by integrating out $p$, but
can also be realized by a system in which the fermions $(p,\theta)$
are both retained and have a first order kinetic energy.  
We can argue more specifically that in this description, $p$ and $\theta$
are simply free fields, just as in the $\R^6$ case.
In fact, in the RNS approach, the ${\rm AdS}_3\times \S^3$ background with
NS flux only is described \gks\
by an $SU(1,1)\times SU(2)$ current algebra with free fermions.  Such
a current algebra is constructed from bosonic $SU(1,1)\times SU(2)$
current algebra plus free RNS fermions $\psi^i$.  
The RNS approach is mapped
to the supersymmetric approach by a change of variables described in detail
for the $\R^6$ case in section 4.  This change of variables maps
$\psi^i$ plus ghosts to $p_a,\theta^b$ plus some ghost-like fields.
Going from $\R^6$ to ${\rm AdS}_3\times \S^3$ should not really change that
story; in the RNS description the $\psi^i$ are still free, so the same
change  of variables can be made in the same way and gives free
fields $p,\theta$.
Thus, we have a prediction  that the current
algebra of $SU'(2|2)$ can be realized by $SU(2)\times SU(2)$ bosonic
current algebra plus free fermions $p,\theta$ of spins $(1,0)$
(with a similar statement relating $SU'(1,1|2)$ to $SU(1,1)\times SU(2)$).
In
fact, this has already been shown  \ref\barsup{I. Bars,
``Free Fields And New Cosets of Current Algebras,'' Phys. Lett. {\bf B255}
 (1991) 353.}, and we will review it below.

The $SU'(2|2)$ current algebra, restricted to $SU(2)\times
SU(2)$, is at levels $(-k,k)$, where the minus sign for one of the $SU(2)$'s
is familiar from section 7.  Since inclusion of $p,\theta$ will shift
the level by $+2$ for each $SU(2)$, we must start with a purely
bosonic current algebra at level $(-k-2,k-2)$ for the two $SU(2)$'s:
\eqn\hcon{c={3(k-2)\over (k-2)+2}+{3(-k-2)\over (-k-2)+2}=6.}

What central charge do we expect for the $SU'(2|2)$ current algebra?
For any group at level $k$, the central charge is
$$c={k{\rm dim \,G}\over k+c_2(G)}.$$
Here $c_2(G)$ denotes the dual coxeter number of the group $G$, and 
in the case
of a supergroup, ${\rm dim}\,{ G}$ is interpreted as the super-dimension
of the group.  For $SU'(2|2)$, as we know from section 7, $c_2(G)=0$, so
the above formula becomes $c={\rm dim}\, G$.   
The superdimension of $SU'(2|2)$ is
$-2$, so we expect $c=-2$.  This agrees with the contribution $6$
from the bosonic current algebra plus $4(-2)=-8$ from four pairs
of spin $(1,0)$ free fermions.

We will now complete the dictionary between the currents $K^{ab}$ and
$S^a_{\alpha}$ of $SU'(2|2)$ and the bosonic
 currents $j^{ab}$ of $SU(2)\times SU(2)$
of WZW and the free fermions $(p_a,\theta^a)$.  
(In writing the bosonic currents as $j^{ab}$, we identify the adjoint
representation of $SU(2)\times SU(2)=SO(4)$ with the antisymmetric
tensor representation of $SO(4)$.)
The $SO(4)$  OPE structure of the $SO(4)$ currents together with
$p$ and $\theta$ is given
(for the left-movers) by\foot{If the levels of the $SU(2)$'s
were equal and opposite, the double pole in the $j_{ab}\cdot j_{cd}$ 
operator product would be proportional to $\epsilon_{abcd}$.  Because
of the shifts $(-k,k)\to (-k+2,k+2)$, this is not quite so and there
is a term proportional to $\delta_{ac}\delta_{bd}-\delta_{ad}\delta_{bc}$.} 
$$j_{ab}(z)j_{cd}(0)\sim{{k} \epsilon_{abcd}+2\delta_{ac}\delta_{bd}
-2\delta_{ad}\delta_{bc}
\over z^2}+
{\delta_{ac}j_{bd}(0)-\delta_{ad}j_{bc}(0)-\delta_{bc}j_{ad}(0)+\delta_{bd}
j_{ac}(0)\over z}$$
$$p_a(z)\theta_b(0)\sim {\delta_{ab}\over z}$$
$$p_a(z) j_{bc}(0)\sim 0 \qquad \theta_a(z) j_{bc}(0)\sim 0$$
The $SU'(2|2)$ currents can be constructed out of these fields as
follows:
\eqn\mapwzw{K^{ab}(z)=j^{ab}(z)-(p^a(z)\theta^b(z)-p^b(z)\theta^a(z))}
\eqn\morwzw{S_{a}^1 (z)=k 
\partial \theta_a(z)+{1\over 2}\epsilon_{abcd}\theta^b(z)
 (j^{cd}(z)+
\theta^c(z)p^d(z))}
\eqn\furwzw{S_{a}^2 (z)=p_a(z)}
It is straightforward to check that these currents satisfy
the expected WZW OPE's for the
affine Kac-Moody $SU'(2|2)$ given above.  
As explained in \gotubo\ and \mutto, 
the left and right
spacetime-supersymmetry generators are related to $S_a^\a$ and 
$\bar S_a^\a$ by 
$$q_a^+ =-i( S_a^1 + i e^\phi S_a^2) ,\quad  q_a^- = S_a^2,$$
$$\bar q_a^+ = -i(\bar S_a^1 + i e^{\bar\phi}\bar S_a^2) ,
\quad  \bar q_a^- = \bar S_a^2.$$

For application to string theory, we also must describe
the global $N=4$ algebra, introduced for $\R^6$ times K3 or ${\bf T}^4$
 in section 4.  To replace $\R^6$ by ${\rm AdS}_3\times {\bf S}^3$, we make
 a relatively simple modification using the above-described fields.
Thus, we write the $N=4$ algebra as follows:
$$T= T_{SU'(2|2)}
+\half\partial\rho\partial\rho
+\half\p\sigma\p\sigma
+{3\over 2}\p^2 (\rho+i\sigma)+T^{GS}_{C}$$
$$ G^{+} = 
- e^{-2\rho-i\sigma} (p)^4 + ~{i\over 2}
e^{-\rho} \left(p_a p_b K^{ab}+{p_a\partial p_a\over 2k}\right) $$
$$
+  e^{i\sigma}(T_{SU'(2|2)} +\half\p(\rho+i\sigma)\p(\rho+i\sigma)
-\half\p^2(\rho+i\sigma))
+ G^{+~GS}_{C}, $$
$$G^{-}  =
e^{-i\sigma} ~+ G^{-~GS}_{C},$$
$$J=\p(\rho+i\sigma)~+J_{C}^{GS}, $$
$$\tilde G^{+}  =
e^{J_{C}^{GS}+\rho} +e^{\rho+i\sigma}\tilde G^{+~GS}_{C},$$
$$\tilde G^{-}=
e^{-J_{C}^{GS}}( -e^{-3\rho-2i\sigma} (p)^4 ~ +{i\over 2}
e^{-2\rho-i\sigma}\left(p_a p_b K^{ab}+{p_a\partial p_a\over 2k}\right)
$$
$$ + e^{-\rho}(
T_{SU'(2|2)} +\half\p(\rho+i\sigma)\p(\rho+i\sigma)-\half\p^2(\rho+i\sigma))) +
e^{-\rho-i\sigma}\tilde G_{C}^{-~GS},$$
$$J^{++}=e^{\rho+i\sigma}~J_{C}^{++~GS}, $$
$$J^{--}=e^{-\rho-i\sigma}~J_{C}^{--~GS}  $$
where $T_{SU'(2|2)}$ is the stress tensor for the $SU'(2|2)$ 
WZW model
$$T_{SU'(2|2)}={\epsilon_{abcd}K^{ab}K^{cd}\over 8k}+{\epsilon_{\alpha
\beta}S^{a\alpha}
S^{a\beta}\over 2k}.$$
Most of these formulas are obtained from the $\R^6$ case by replacing
$dx$ by $K$ in the obvious way; however, the terms involving $p\partial p/2k$
in $G^+$ and $\tilde G^{-}$ are new.  These terms arise
 from a multiple contraction in the
current algebra.  The verification of supersymmetry and holomorphy
of the constraints is analogous to the flat space case reviewed at the
beginning of section 9.

\vskip 15pt

{\bf Acknowledgements:} We would like to thank M. Bershadsky, L. Dolan,
P. Etingof, V. Kac, G. Sotkov and 
A. Strominger for valuable discussions.
NB would also like to thank Harvard University,
the Institute for Advanced Study, and Rutgers University for their
hospitality, and would like to thank CNPq grant 300256/94-9 and
FAPESP grant 98/04086-5 for partial financial support.
Research of CV was supported in part by NSF grant PHY-98-02709.
Research of EW was partly supported by NSF Grant PHY-9513835.
\listrefs
\end